\documentclass[a4paper,10pt]{article}
\usepackage{tikz}
\usepackage{textcomp}
\usepackage{gensymb}
\usepackage[labelformat=simple]{subcaption}

\usepackage{multirow}
\usepackage{makecell}
\usepackage{authblk}
\usepackage{xurl}
\usepackage{hyperref}
\usepackage{amsthm}
\usepackage{amsmath}
\usepackage{amssymb}
\usepackage{geometry}
\usepackage{rotating}
\newtheorem{definition}{Definition}

\title{Analysis of Spatial-temporal Behavior Pattern of the Share Bike Usage during COVID-19 Pandemic in Beijing}
\author{}
\author[a,b,c]{Xinwei Chai}
\author[a,*]{Xian Guo}
\author[c]{Jihua Xiao}
\author[a]{Jie Jiang}
\affil[a]{Beijing University of Civil Engineering and Architecture, 102616 Beijing, China; guoxian@bucea.edu.cn (X.G.); jiangjie@bucea.edu.cn (J.J.)}
\affil[b]{BeiDou Navigation \& LBS (Beijing) Co., Ltd, 100191 Beijing, China; xw.chai@chinalbs.org (X.C.); jh.xiao@chinalbs.org (J.X.)}
\affil[c]{Beijing Unistrong Science \& Technology Co., Ltd, 100176 Beijing, China; xw.chai@unistrong.com (X.C.)}
\date{}
\graphicspath{{Figures/base/}}
\begin{document}
\maketitle
\begin{abstract}
During the epidemics of COVID-19, the whole world is experiencing a serious crisis on public health and economy.
Understanding the human mobility during the pandemic helps one to design intervention strategies and resilience measures.
The widely used Bike Sharing System (BSS) can characterize the activities of urban dwellers over time \& space in big cities but is rarely reported in epidemiological research.
In this paper, we present a \textit{human mobility analyzing framework} based on BSS data, which examines the spatiotemporal characteristics of share bike users, detects the key time nodes of different pandemic stages, and demonstrats the evolution of human mobility due to the onset of the COVID-19 threat and administrative restrictions. 
We assessed the \textit{net} impact of the pandemic by using the result of co-location analysis between share bike usage and POIs (Point Of Interest).
Our results show the pandemic reduced the overall bike usage by 64.8\%, then an average increase (15.9\%) in share bike usage appeared afterwards, suggesting that productive and residential activities have partially recovered but far from the ordinary days.
These findings could be a reference for epidemiological researches and inform policymaking in the context of the current COVID-19 outbreak and other epidemic events at city-scale.

\end{abstract}


\newcommand{\mainBasic}[1]{\node (a) {\includegraphics[width=\textwidth]{#1}};}
\newcommand{\legendBar}[1]{\node at (a.south west)[anchor=south west]{\includegraphics[scale=0.2]{#1}};\northArrow}
\newcommand{\northArrow}{\node at (a.north east)[anchor=north east, xshift=-1.5mm, yshift=-1.5mm]{\includegraphics[width=0.05\textwidth]{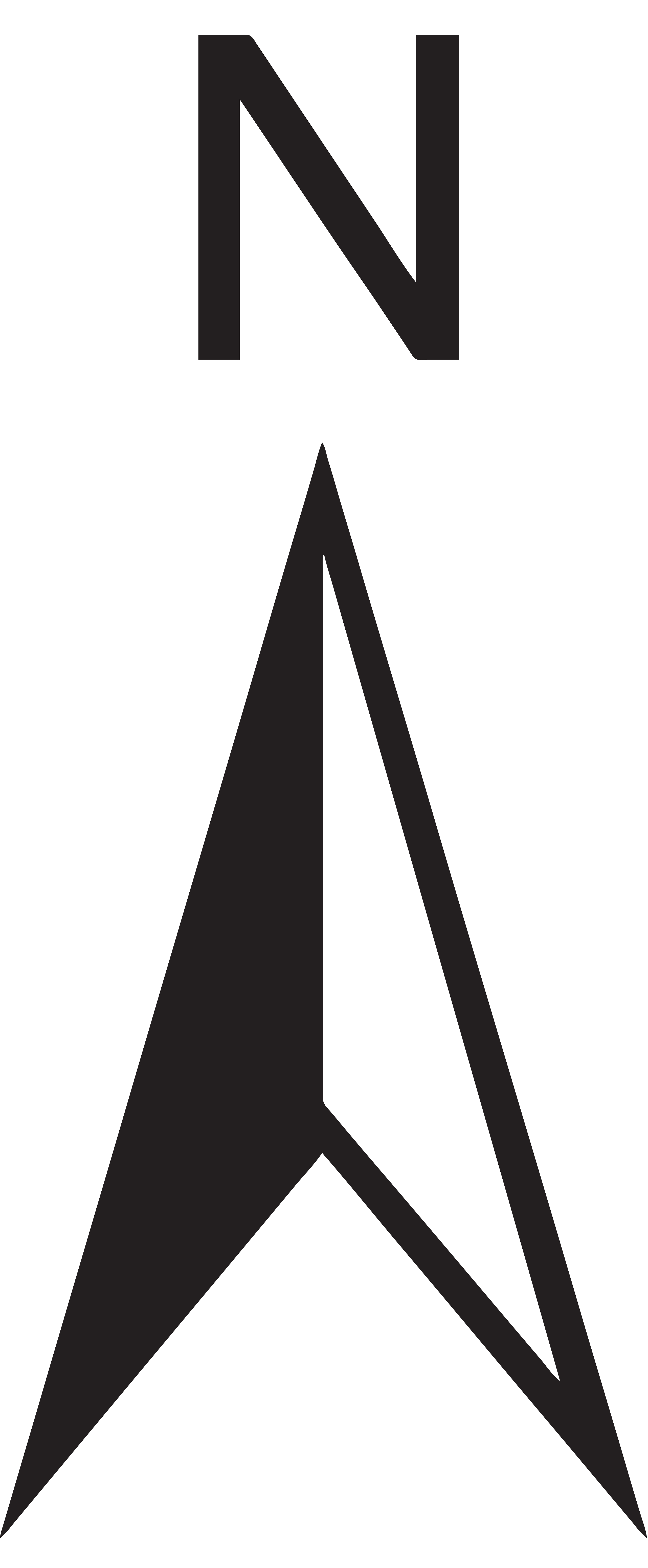}};}
\section{Introduction}
COVID-19 is a rapidly spreading infectious disease caused by the novel coronavirus SARS-COV-2, which has now triggered a global pandemic\footnote{\url{https://www.who.int/emergencies/diseases/novel-coronavirus-2019/technical-guidance/naming-the-coronavirus-disease-(covid-2019)-and-the-virus-that-causes-it}}.
According to a situation report of WHO\footnote{\url{https://www.who.int/docs/default-source/coronaviruse/situation-reports/20200422-sitrep-93-covid-19.pdf}}, the worldwide total confirmed cases have reached 2,471,136 including 84,287 that of China as of 22 April 2020.
COVID-19 pandemic inflicts a huge impact on public health and most economic sectors, from the beginning of Chinese New Year holiday till May 2020.
Under the threat of the pandemic, productive and social activities of residents were inevitably influenced \cite{Akter2019Big, Wang2016Research}. 
Measuring the changes in human mobility dynamics is essential for transmission prediction, control measure design, and post-pandemic recovery.

Current COVID-19-related epidemiological studies mainly focus on transmission dynamics \cite{li2020early, pitzer2009demographic} and preventive measures \cite{boulos2020geographical, chinazzi2020effect, van2006today}, from a local or global perspective \cite{mollalo2020gis, yang2020taking}.
Nevertheless, few studies investigate the spatiotemporal dynamics quantitatively during the pandemic. 
Ferguson \textit{et al.}\ \cite{ferguson2020report} studied two main strategies of non-pharmaceutical interventions in infectious disease prevention: \textit{suppression}, case of China and South Korea, with enormous social and economic costs which may cause secondary disasters on health and well-being in short and longer-term;
\textit{mitigation}, case of Great Britain and the United States, may not be able to protect residents at risk from severe diseases and resulting in high mortality.
Fang \textit{et al.}\ \cite{Fang2020human} provided a causal interpretation of the impact of city lockdown on human mobility and the spread of COVID-19 in China.
Mollalo \textit{et al.}\ \cite{mollalo2020gis} developed nationwide geographic modeling of COVID-19 and investigated the county-level variations of COVID-19 incidence across the United States.
With the help of the mobile internet user data from Baidu Mobility platform, Mu \textit{et al.}\ \cite{mu2020interplay} examined the interplay between disease spread of COVID-19 and inter- \& intra-city mobility in 319 Chinese cities.
However, few studies focus on the dynamics of human mobility at a local scale during a long-term pandemic.

Geospatial big data have great potential to improve disease surveillance and disaster response \cite{Goodchild2010Crowdsourcing, Horanont2013Auto, Huang2015Predicating, Yu2018Big}.
In this paper, we focus on the spatiotemporal changes in human mobility under the effect of the COVID-19 pandemic at city-scale.
Even though online surveys and mobile phone positioning data could be referential \cite{CHAE2013Public, Wesolowski2012Quantifying, Wilson2016Rapid}, one can hardly distinguish purposive movements from random wandering/indoor movements in these data.
Moreover, they do not cover all the population especially those who care about their privacy and are thus not willing to offer their precise location \cite{Li2016ISPRS}.
In addition, after the outbreak of COVID-19, social distancing and home quarantine, were imposed for pandemics prevention. 
Also, there was a suspension of buses and taxis for a short time after the outbreak, because they form a public enclosed space.
These strict control measures inevitably narrowed the options of public transit.
For the above reasons, the wide-spread Bike Sharing System (BSS) in China became an alternative to fulfill people's need in short-distance transportation and it became a data source to analyze human mobility \textit{at city-scale} during the pandemic period.

Thanks to the rapid development of GIS- and IoT-based system, the 3rd-generation BSS (\textit{a.k.a} free-floating/dockless BSS) emerges in China in 2015.
Compared to its predecessors, the 3rd-generation BSS (from now on referred to as BSS) is no longer constrained by docking stations.
They are often spread along roads and cover most of the urban residents. 
In the city of Beijing, for example, the number of share bikes reached its peak in 2017 and the municipal government began removing excess bike supply afterwards\footnote{\url{http://ebma-brussels.eu/bike-sharing-in-china/}}.
After two years of rapid development and regulation, the demand and supply of BSS reached a balance in 2019, which made it a stable data source.

BSS records contain OD (origin-destination) information and timestamps from anonymous users, offering a promising alternative data source revealing the spatial and temporal information of outdoor activities of residents.
According to Daxue Consulting\footnote{\url{https://daxueconsulting.com/mobike-and-ofo-bike-sharing/}}, in Beijing, 93\% of travels less than 5 km are quicker done by bike and public transport than by car. 
Chen \textit{et al.}\ \cite{chen2020exploring} stated that approximately half of the population in Beijing were registered as dockless bike share members in 2017, suggesting that this large user base and the easy access to share bikes have made BSS data suitable for characterization of human mobility.
Du \textit{et al.}\ \cite{du2018better} studied the travel patterns of BSS in Nanjing \textit{via} limited questionnaires.
Xu \textit{et al.}\ \cite{xu2019unravel} characterized the temporal flow and spatial distribution of share bikes in Singapore.
Kaggle organized a competition of predicting share bike demands\footnote{\url{https://www.kaggle.com/c/bike-sharing-demand}} based on limited entries.
There are also studies focusing on BSS rebalancing strategies \cite{ai2019deep, chen2016dynamic, pal2017free}.
However, the use of BSS data in the studies of pandemic response is still at its early stages. 

To address the lack of understanding of the spatiotemporal dynamics of city-scale human mobility influenced by COVID-19 pandemic, we put up the following objectives: 
\begin{enumerate}
\item Determining the pandemic period and measuring period-wise changes in human mobility;
\item Assessing the net pandemic impact and the rehabilitation progress.
\end{enumerate}

We achieved these objectives by constructing a \textit{human mobility analyzing framework}.
It managed to demonstrate spatiotemporal patterns \textit{intuitively} and \textit{computationally}.

\begin{itemize}
    \item The intuitional part helps one to understand quickly the situation by showing the timeline, different phases of the whole pandemic, and basic statistics of POI categories revealing roughly the severity of the pandemic.
    \item The computational part gives a more quantitative vision.
    We first divided the study period (\textit{i.e.}, Jan, 2020 to Mar, 2020) into several pandemic periods \textit{via} a \textit{k-segmentation} approach, which inferred the epidemic stages based on temporal characteristics of human mobility from BSS dataset.
    Afterward, we assessed the net impact of the COVID-19 pandemic \textit{via} a \textit{DID (Difference-In-Differences) model} with the long-time-sequenced BSS data dating from 2019. 
    Among numerous candidate explanatory variables, we removed the effect of the Chinese New Year and weather factor and quantified the impact of the epidemic on human mobility. 
    Finally, in different pandemic periods, we implemented a \textit{co-location analysis} between the share bike usage and different types of POIs, which demonstrated the evolution of mobility due to the onset of the COVID-19 threat and assessed the rehabilitation progress corresponding to different urban functions.
\end{itemize}

The main novelty of our work is: quantifying and assessing the \textit{net} impact of COVID-19 pandemic on general cities (outside the outbreak zone) from spatiotemporal perspective \textit{via} the human mobility extracted from long-time-sequenced share bikes records. 
As far as we know, no previous study has investigated the impact of epidemics based on BSS data.
The results are at city-level, which could provide fine-scaled references for policymaking and epidemiological researches.

The rest of this paper is organized as follows. 
Section \ref{sec:data} \& \ref{sec:method} describe the study area and research methods.
Section \ref{sec:experimental results} reports the spatiotemporal characteristics of the share bike usage and shows DID results.
Section \ref{sec:discussions} presents a co-location analysis with POIs, which quantifies the pandemic influence and degree of rehabilitation towards different urban functions.
Section \ref{sec:conclusion} concludes with some remarks and Section \ref{sec:future} states the outline of our future research.

\section{Study Area and Data}\label{sec:data}

\begin{figure}[ht]
    \centering
    \includegraphics[width=\textwidth]{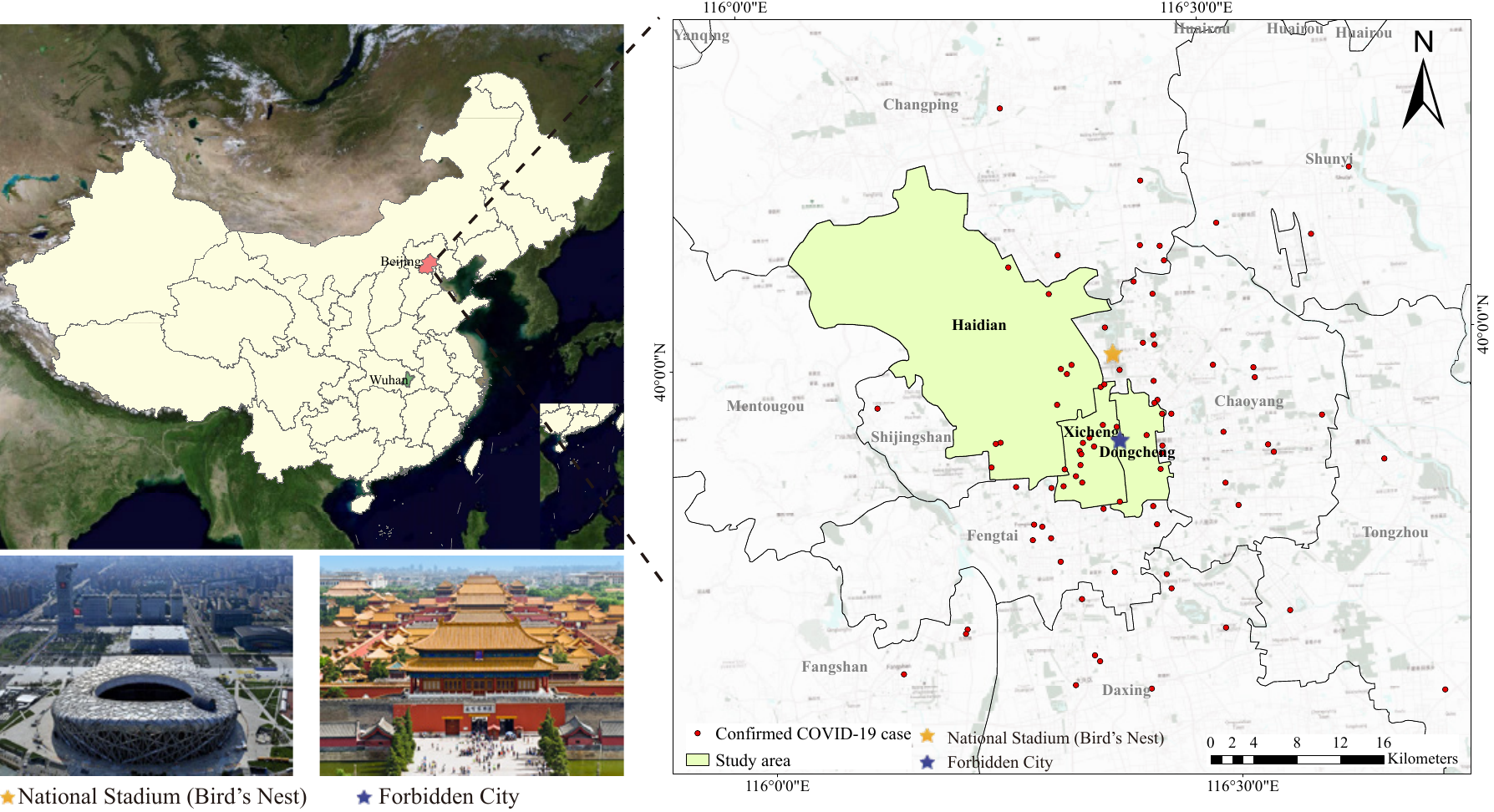}
    \caption[Study area]{Study area (left) and 87 confirmed infected cases in Beijing by 05 Mar, 2020 (right).}
    \label{fig:study_area}
\end{figure}

\subsection{Study Area}
This study was conducted in the city of Beijing, the capital of China.
As shown in Figure~\ref{fig:study_area}, Beijing is located at the North China Plain, occupying an area of 16,411 km$^2$ (39.4{\degree}-41.6{\degree}N, 115.7{\degree}-117.4{\degree}E).
In 2019, the municipal population of Beijing has reached 21.53 million.
The cumulative total COVID-19 confirmed cases in Beijing has reached 418 by 05 Mar, 2020\footnote{\url{http://wb.beijing.gov.cn/home/ztzl/kjyq/fk_yqtb/202003/t20200305_1681121.html}}.
As a metropolis with a huge population of immigrants, Beijing has taken measures in response to the outbreak, such as holiday extension, executive orders like ``stay at home'' and  ``working from home''.
Under this circumstance, the human mobility was influenced, showing spatiotemporal patterns different from ordinary days.

\subsection{Data Sets}\label{subsec:dataset}
Four data sets were used in our study: BSS records, Points Of Interest (POIs), confirmed COVID-19 cases, and weather records.
\begin{enumerate}
\item BSS records

This OD dataset came from 1.02 million share bikes belonging to 4 main BSS operators (Mobike, DiDi Bike, Hellobike, and Ofo) in Beijing.
The records date from Mar, 2019 to Mar, 2020 (66.8 GB) and cover 1.5 million uses per day contributed by 11 million users, which account for half of the total population of Beijing.
They were created when users locked/unlocked their share bikes, excluding that of rebalancing operations.
This exclusion guarantees that the records come purely from users.
Moreover, the BSS data is anonymous, which does not cause privacy concerns.
It should be noted that the BSS records in certain districts (Chaoyang, Fengtai and Shijingshan) are not available due to different policies of local governments.

\item Points Of Interest (POIs)

POIs of Beijing are collected from AutoNavi API provided by Gaode Maps\footnote{\url{https://opendata.pku.edu.cn/dataset.xhtml?persistentId=doi:10.18170/DVN/WSXCNM}}, one of the most popular web mapping platforms in China.
Each entry has a unique POI ID \texttt{object\_id}, an address including longitude/latitude information, and a three-level category \texttt{large\_category}, \texttt{mid\_category} and \texttt{sub\_category}.
Among hundreds of categories, we chose seven mid ones: residential area (RA), high-tech company (HC), other company (OC), subway station (SS), shopping plaza (SP), supermarket (SM) and tertiary hospitals (TH)\footnote{Tertiary hospitals are considered as the top-class hospitals in China.}, as HC and OC reflect productive activities, SP, SM, TH reflects social activities, and RA, SS reflect both activities.

\begin{figure}[ht]
    \centering
    \begin{subfigure}{.23\textwidth}
        \frame{
        \begin{tikzpicture}[inner sep = 0pt]
        \node (a) {\includegraphics[width=\textwidth]{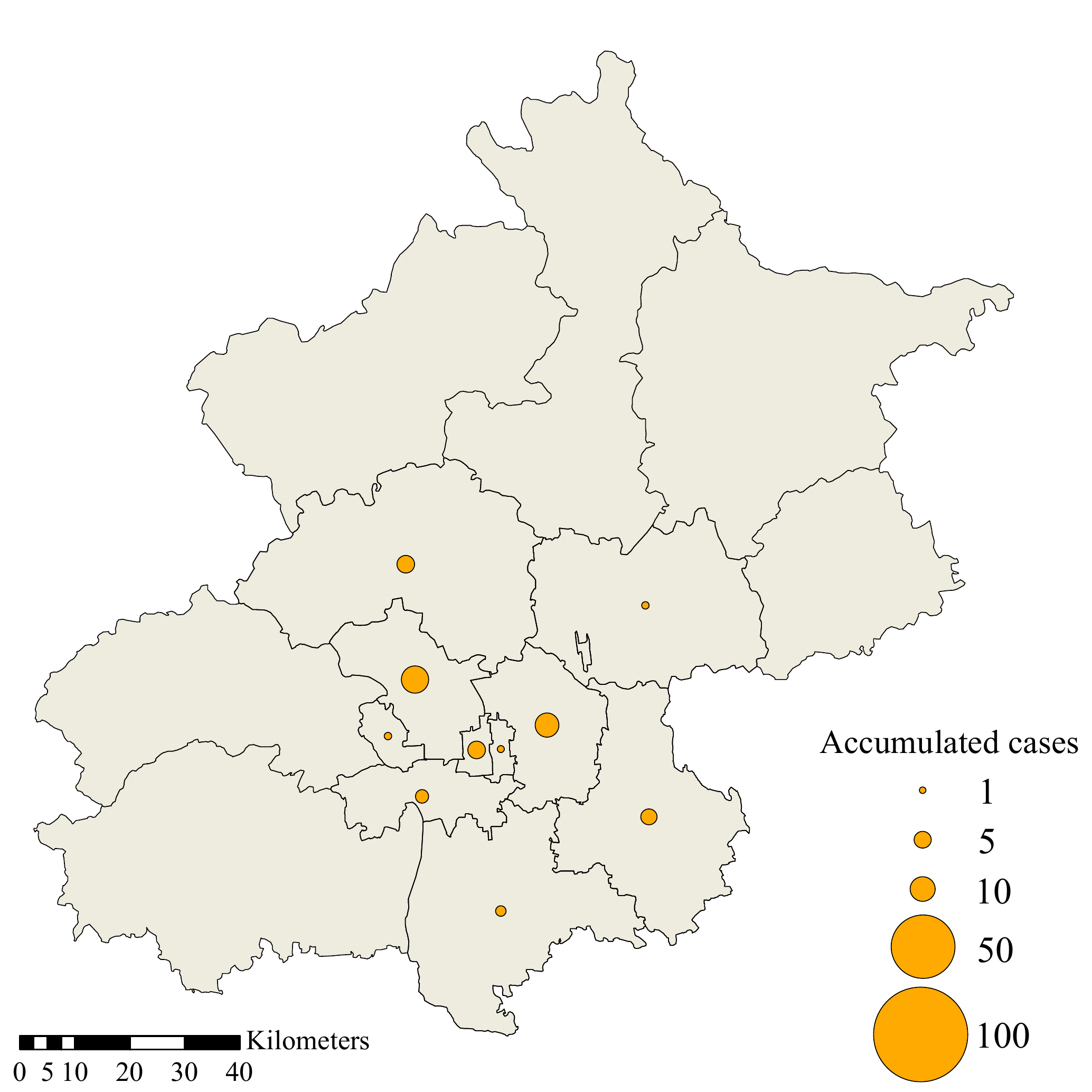}};
        \northArrow
        \end{tikzpicture}}
        \caption{25 Jan, 2020}
    \end{subfigure}
    \begin{subfigure}{.23\textwidth}
        \frame{
        \begin{tikzpicture}[inner sep = 0pt]
        \node (a) {\includegraphics[width=\textwidth]{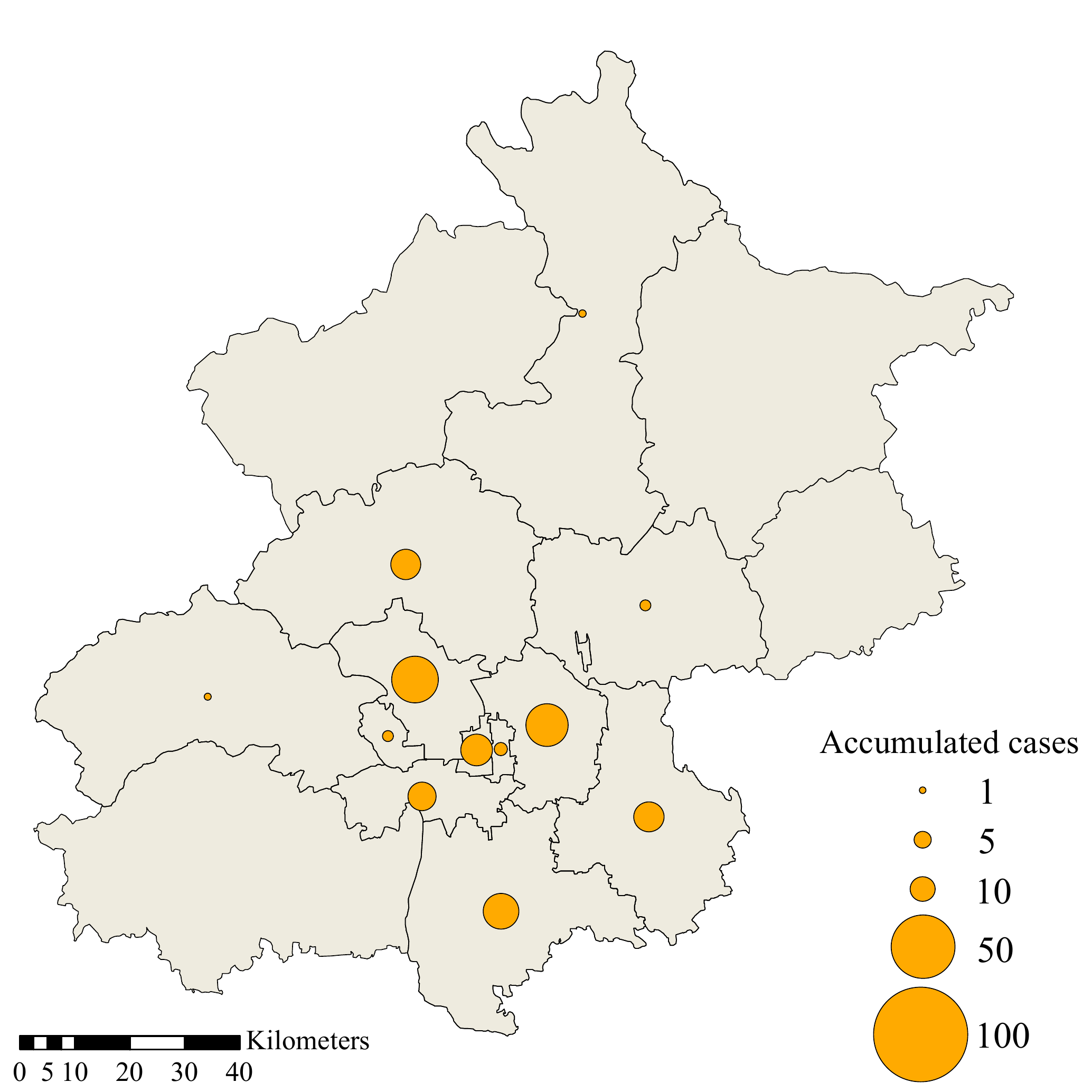}};
        \northArrow
        \end{tikzpicture}}
        \caption{30 Jan, 2020}
    \end{subfigure}
    \begin{subfigure}{.23\textwidth}
        \frame{
        \begin{tikzpicture}[inner sep = 0pt]
        \node (a) {\includegraphics[width=\textwidth]{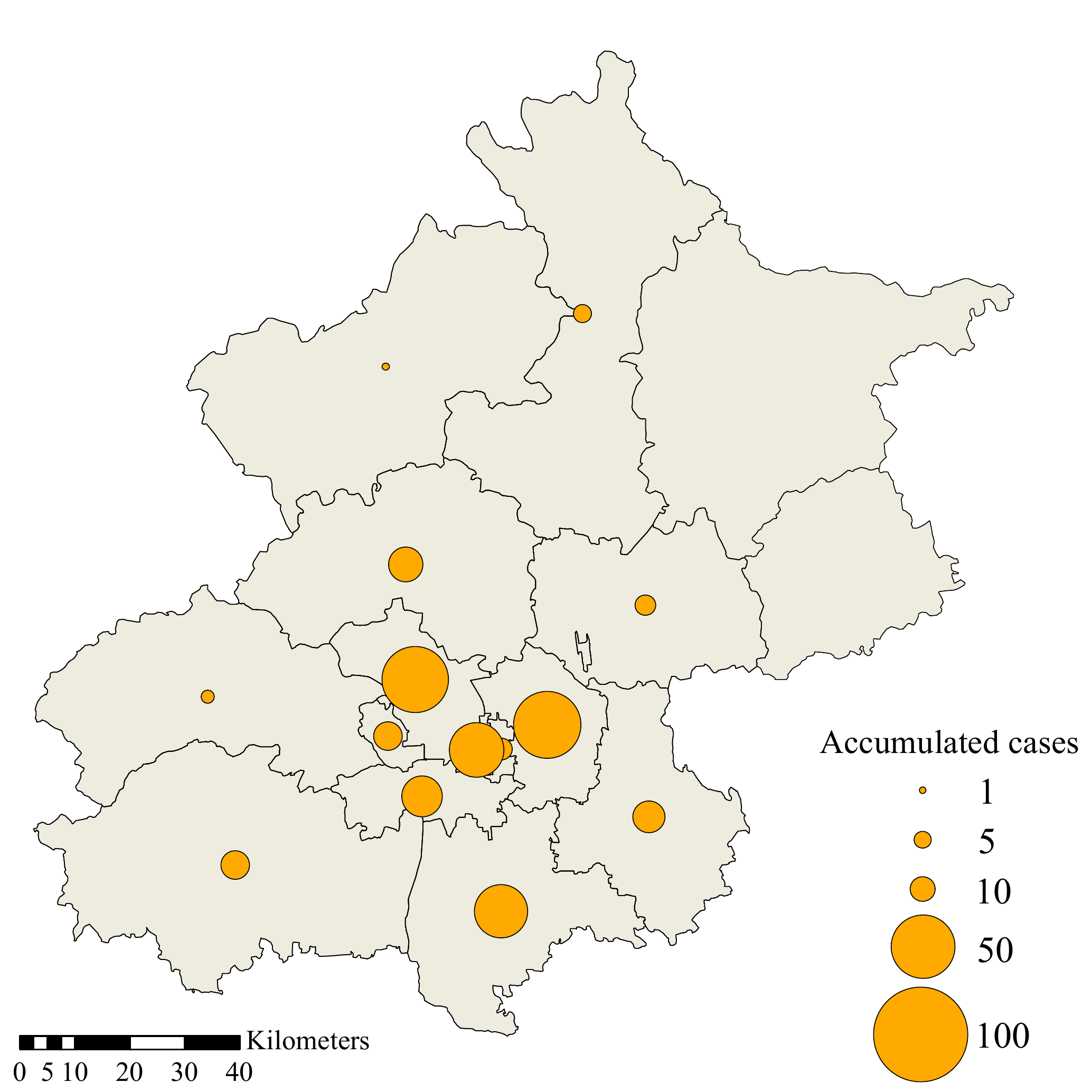}};
        \northArrow
        \end{tikzpicture}}
        \caption{05 Feb, 2020}
    \end{subfigure}
    \begin{subfigure}{.23\textwidth}
        \frame{
        \begin{tikzpicture}[inner sep = 0pt]
        \node (a) {\includegraphics[width=\textwidth]{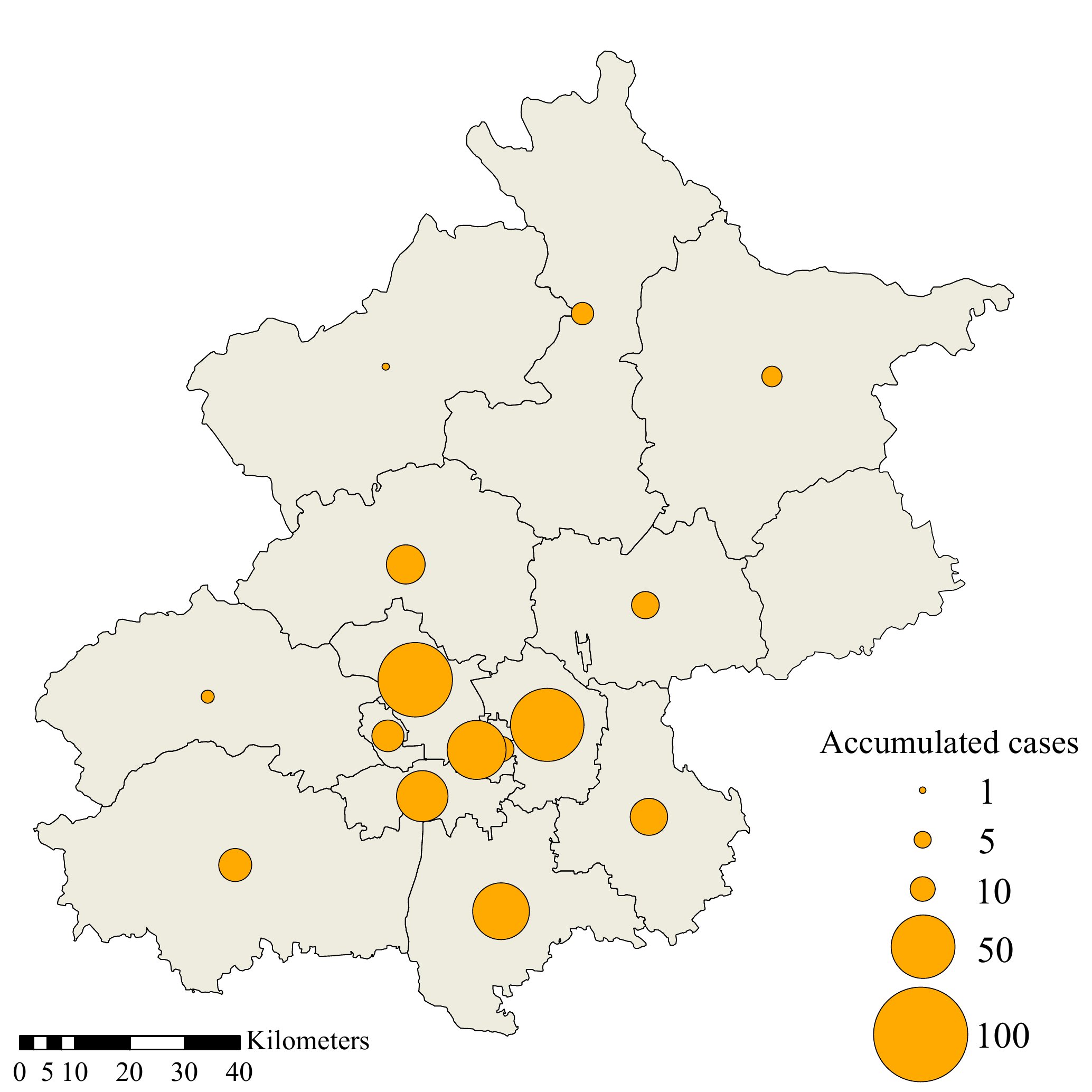}};
        \northArrow
        \end{tikzpicture}}
        \caption{10 Feb, 2020}
    \end{subfigure}

    \caption[Accumulated confirmed cases]{Snapshots of accumulated confirmed cases in Beijing from 25 Jan to 10 Feb, 2020.}
    \label{fig:number_of_confirmed_cases}
\end{figure}

\item Confirmed COVID-19 Cases

The cumulative daily counts of clinically diagnosed cases in each district from 20 Jan, to 05 Mar, 2020 were collected from the daily update on the COVID-19 outbreak dashboard provided by Foreign Affairs Office of the People's Government of Beijing Municipality\footnote{\url{http://wb.beijing.gov.cn/home/ztzl/kjyq/}}.
We picked out a total of 87 infected residential areas (see Figure~\ref{fig:study_area}).
According to the timeline \cite{li2020early} of the outbreak, we visualized the evolution of the overall pandemic situation of Beijing in Figure~\ref{fig:number_of_confirmed_cases}.

\item Weather data 

Weather data came from China Meteorological Data Service Center\footnote{\url{http://data.cma.cn/en}}, containing daily weather information such as temperature, wind speed, and precipitation, dating from 01 Jan, to 05 Mar of 2019 and 2020.

\end{enumerate}


\section{Tools and Methodology}\label{sec:method}

\subsection{Tools}
To deal with large-scale spatial queries on the huge BSS dataset (66.8 GB), we need to apply parallel computing all along this study:
\begin{enumerate}
    \item HDFS (Hadoop Distributed File System)\footnote{\url{https://hadoop.apache.org/docs/r1.2.1/hdfs_design.html}}: a distributed file system which is suitable for parallel computing.
    \item Spark\footnote{\url{https://spark.apache.org/}}: a parallel analytic engine for big data, and can invoke SQL to process temporal queries in BSS data, performing denoising and statistical analysis.
    \item GeoSpark \cite{huang2017geospark}: a GIS-based engine based on Spark, capable of performing spatial analysis and visualization of geo-based data.
\end{enumerate}

Python is our primary programming language, and cartographic visualization is done by ESRI ArcGIS 10.7\footnote{\url{https://www.esri.com/en-us/arcgis/about-arcgis/overview}}.
All computation in this paper was run on a computer cluster consisting of 7 machines with Intel\textregistered Xeon\textregistered, CPU E5-2640 v2 @2.00 GHz, 8 cores, 61.7 GB RAM, 20 MB cache. 

\subsection{Methodology}
\begin{figure}[ht]
    \centering
    \includegraphics[width=\textwidth]{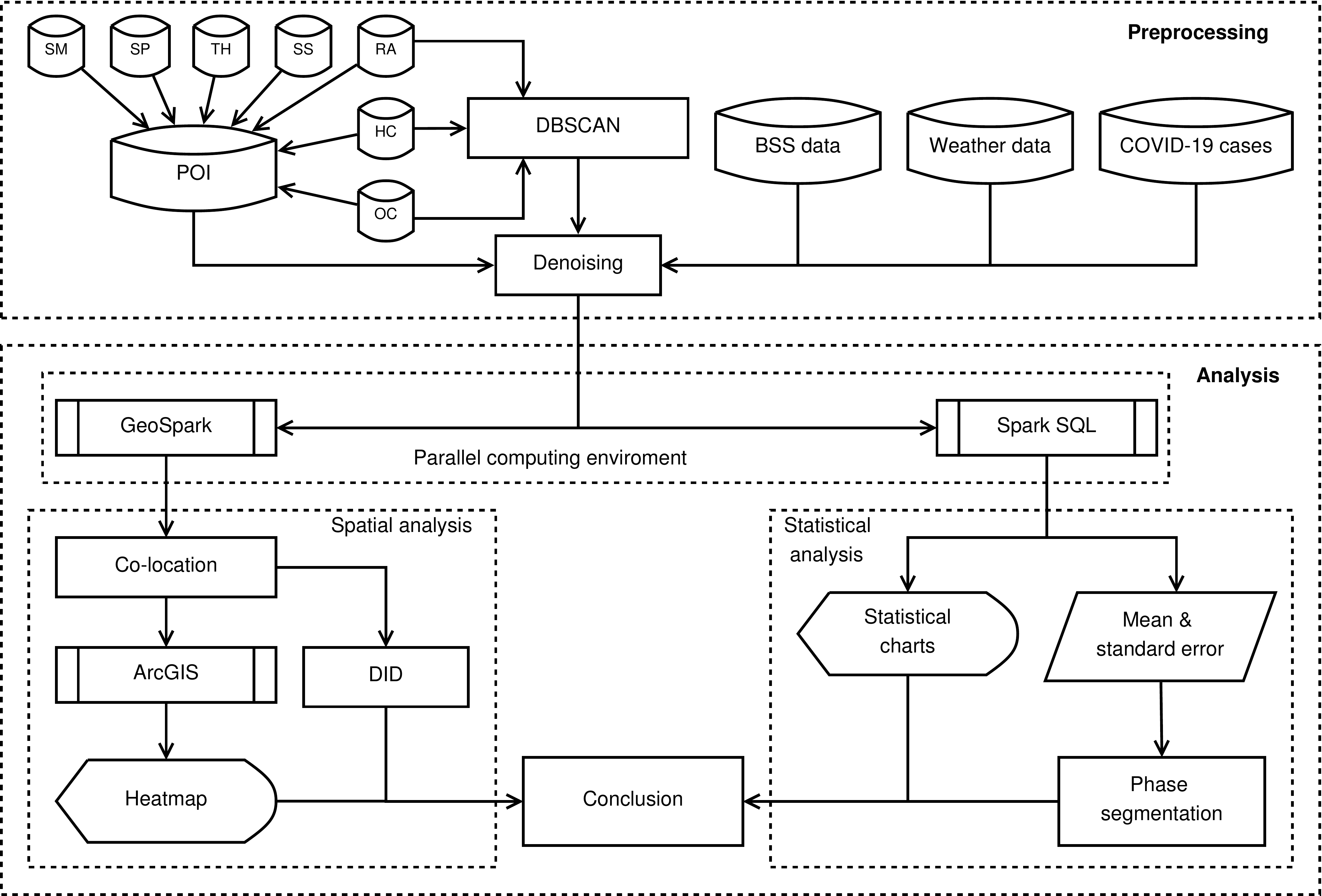}
    \caption[Workflow of the human mobility analyzing framework]{Workflow of the human mobility analyzing framework. After clustering and denoising the raw data, the main process walks into two branches, spatial analysis and statistical analysis with the help of parallel computing tools. Spatial analysis contains a first-step co-location analysis, giving results to ArcGIS for visualization and to DID for quantitative analysis. The branch of statistical analysis also runs with the visual + numerical logic: statistical charts for visualization and phase segmentation for quantitative analysis. We then combine the results of the two branches and reach the final conclusion.}
    \label{fig:diagram}
\end{figure}

This study aims at analyzing quantitatively the impacts of COVID-19 on Beijing from spatiotemporal perspective \textit{via} the human mobility extracted from long-time-sequenced BSS data.
Figure~\ref{fig:diagram} shows the workflow of human mobility analyzing framework.
In the Preprocessing block, BSS data and other data sets were stored in a spatial database and denoised before use.
Certain POI types were clustered using DBSCAN (Density-Based Spatial Clustering of Applications with Noise) algorithm~\cite{ester1996density}, as they may be located close to each other, causing duplicate patterns in further analysis. 

In the Analysis block, there are mainly five tasks:

\begin{enumerate}
    \item Statistical charts: a first-step visualization of the statistics of share bike data; 
    \item Phase segmentation: using k-segmentation approach to divide the whole study period into logical phases, \textit{i.e.}, pandemic phases;
    \item Co-location analysis: measuring share bike usage near different POIs in different phases;
    \item Heatmap: visualization of the share bike data in the aspect of space + time;
    \item DID (Difference-In-Differences): quantitative analysis of the share bike data.
\end{enumerate}


\subsection{Data Preprocessing}\label{sec:preprocessing}
Our raw datasets need preprossessing to fit further analysis.
We first removed the outliers, \textit{e.g.} bikes ``in the water'', temperature of 999 \textcelsius...
We also had to deal with the duplicate positions.
HC, and OC are of high spatial density because multiple companies may co-locate in the same/adjacent office building.
To avoid repeated counting of POI-bike co-located pairs, we used POI clusters for these three categories instead.
We clustered nearby POIs using DBSCAN algorithm, which has three merits:
\begin{itemize}
    \item No need to specify the number of clusters;
    \item Allowing clusters of all forms and sizes;
    \item Dealing with noise and outliers.
\end{itemize}

We configured the DBSCAN parameter as follows: $\varepsilon=100$ m, the maximum distance between two samples and $min\_samples=10$, the number of samples in a neighborhood for a point to be considered as a core point of a cluster.

Among the seven chosen POI categories, RA is the only ``area of interest'' (not ``point of interest'').
The records of RA contain not only their geometric centers but also their contours which is a key factor of further results.
To solve such problem and unify the input of the whole procedure, we used the buffer zones of POIs to replace the original POIs in the co-location analysis.

\subsection{Co-location Analysis}\label{sec:co-location}
Co-location analysis depicts the co-location patterns of share bike usage with adjacent urban functions.
We proceeded share bike data and POIs with GeoSpark, which performs spatial queries \textit{via} parallel computing on Spark.
Among numerous spatial queries, we utilized \texttt{spatial\_join(geom1, geom2)}, which queries if object \texttt{geom1} is inside \texttt{geom2} and \texttt{distance\_join(geom1, geom2, dist)} which queries if the distance of \texttt{geom1} and \texttt{geom2} is less than \texttt{dist}.
As these queries are partition-independent except for the points near the borders of partitions, the parallel portion is high if the data are well indexed.

\subsection{Phase Classification}\label{sec:phase_classification}
Generally, the evolution of an epidemic can be empirically divided into several stages according to the timeline of the outbreak.
Apart from empirical division, we segmented the study period (two months) in a more logical way with a phase classification strategy based on temporal characteristics of human mobility from BSS dataset.
In machine learning tasks, one often aims at minimizing the predefined loss function to perform the best classification such that elements within the same cluster are similar, and elements across clusters are different.
Likewise, we tried to segment the study period into phases with the most resembling patterns using k-segmentation.
\begin{definition}[k-segmentation]\label{def:k-seg}
    Let $X=\{x_1,x_2,\ldots,x_N\}$ be a time series of length $N$.
    Given $k\in \mathbb{N}$, $k<N$ and index set $\mathbf{T}=\{n_0,\ldots,n_k\}$ with $n_0=0$, $n_k=N$ and $\forall i$, $n_i<n_{i+1}$, a k-segmentation of $X$ is the set of time series $X_i=\{x_{n_i+1},\ldots,x_{n_{i+1}}\}$ where $0\leq i\leq k-1$.
\end{definition}
To evaluate a k-segmentation, we use $\sigma=\sum_{i=1}^{k}{\sigma_i}$ as the loss function where $\sigma_i$ is the standard deviation of division $X_i$. 
The goal is to find the best $\mathbf{T}$ to minimize $\sigma$, \textit{i.e.,}\ $\arg\min_{\mathbf{T}}\sigma(\mathbf{T})$.
This problem can be solved at the complexity level of $O(N^2k)$ \cite{terzi2006efficient}.
In case $k$ and $N$ are small, the optimum can be found \textit{via} exhaustive search.

\subsection{Difference-In-Differences (DID)}\label{sec:did_method}
Usually, it is difficult to find and quantify all the factors of a certain event.
DID is a techinique which tries to mimic an experimental research design using observational study data by constructing a ``treatment group'' and a ``control group'', under the assumption of ``common trend'' (the two groups will follow the same trend if no treatment is done) \cite{card1994minimum}.
The treatment (effect) can be extracted because all the effect of common factors are included in the ``common trend''. 
We applied DID to distinguish the impact of epidemic effect from other effects, \textit{e.g.}, the effect of Chinese New Year.
Table~\ref{tab:didConfig} shows the configuration of the DID analysis.
Assume we have a ``virtual pandemic'' just after Chinese New Year of 2019, we set $T$ and $D$ as binary variables, with $T$ indicating whether the year is 2020, and $D$ indicating whether the study period is during the pandemic.
The ``real'' pandemic is present only when $T=1$ and $D=1$. 

\begin{table}[ht]
\centering
    \begin{tabular}{|l|c|c|}
        \hline
        $T\times D$ & before pandemic ($D=0$) & during pandemic ($D=1$)\\
        \hline
        2019 ($T=0$) & 0 & 0\\
        \hline
        2020 ($T=1$) & 0 & 1\\
        \hline
    \end{tabular}
    \caption[DID configuration]{DID configuration.}
    \label{tab:didConfig}
\end{table}

By including BSS dataset of 2019 as a control, the DID regression function is as follows:
\begin{equation}\label{did_eq}
\log(U_t)= \alpha + \beta_1 \cdot \mathit{Before_{2020,t}} + \beta_2 \cdot \mathit{During_{2020,t}} + \theta_t + \varepsilon_t
\end{equation}
where $t$ is the date, $U_t$ is the share bike usage on date $t$. $\mathit{Before_{2020,t}}$ and $\mathit{During_{2020,t}}$ are dummy variables.
$\mathit{Before_{2020,t}}=1$ if $t$ is 4 to 11 days before the outbreak of pandemic. This term is set to verify the common trend assumption in DID analysis.
$\mathit{During_{2020,t}}=1$ when $t$ is during the pandemic or in mitigation period (corresponding to $T\times D$ in Table~\ref{tab:didConfig}).
$\alpha$ is a constant term, $\beta_1$, $\beta_2$ are fitted coefficients, $\theta_t$ is the date fixed effects (weather, temperature, weekday/weekend, Chinese New Year, \textit{etc.}), and $\varepsilon_t$ is the residual term.
The effect of holiday and pandemic can be evaluated by $\beta_1$ and $\beta_2$.

\section{Results}\label{sec:experimental results}

\subsection{Temporal Characteristics of Share Bike Usage}
We chose several time points from WHO statements\footnote{\url{https://www.who.int/news-room/detail/27-04-2020-who-timeline---covid-19}} as shown in Table~\ref{tab:important_dates}. 
These important dates could track the virus transmission in Beijing: 

\begin{table}[ht]
    \centering
    \begin{tabular}{ll}
    04 Feb, 2019 & Start of Chinese New Year holiday 2019\\
    10 Feb, 2019 & End of new year holiday 2019\\
    07 Jan, 2020 & Identification of COVID-19\\
    22 Jan, 2020 & Shutdown of Wuhan and other 15 cities\\
    24 Jan, 2020 & Start of Chinese New Year holiday 2020\\
    02 Feb, 2020 & End of \textit{extended} New Year holiday 2020\\
    10 Feb, 2020 & Partial restart of productive and social activities
    \end{tabular}
    \caption[Important dates in the COVID-19 timeline]{Important dates before \& during the outbreak of COVID-19 in China}
    \label{tab:important_dates}
\end{table}

\begin{figure}[ht]
    \centering
    \includegraphics[width=\textwidth]{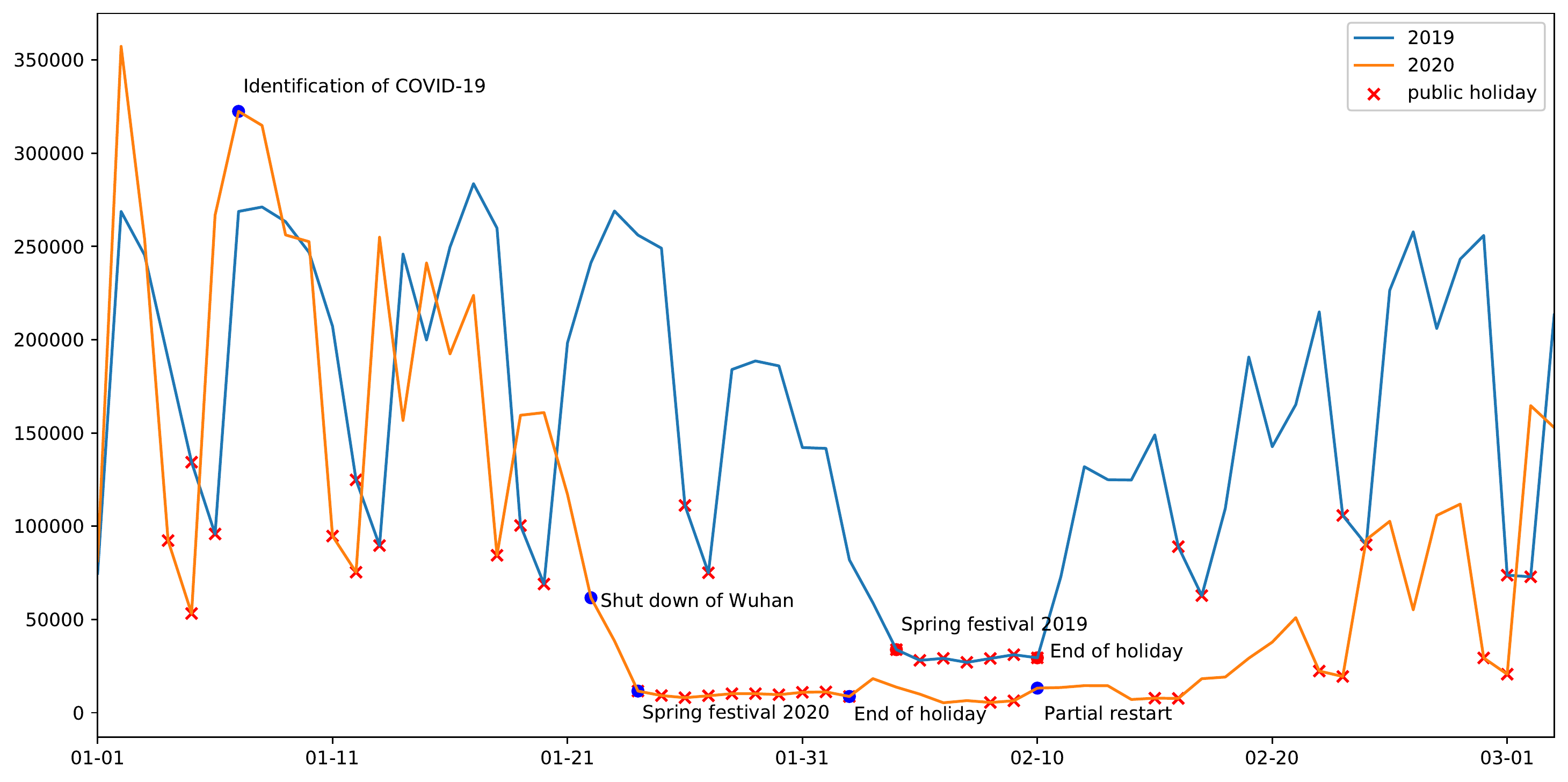}
    \caption[Share bike usage during rush hours]{Share bike usage during 08:00-09:00 of year 2019 and 2020. X-axis represents the date and y-axis denotes the magnitude of the bike usage.}
    \label{fig:hour_comparison_8}
\end{figure}

As the outbreak of COVID-19 coincided with the Chinese New Year holiday 2020, we used the Chinese New Year holiday of 2019 as a comparison to assess the influence of this period on share bike usage.
It is worth noticing that the Chinese New Year holiday of 2019 and 2020 are not equal-length, because that of 2020 was \textit{extended} by executive orders.
We compared the share bike usage on rush hours (8:00-09:00) during 64 days from 01 Jan to 02 Mar, 2019 and 01 Jan to 01 Mar, 2020 respectively (data of 29 Feb and 01 Mar, 2020 were brought backward by one day due to the leap year 2020).
We chose rush hours because this time interval corresponds to the highest bike usage frequency of the day and it reflects productive activities.
Figure~\ref{fig:hour_comparison_8} illustrates temporal evolution of share bike usage during the selected periods of 2019 and 2020 (POI-wise graphs are in Figure~\ref{fig:poi_wise} of Appendix~\ref{sec:figures}).

As shown in Figure~\ref{fig:hour_comparison_8}, share bike usage sharply dropped when the Chinese New Year holiday started.
During these periods, schools and workplaces were closed.
In 2020, the overall closure was forced to be extended to mitigate the pandemic.

The share bike usage on rush hours exhibits a periodicity in productive activities: high on weekdays and low on weekends, which fits our general knowledge.
However, this pattern does not match the week of 10 Feb, 2020 when the government declared the partial restart of certain productive and social activities.
This anomaly suggests that the activities were not resumed at least until 17 Feb, one week after the partial restart, corresponding to the fact that the impact of the pandemic was lasting till the end of our study period.

\textbf{Remark:} In this paper, we assume that share bike usage follows normal distribution, whose $95\%$ confidence interval is $[\bar{x}-2\sigma,\bar{x}+2\sigma]$, and data are presented in the form $\bar{x}\pm2\sigma$.

\begin{table}[ht]
    \centering
    \begin{tabular}{|l|l|l|l|}
        \hline
        \multirow{2}{*}{Phase} &\multicolumn{3}{c|}{Daily average share bike usage ($\times 10^{5}$)}\\
        \cline{2-4}
        & 08:00-09:00 (on weekdays) & 18:00-19:00 (on weekdays) & All-day\\
        \hline
        02 Jan-20 Jan, 2019 & $2.46\pm0.59$ & $1.30\pm0.39$ & $15.0\pm4.5$\\
        \hline
        02 Jan-20 Jan, 2020 & $2.58\pm1.10$ & $1.37\pm0.74$ & $12.7\pm6.3$\\
        \hline
        \hline
        & 08:00-09:00 (all week) & 18:00-19:00 (all week) & All-day\\
        \hline
        04 Feb-10 Feb, 2019 & $0.30\pm0.05$ & $0.24\pm0.08$ & $4.90\pm1.56$\\
        \hline
        24 Jan-02 Feb, 2020 & $0.10\pm0.02$ & $0.12\pm0.03$ & $1.72\pm0.35$\\
        \hline
    \end{tabular}
    \caption[Share bike usage in different time intervals]{Share bike usage in different time intervals, shown in the form of $\bar{x}\pm2\sigma$.}
    \label{tab:overall_comparison}
\end{table}

Table~\ref{tab:overall_comparison} gives a statistical view of aggregated bike usages during different time-intervals.
The upper part shows that on the rush hours of ordinary days, the share bike usage in 2020 is of the same order of magnitude as that of 2019, suggesting that share bike demand follows common trends.
However, the lower part shows the case of the Chinese New Year holiday, where the overall share bike usage dropped to less than 40\% compared with the same period in 2019, suggesting more companies stopped working due to the Chinese New Year holiday in 2020. 

However, the dates in Table~\ref{tab:important_dates} may not be a proper segmemtation, as there is no noticeable share bike usage change after 10 Feb.
We obtained a time series by computing the share bike usage within 100 m of each POI every day from 02 Jan to 02 Mar, 2020.
Then we segmented the time series into three phases in accordance with k-segmentation (Definition \ref{def:k-seg} in Section \ref{sec:phase_classification}).
We found the best-classified phases at $(k,N)=(3,62)$ with the minimum sum of standard deviation as metric \textit{via} exhaustive search shown in Table~\ref{tab:segment}.
The COVID-19 transmission phases in Beijing can be identified as: before pandemic (\textbf{phase a}, before 23 Jan), during pandemic (\textbf{phase b}, 24 Jan-24 Feb), pandemic mitigated (\textbf{phase c}, after 25 Feb).

\begin{table}[ht]
    \centering
    \begin{tabular}{|*{9}{l|}}
        \hline
        Category & HC & OC & RA & SS & SP & SM & TH & Overall\\
        \hline
        Split point 1 & 23 Jan & 23 Jan & 24 Jan & 24 Jan & 24 Jan & 24 Jan & 24 Jan & 23 Jan\\
        \hline
        Split point 2 & 24 Feb & 28 Feb & 24 Feb & 24 Feb & 24 Feb & 24 Feb & 24 Feb & 24 Feb\\
        \hline
    \end{tabular}
    \caption[Period segmentation]{Period segmentation of 02 Jan to 02 Mar, 2020}\label{tab:segment}
\end{table}

We also noticed a minor difference between the share bikes around HC,OC and that of others.
This difference can be explained by the lag between the end of work and the start of vacation.
Social and productive activities were not resumed until 24 Feb, two weeks after the official declaration of the partial restart.

\subsection{Spatiotemporal Evolution of Share Bike Usage}
In this section, we first demonstrated the spatiotemporal change in share bike usage during the pandemic.
Furthermore, a comparison of share bikes usage between 2019 and 2020 was conducted to show the influence of the pandemic roughly in different phases.
To evaluate its net impact, we applied DID method to remove the influence of other factors, \textit{e.g.}, Chinese New Year, weather and common trend of BSS data.

\begin{figure}[ht]
    \centering
    \begin{subfigure}{.23\textwidth}
        \begin{tikzpicture}[inner sep = 0pt]
            \mainBasic{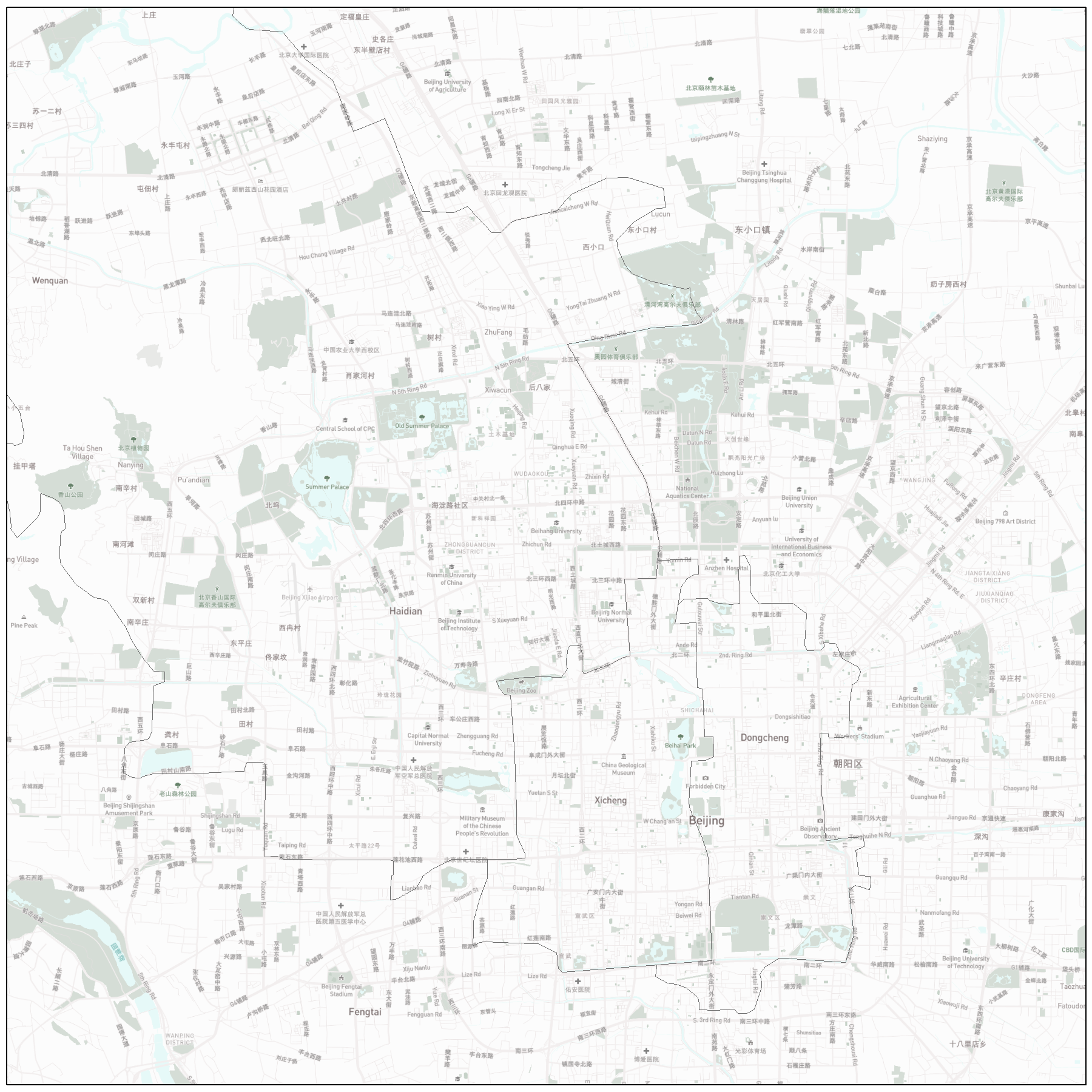}
            \node at (a) {\includegraphics[width=\textwidth]{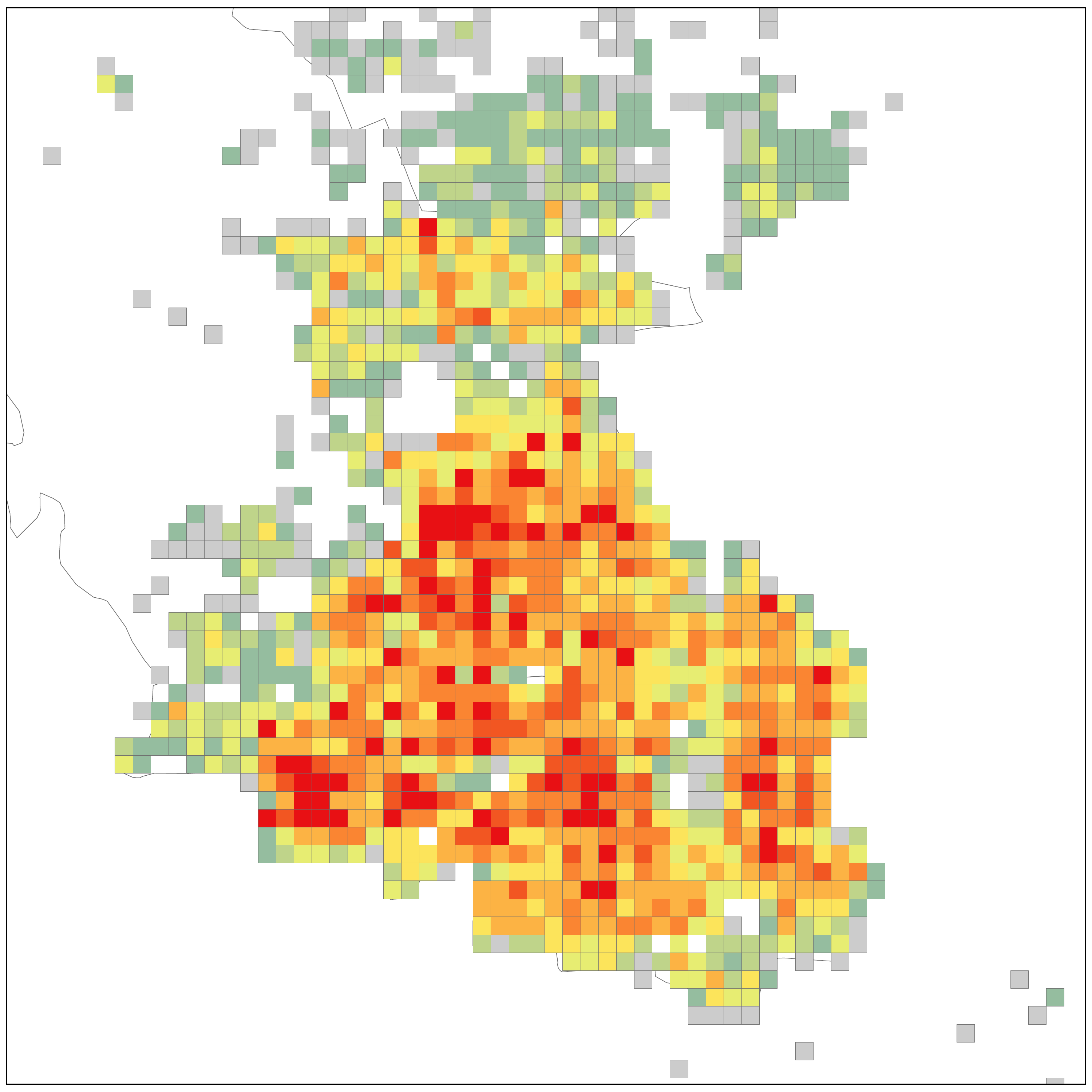}};
            \legendBar{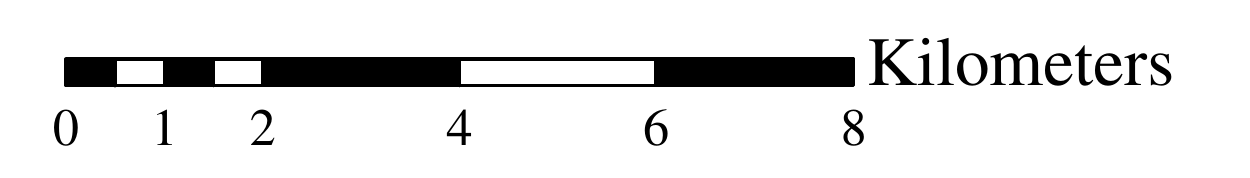}
        \end{tikzpicture}
        \caption{21 Jan}
    \end{subfigure}
    \begin{subfigure}{.23\textwidth}
        \begin{tikzpicture}[inner sep = 0pt]
            \mainBasic{Main_basic1.png}
            \node at (a) {\includegraphics[width=\textwidth]{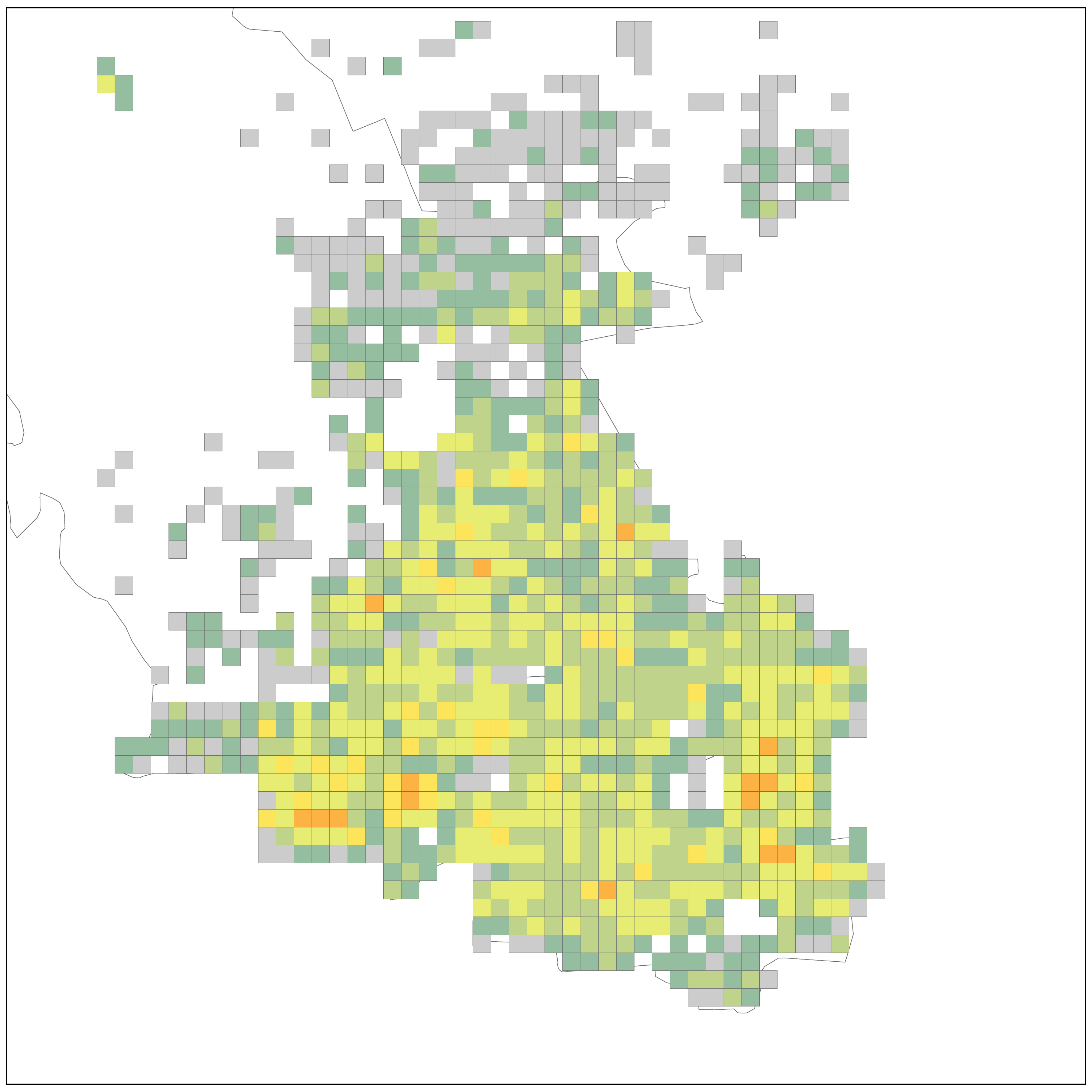}};
            \legendBar{Scalebar_basic1.pdf}
        \end{tikzpicture}
        \caption{25 Jan}
        \label{fig:pattern_01_25}
    \end{subfigure}
    \begin{subfigure}{.23\textwidth}
        \begin{tikzpicture}[inner sep = 0pt]
            \mainBasic{Main_basic1.png}
            \node at (a) {\includegraphics[width=\textwidth]{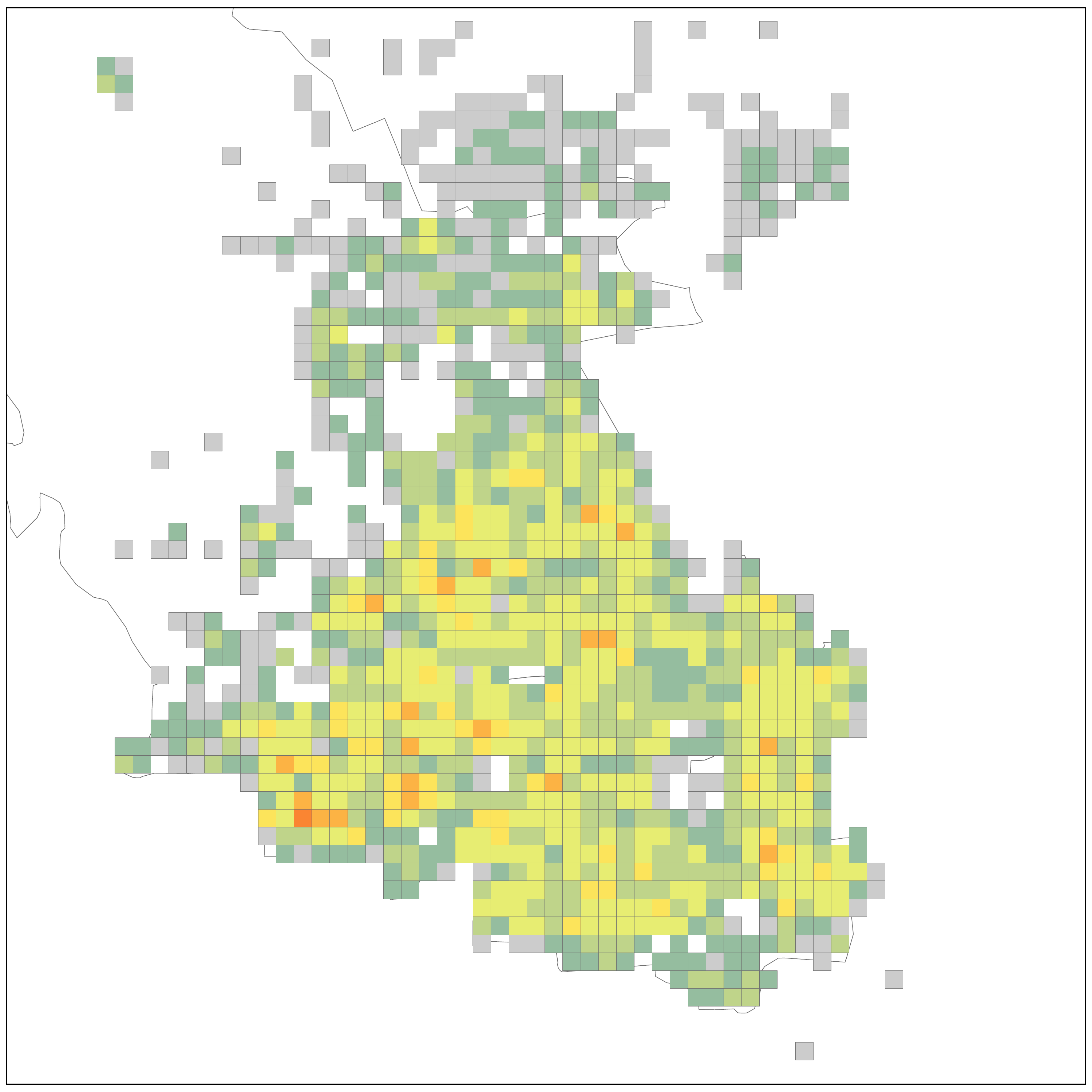}};
            \legendBar{Scalebar_basic1.pdf}
        \end{tikzpicture}
        \caption{29 Jan}
    \end{subfigure}
        \begin{subfigure}{.23\textwidth}
        \begin{tikzpicture}[inner sep = 0pt]
            \mainBasic{Main_basic1.png}
            \node at (a) {\includegraphics[width=\textwidth]{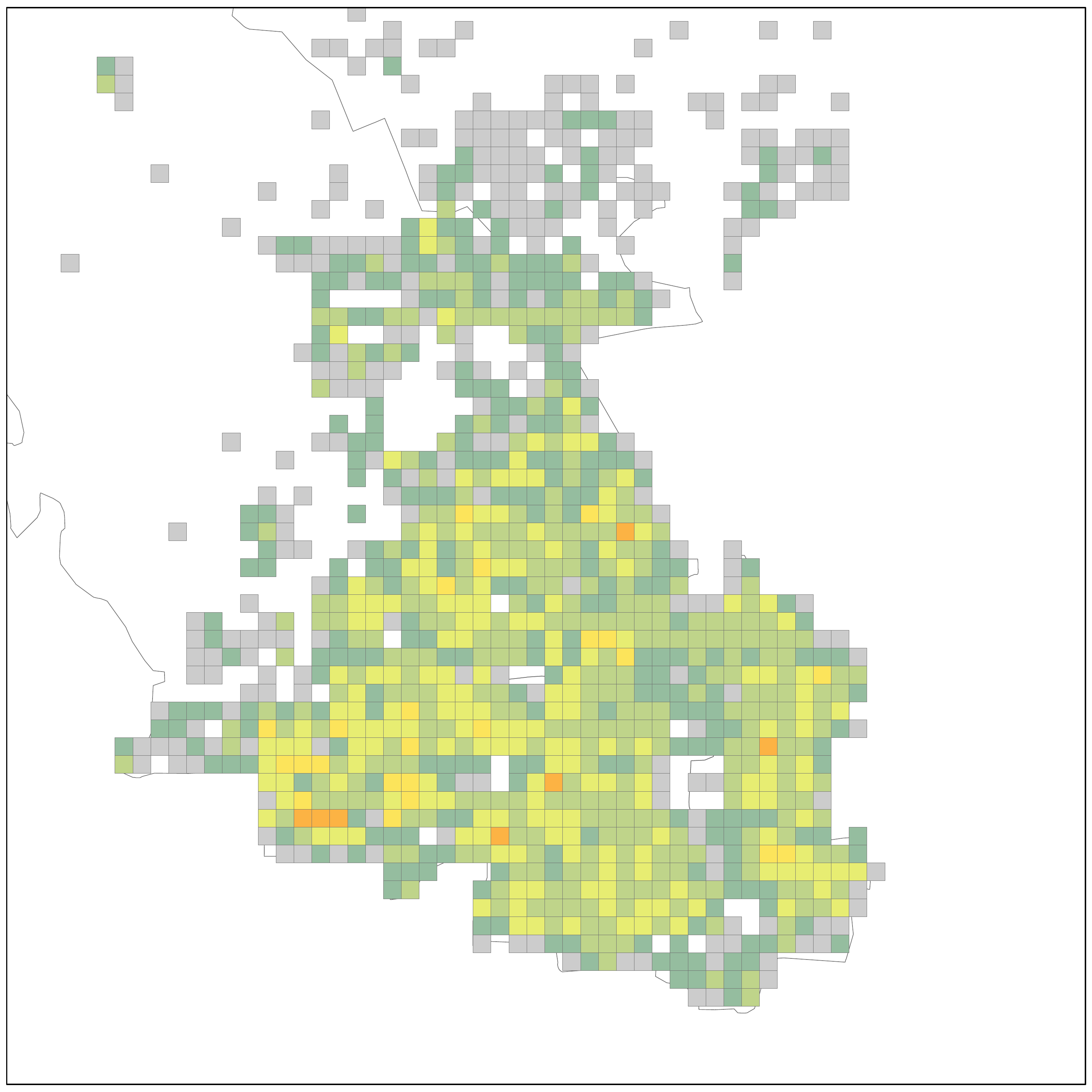}};
            \legendBar{Scalebar_basic1.pdf}
        \end{tikzpicture}
        \caption{02 Feb}
    \end{subfigure}
    
    \begin{subfigure}{.23\textwidth}
        \begin{tikzpicture}[inner sep = 0pt]
            \mainBasic{Main_basic1.png}
            \node at (a) {\includegraphics[width=\textwidth]{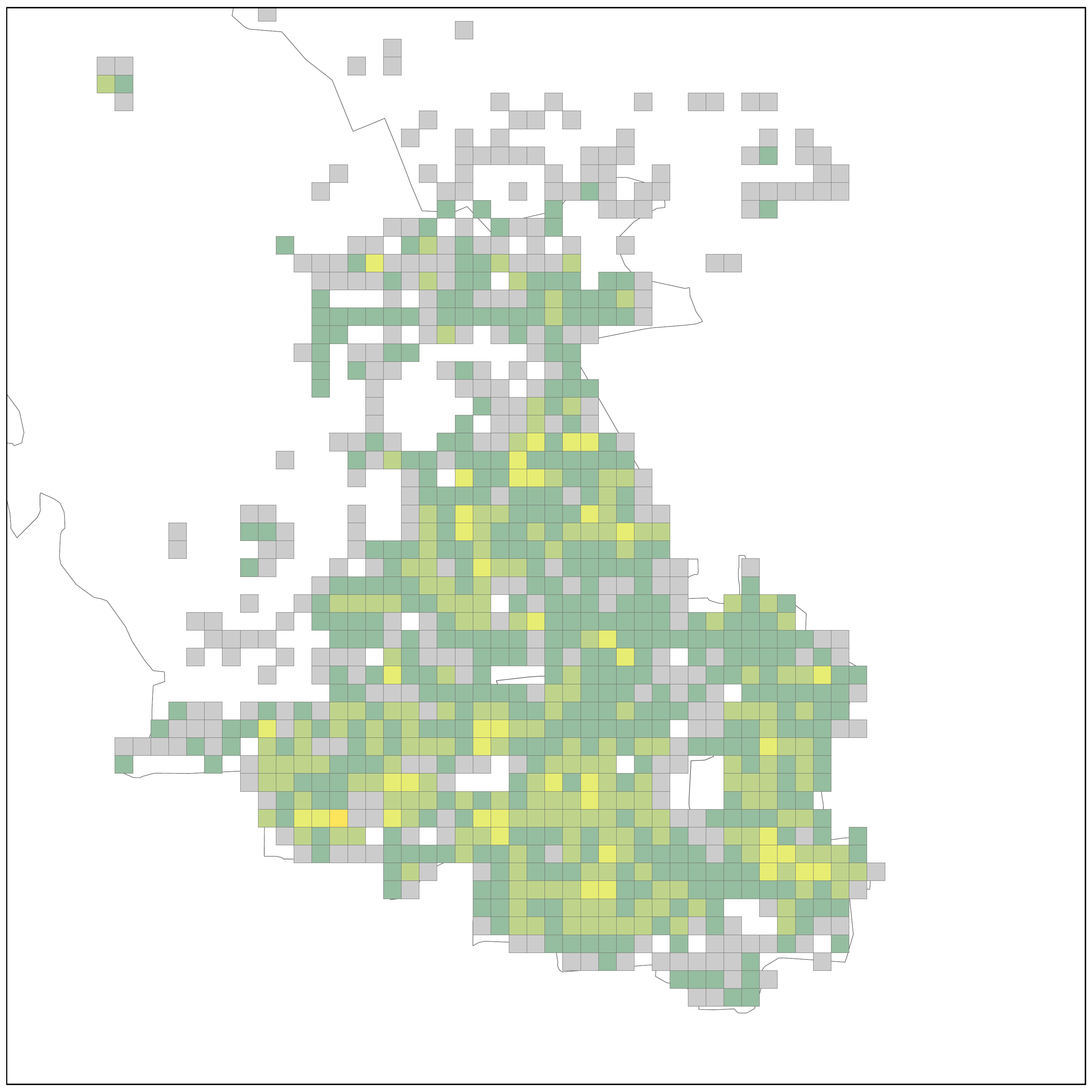}};
            \legendBar{Scalebar_basic1.pdf}
        \end{tikzpicture}
        \caption{06 Feb}
        \label{fig:pattern_02_06}
    \end{subfigure}
    \begin{subfigure}{.23\textwidth}
        \begin{tikzpicture}[inner sep = 0pt]
            \mainBasic{Main_basic1.png}
            \node at (a) {\includegraphics[width=\textwidth]{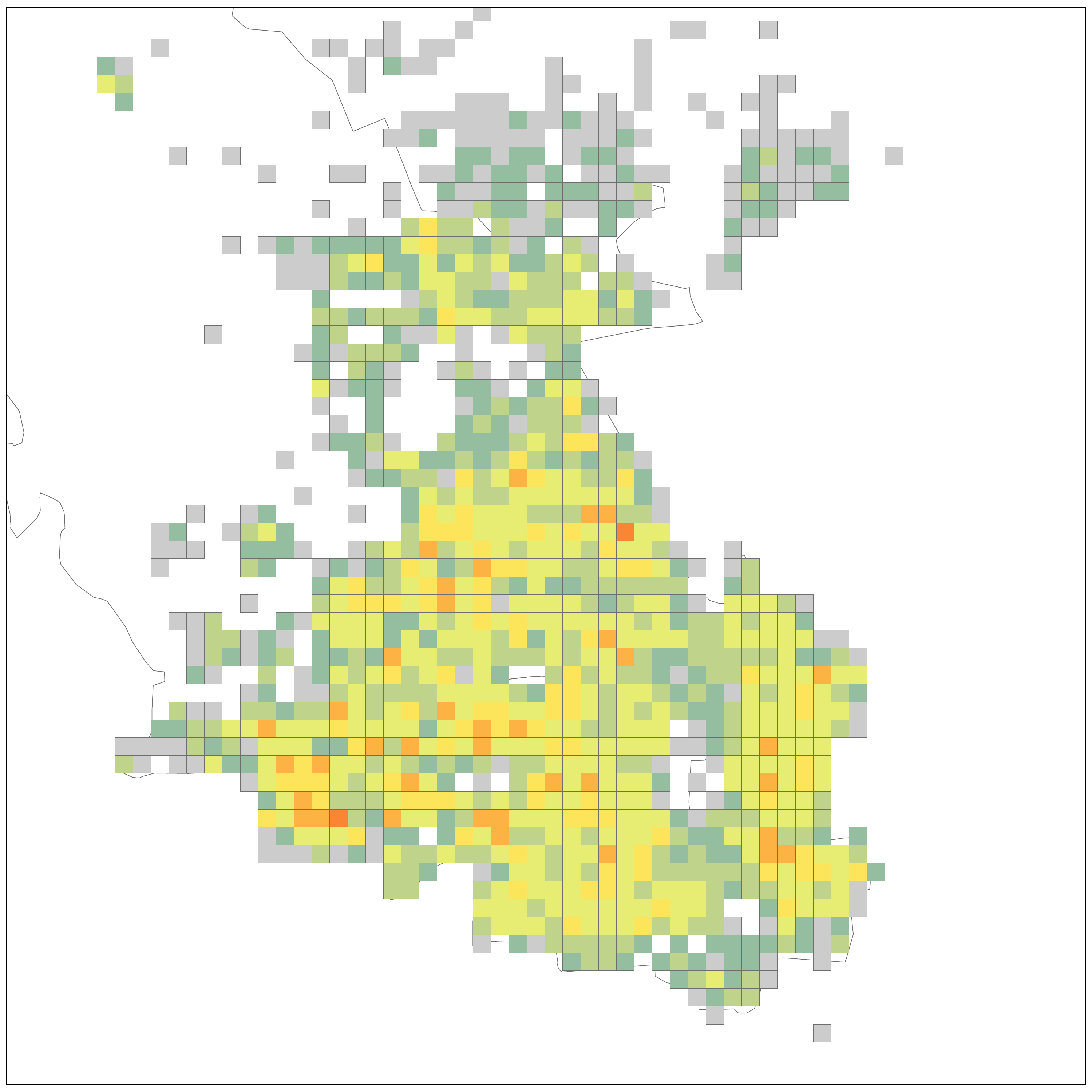}};
            \legendBar{Scalebar_basic1.pdf}
        \end{tikzpicture}
        \caption{10 Feb}
        \label{fig:pattern_02_10}
    \end{subfigure}
    \begin{subfigure}{.23\textwidth}
        \begin{tikzpicture}[inner sep = 0pt]
            \mainBasic{Main_basic1.png}
            \node at (a) {\includegraphics[width=\textwidth]{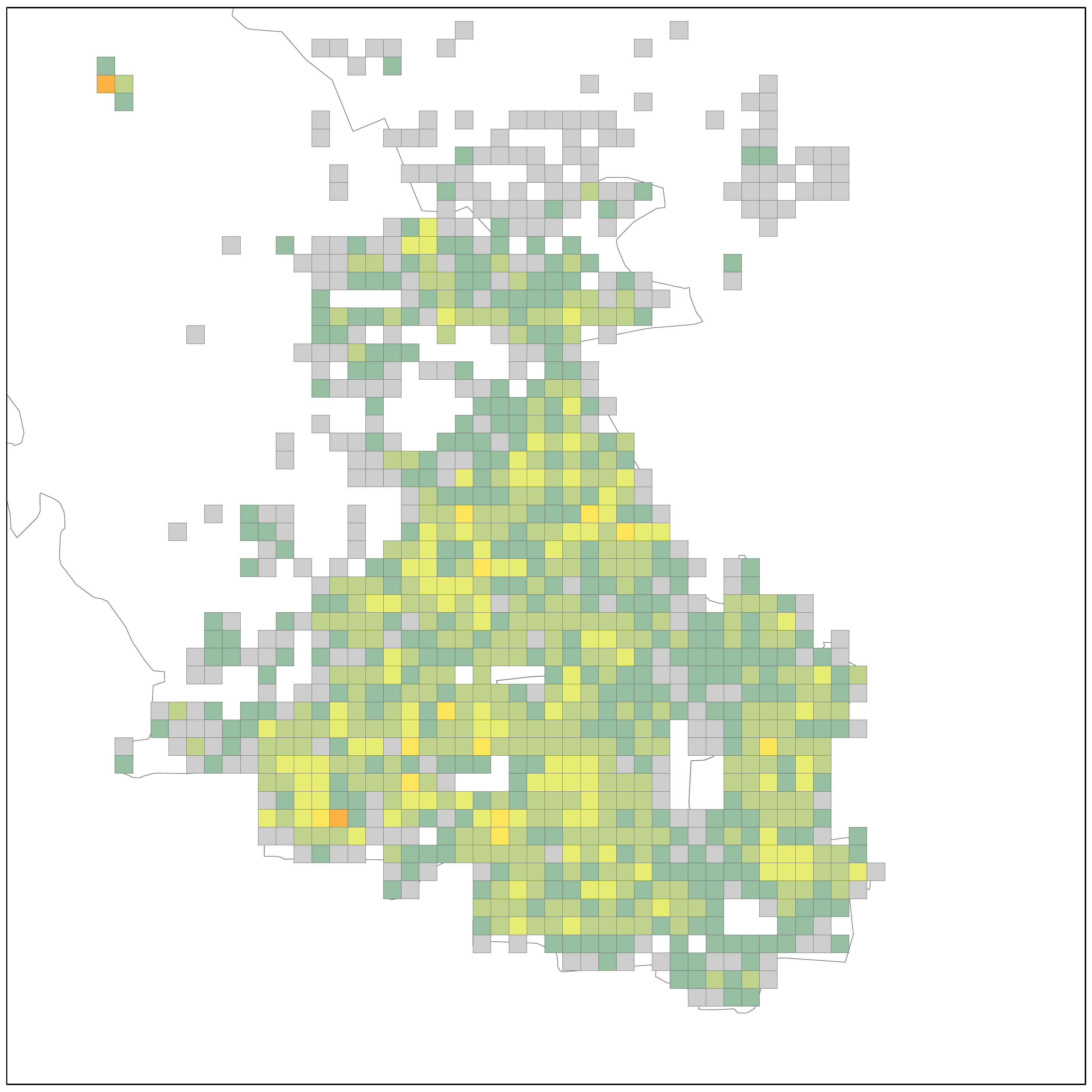}};
            \legendBar{Scalebar_basic1.pdf}
        \end{tikzpicture}
        \caption{14 Feb}
    \end{subfigure}
        \begin{subfigure}{.23\textwidth}
        \begin{tikzpicture}[inner sep = 0pt]
            \mainBasic{Main_basic1.png}
            \node at (a) {\includegraphics[width=\textwidth]{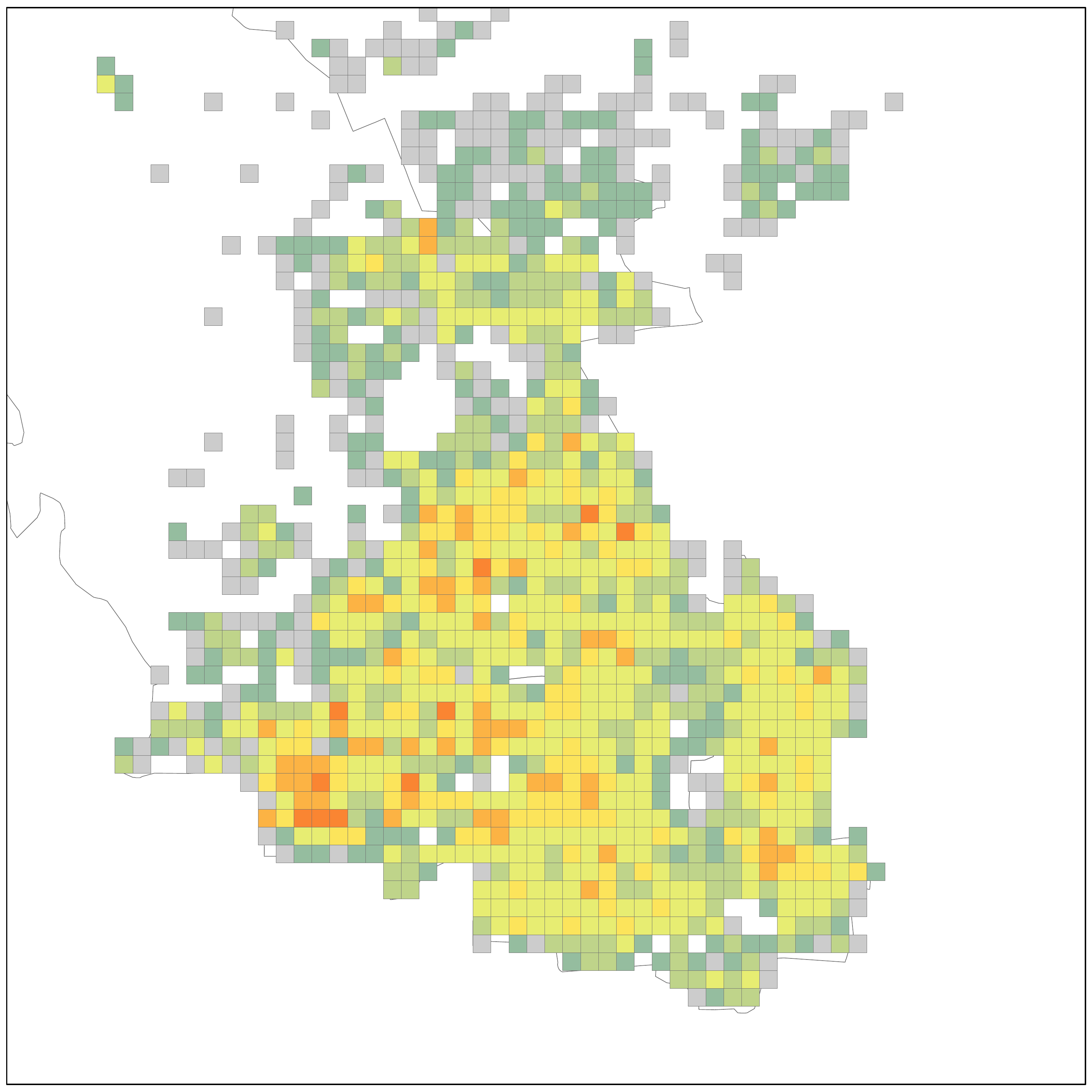}};
            \legendBar{Scalebar_basic1.pdf}
        \end{tikzpicture}
        \caption{18 Feb}
    \end{subfigure}
    
    \begin{subfigure}{.23\textwidth}
        \begin{tikzpicture}[inner sep = 0pt]
            \mainBasic{Main_basic1.png}
            \node at (a) {\includegraphics[width=\textwidth]{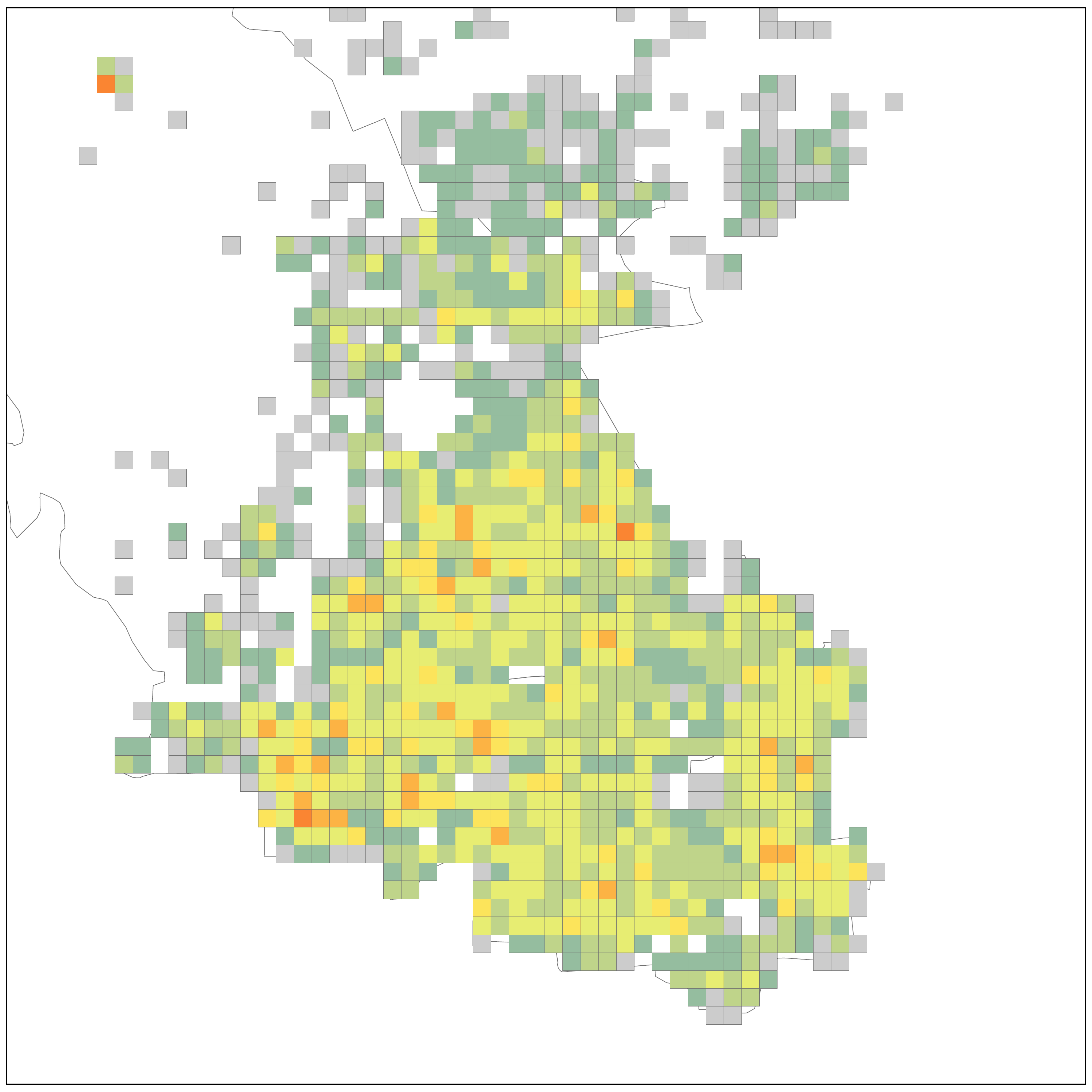}};
            \legendBar{Scalebar_basic1.pdf}
        \end{tikzpicture}
        \caption{24 Feb}
    \end{subfigure}
    \begin{subfigure}{.23\textwidth}
        \begin{tikzpicture}[inner sep = 0pt]
            \mainBasic{Main_basic1.png}
            \node at (a) {\includegraphics[width=\textwidth]{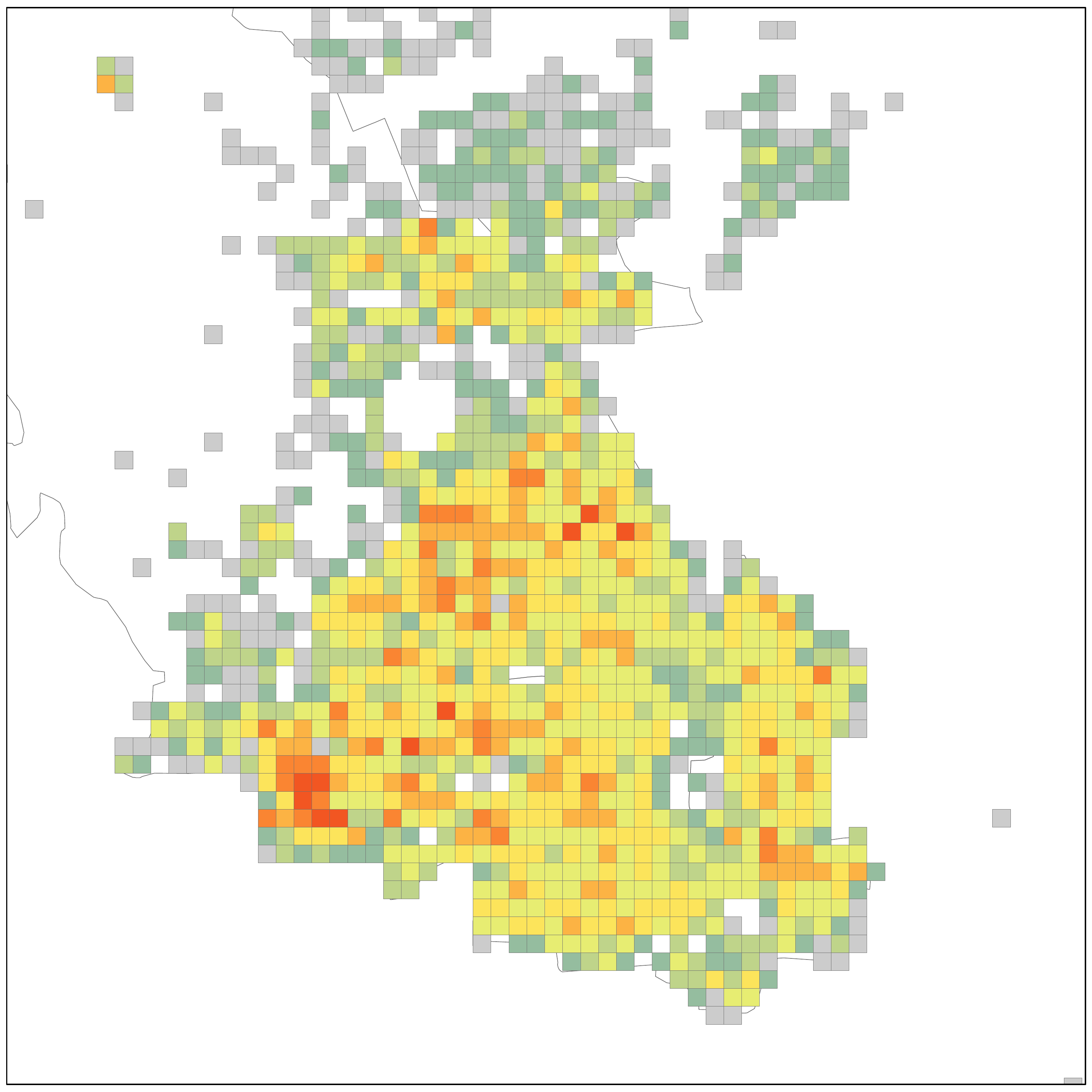}};
            \legendBar{Scalebar_basic1.pdf}
        \end{tikzpicture}
        \caption{26 Feb}
    \end{subfigure}
    \begin{subfigure}{.23\textwidth}
        \begin{tikzpicture}[inner sep = 0pt]
            \mainBasic{Main_basic1.png}
            \node at (a) {\includegraphics[width=\textwidth]{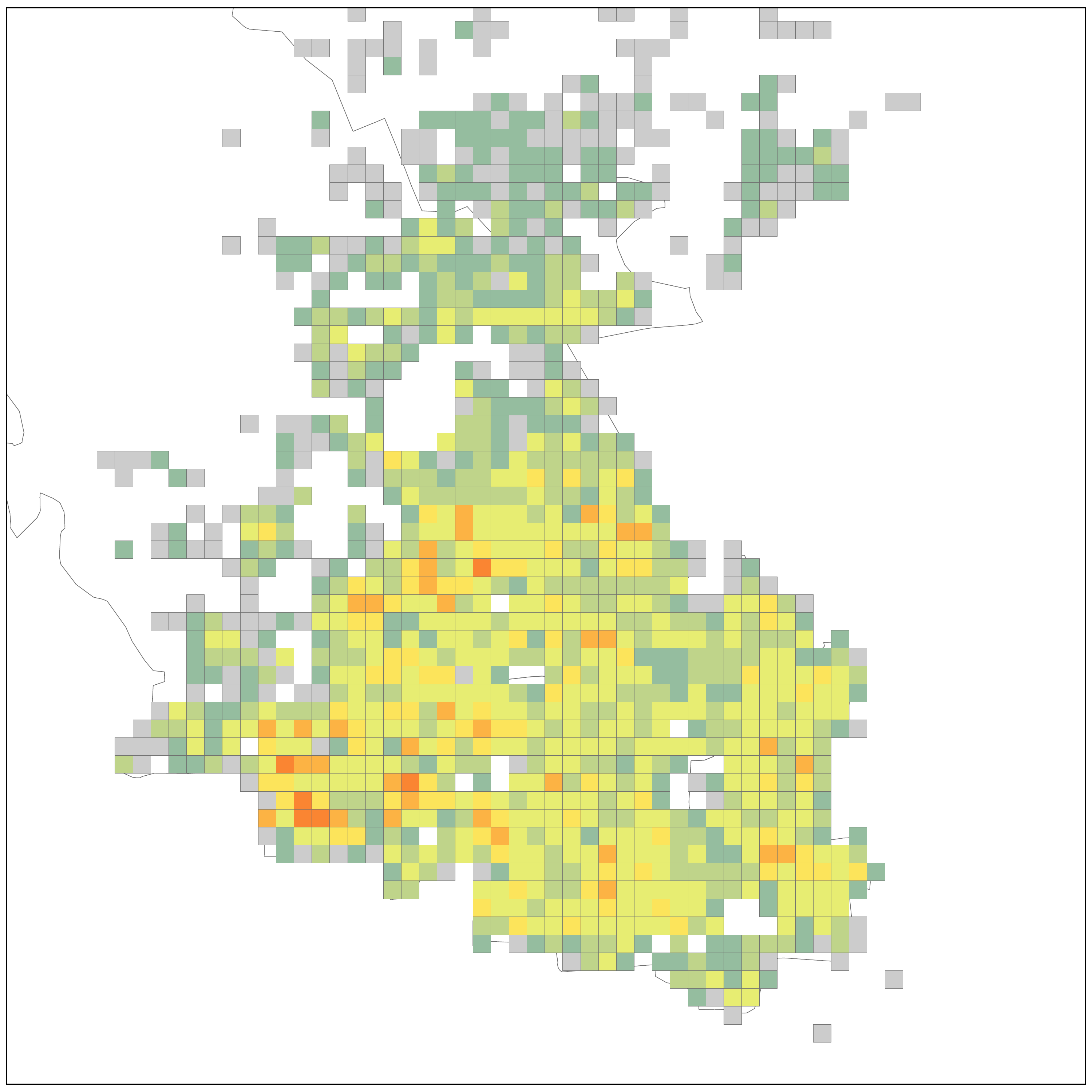}};
            \legendBar{Scalebar_basic1.pdf}
        \end{tikzpicture}
        \caption{01 Mar}
        \label{fig:pattern_03_01}
    \end{subfigure}
    \begin{subfigure}{.13\textwidth}
        \includegraphics[width=\textwidth]{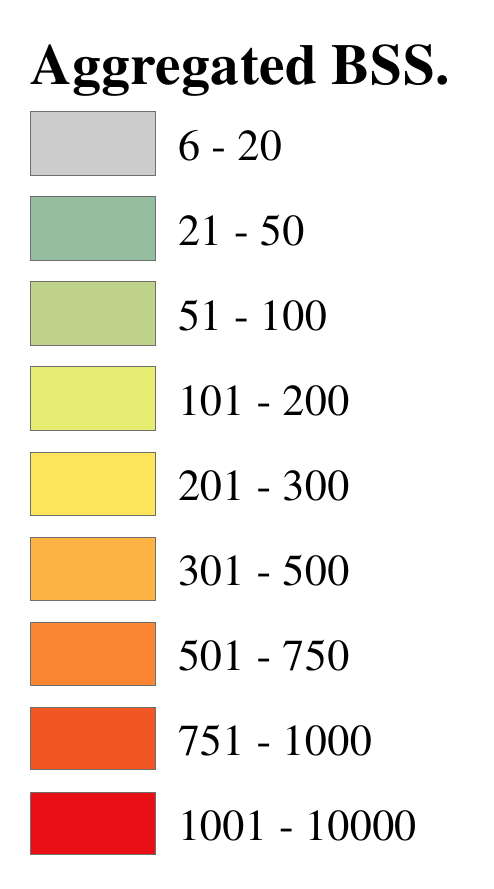}
    \end{subfigure}
    \caption[Evolution of share bike usage]{Changes in distribution of share bike usage from 21 Jan to 01 Mar, 2020. The daily share bike records were aggregated within 500-meter grids and rendered with a color ramp from grey to green to red.}
    \label{fig:full_spatial_pattern_2020}
\end{figure}

\subsubsection{Evolution in 2020}

Figure~\ref{fig:full_spatial_pattern_2020} shows the evolution of BSS activities during a 4-day interval from 21 Jan to 2 Mar.
The trend of human mobility are consistent with Figure~\ref{fig:hour_comparison_8}.

Before 21 Jan, 2020, share bike activities were spread throughout the city limits, with a significant concentration in the downtown area (up to 500-1000 records/hour).
This pattern indicates the spatial distribution of human mobility before the COVID-19 outbreak.

After the outbreak, there was a dramatic drop in mobility since 25 Jan, which was also the beginning of the Chinese New Year holiday.
Figure~\ref{fig:pattern_01_25}-\ref{fig:pattern_02_06} ranging from 25 Jan to 06 Feb are dominated by low intensity, indicating non-essential trips has dropped significantly due to combined effects of holidays and the epidemic.
This situation continued until 09 Feb, when the spread of COVID-19 slowed down and productive and social activities were allowed to restart partially.

Figure~\ref{fig:pattern_02_10}-\ref{fig:pattern_03_01} show a gradual increase in the mobility from 10 Feb to 01 Mar, 2020. However, the human mobility only restored around $30\%$ of the pre-pandemic level.
It is worth noticing that only slight differences could be observed between the weekdays and the weekends before 17 Feb and the weekday-weekend oscillation reappeared afterward.
There was a higher demand for share bikes on workdays in downtown area than the outbreak phase.
However, this high-demand shrank on weekends, implying the residents were more inclined to reduce the risk of increased exposure by going outside under the threat of COVID-19. 

\subsubsection{Comparison between 2019 and 2020}
However, both the Chinese New Year holiday shutdown and pandemic can result in the mobility decrease.
Hence, we compared the share bike usage between 2020 and that of 2019 in the same period. 
This procedure \textit{roughly} removed the impact of the Chinese New Year holiday and reflected the influence of the pandemic in the whole study area. 

\begin{figure}[ht]
    \centering
    \begin{subfigure}{.28\textwidth}
        \begin{tikzpicture}[inner sep = 0pt]
            \mainBasic{Main_basic1.png}
            \node at (a) {\includegraphics[width=\textwidth]{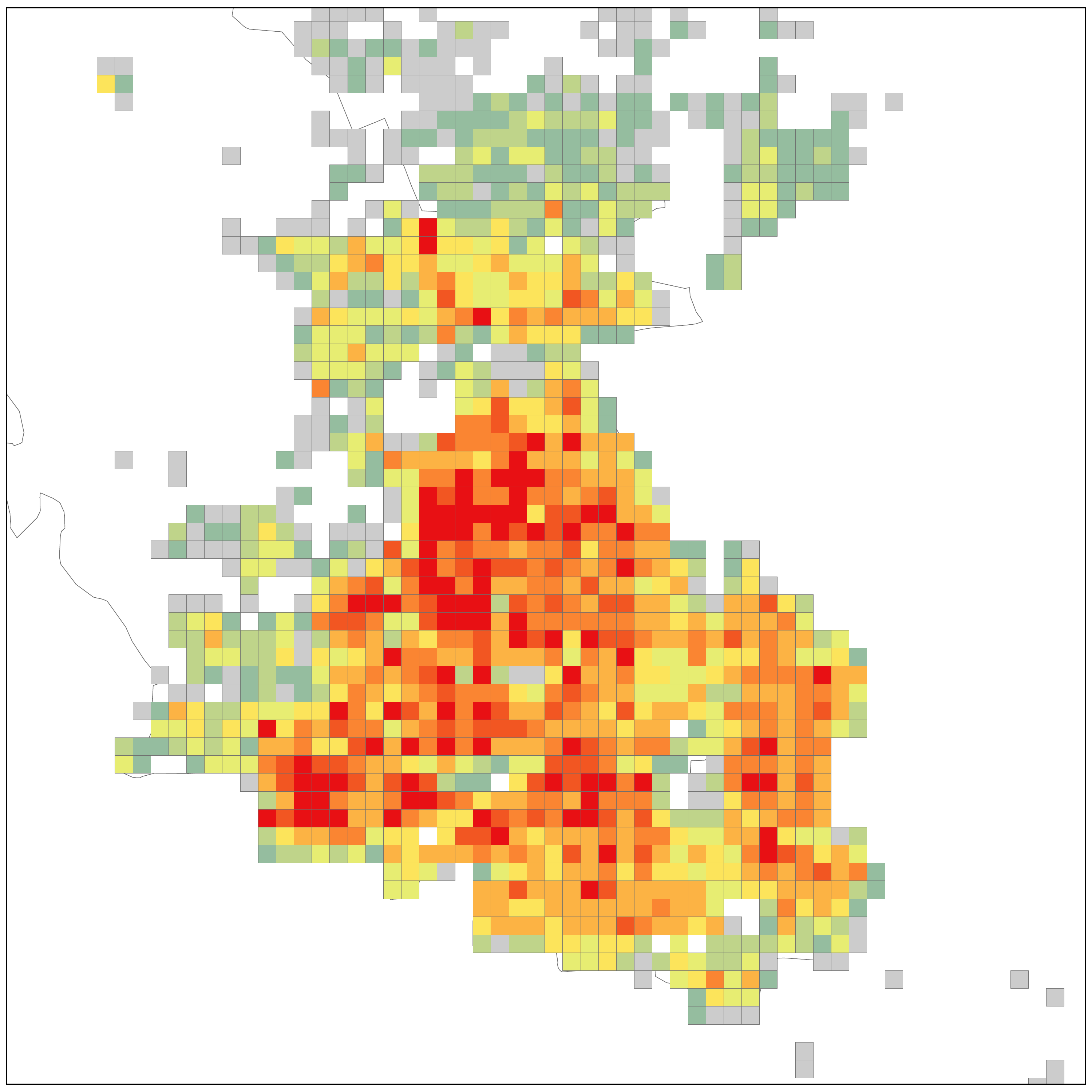}};
            \legendBar{Scalebar_basic1.pdf}
        \end{tikzpicture}
        \caption{\textbf{phase a} of 2020}
        \label{fig:p_a_2020}
    \end{subfigure}
    \begin{subfigure}{.28\textwidth}
        \begin{tikzpicture}[inner sep = 0pt]
            \mainBasic{Main_basic1.png}
            \node at (a) {\includegraphics[width=\textwidth]{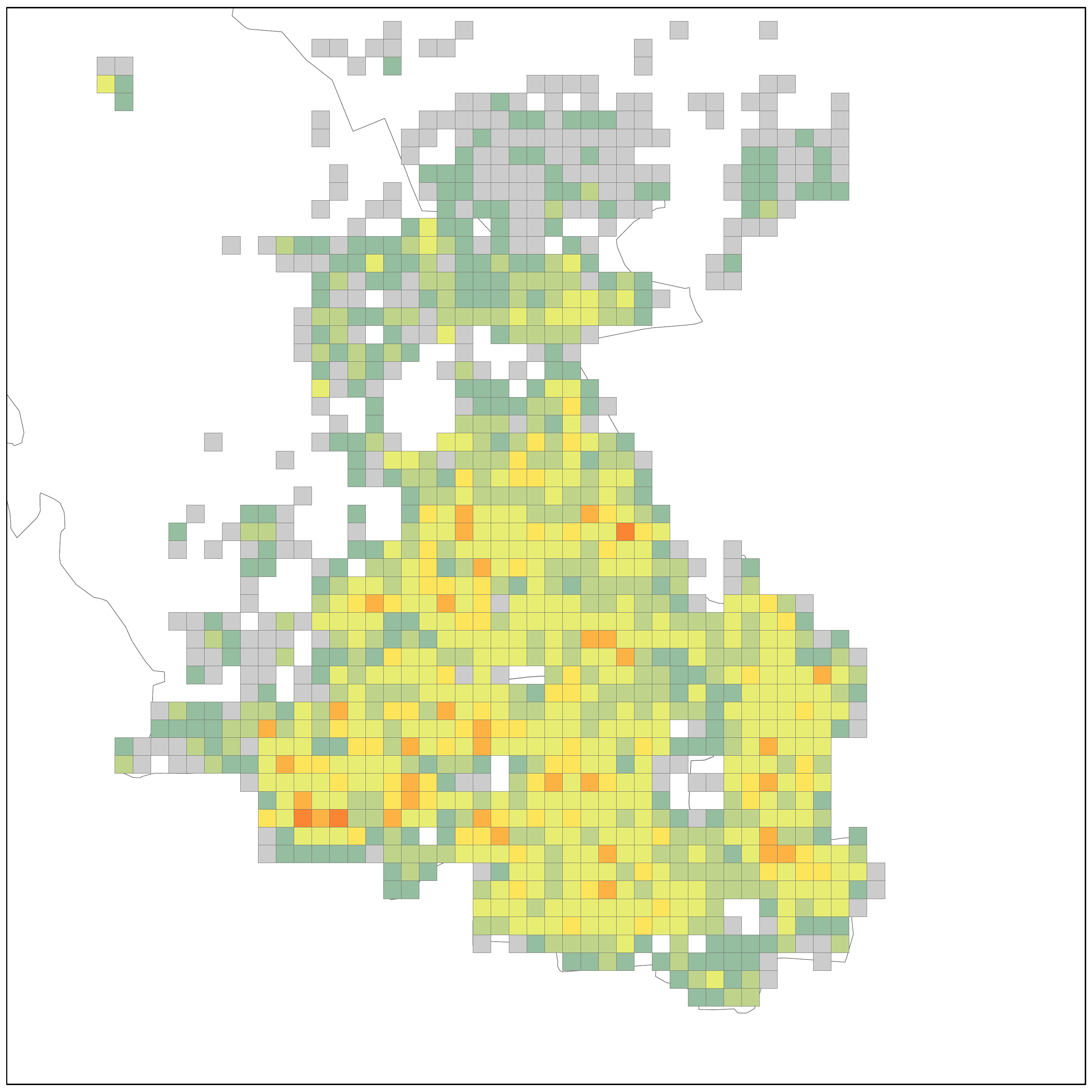}};
            \legendBar{Scalebar_basic1.pdf}
        \end{tikzpicture}
        \caption{\textbf{phase b} of 2020}
        \label{fig:p_b_2020}
    \end{subfigure}
    \begin{subfigure}{.28\textwidth}
        \begin{tikzpicture}[inner sep = 0pt]
            \mainBasic{Main_basic1.png}
            \node at (a) {\includegraphics[width=\textwidth]{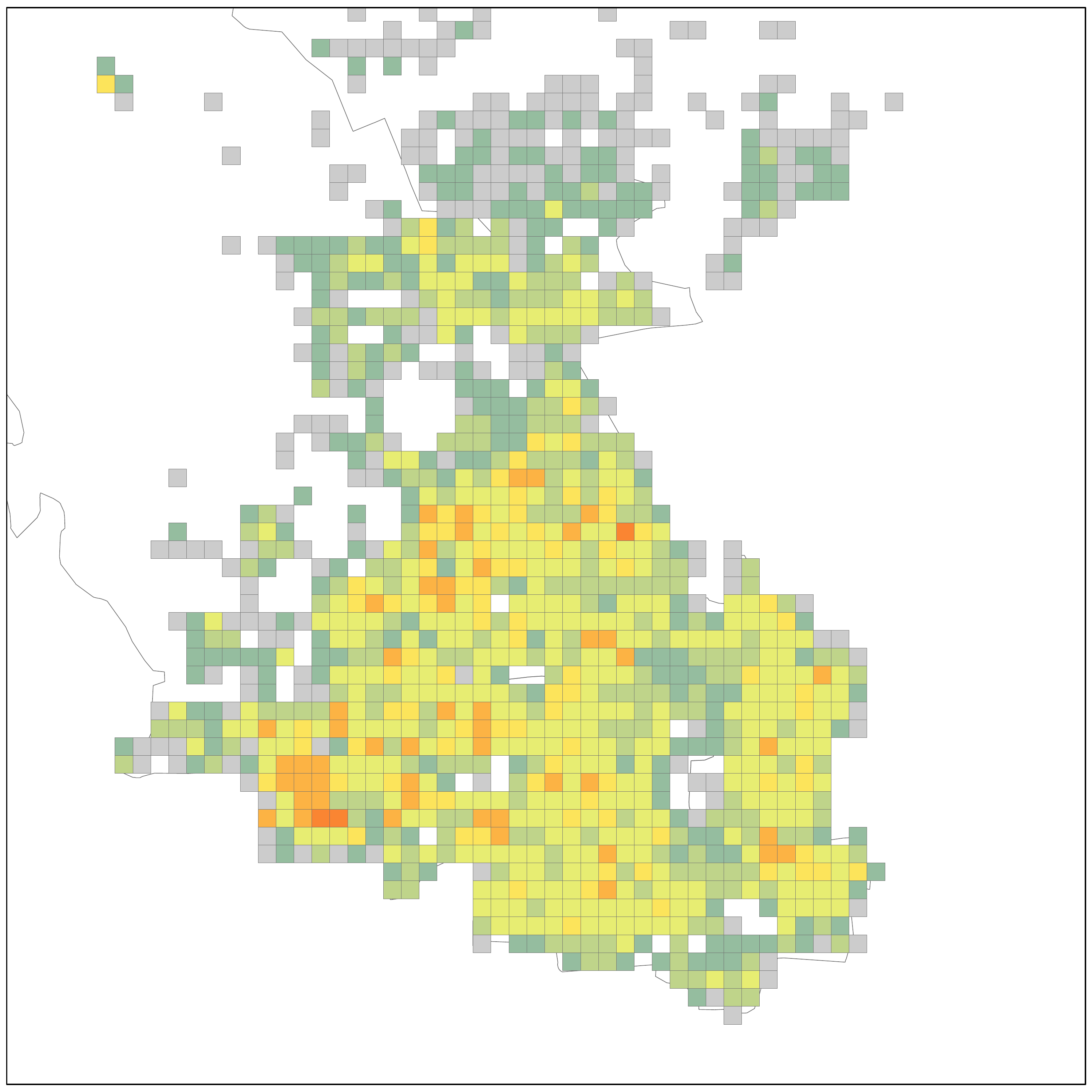}};
            \legendBar{Scalebar_basic1.pdf}
        \end{tikzpicture}
        \caption{\textbf{phase c} of 2020}
        \label{fig:p_c_2020}
    \end{subfigure}
    \begin{subfigure}{.14\textwidth}
        \includegraphics[width=\textwidth]{Figures/Overall_spatial_patterns/legend5.pdf}
    \end{subfigure}
    
    \begin{subfigure}{.28\textwidth}
        \begin{tikzpicture}[inner sep = 0pt]
            \mainBasic{Main_basic1.png}
            \node at (a) {\includegraphics[width=\textwidth]{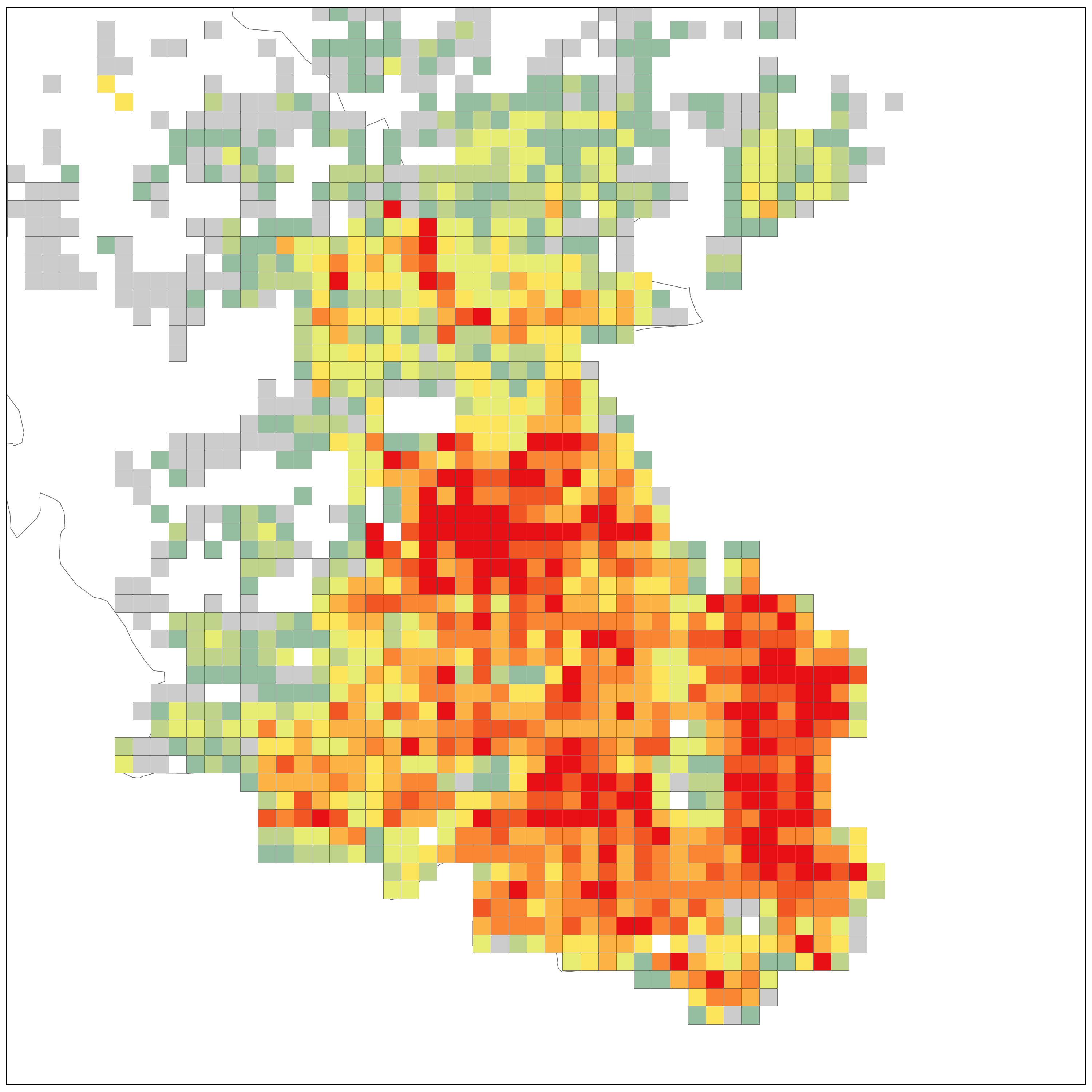}};
            \legendBar{Scalebar_basic1.pdf}
        \end{tikzpicture}
        \caption{\textbf{phase a} of 2019}
        \label{fig:p_a_2019}
    \end{subfigure}
    \begin{subfigure}{.28\textwidth}
        \begin{tikzpicture}[inner sep = 0pt]
            \mainBasic{Main_basic1.png}
            \node at (a) {\includegraphics[width=\textwidth]{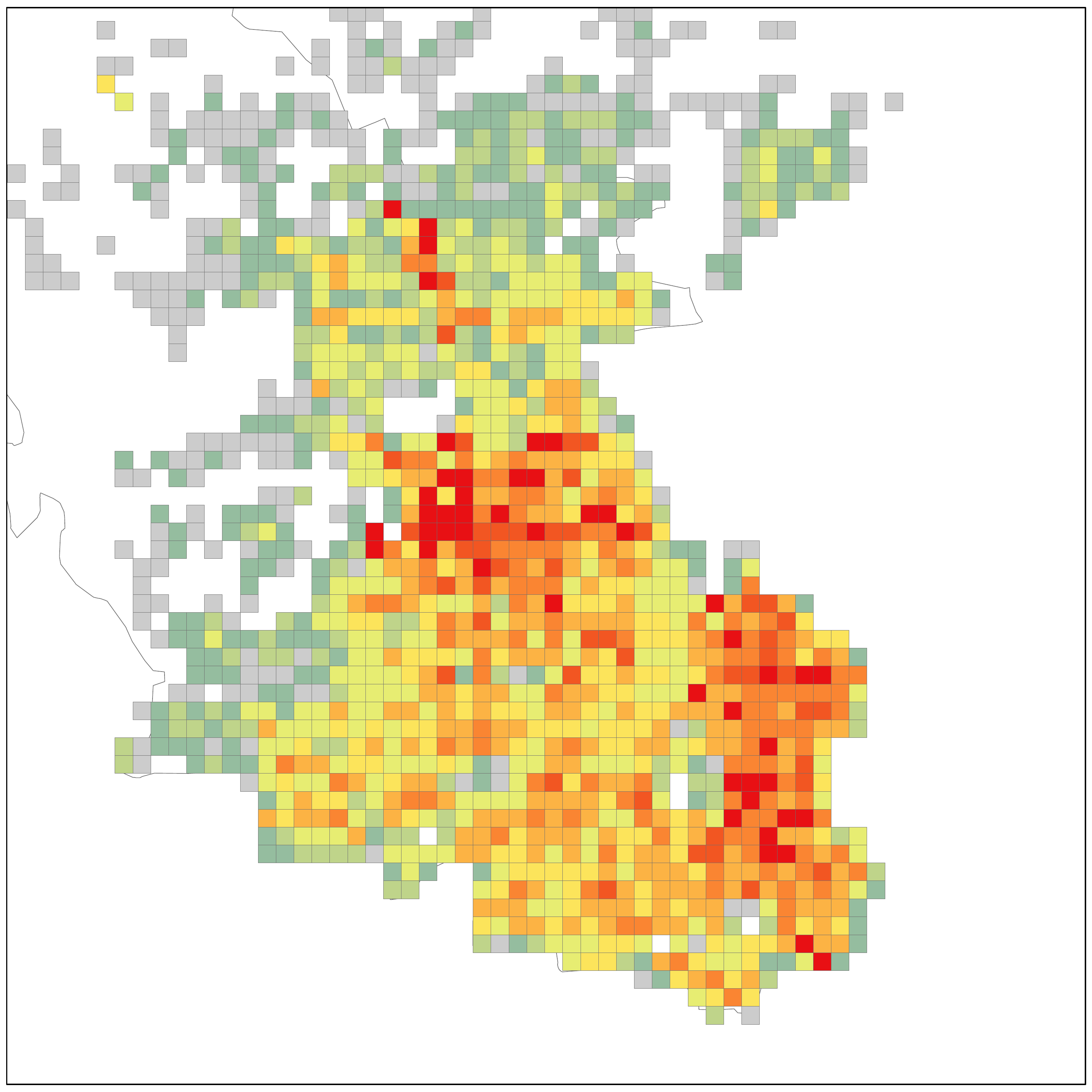}};
            \legendBar{Scalebar_basic1.pdf}
        \end{tikzpicture}
        \caption{\textbf{phase b} of 2019}
        \label{fig:p_b_2019}
    \end{subfigure}
    \begin{subfigure}{.28\textwidth}
        \begin{tikzpicture}[inner sep = 0pt]
            \mainBasic{Main_basic1.png}
            \node at (a) {\includegraphics[width=\textwidth]{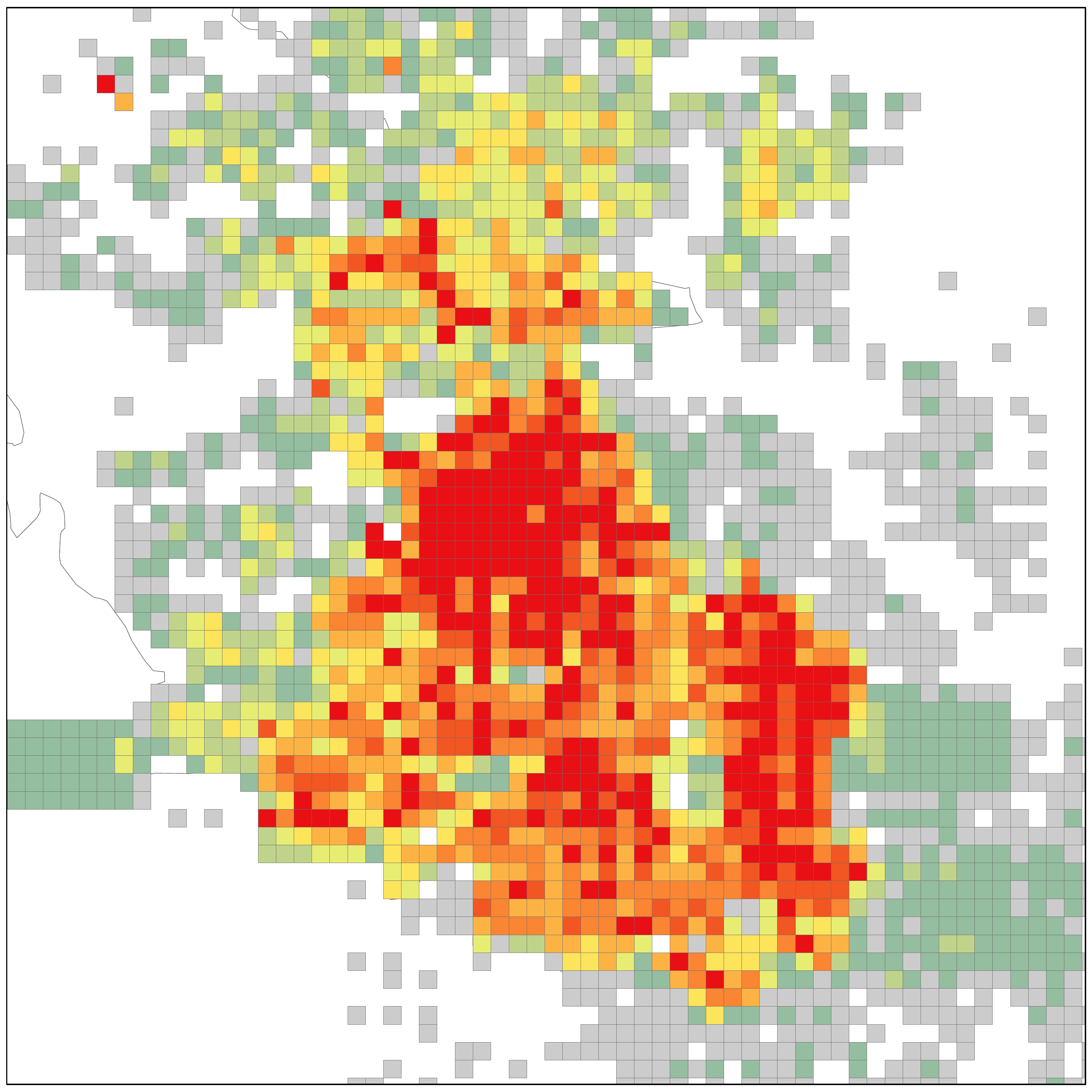}};
            \legendBar{Scalebar_basic1.pdf}
        \end{tikzpicture}
        \caption{\textbf{phase c} of 2019}
        \label{fig:p_c_2019}
    \end{subfigure}
    \begin{subfigure}{.14\textwidth}
        \includegraphics[width=\textwidth]{Figures/Overall_spatial_patterns/legend5.pdf}
    \end{subfigure}
    
    \begin{subfigure}{.28\textwidth}
        \begin{tikzpicture}[inner sep = 0pt]
            \mainBasic{Main_basic1.png}
            \node[opacity=0.8] at (a) {\includegraphics[width=\textwidth]{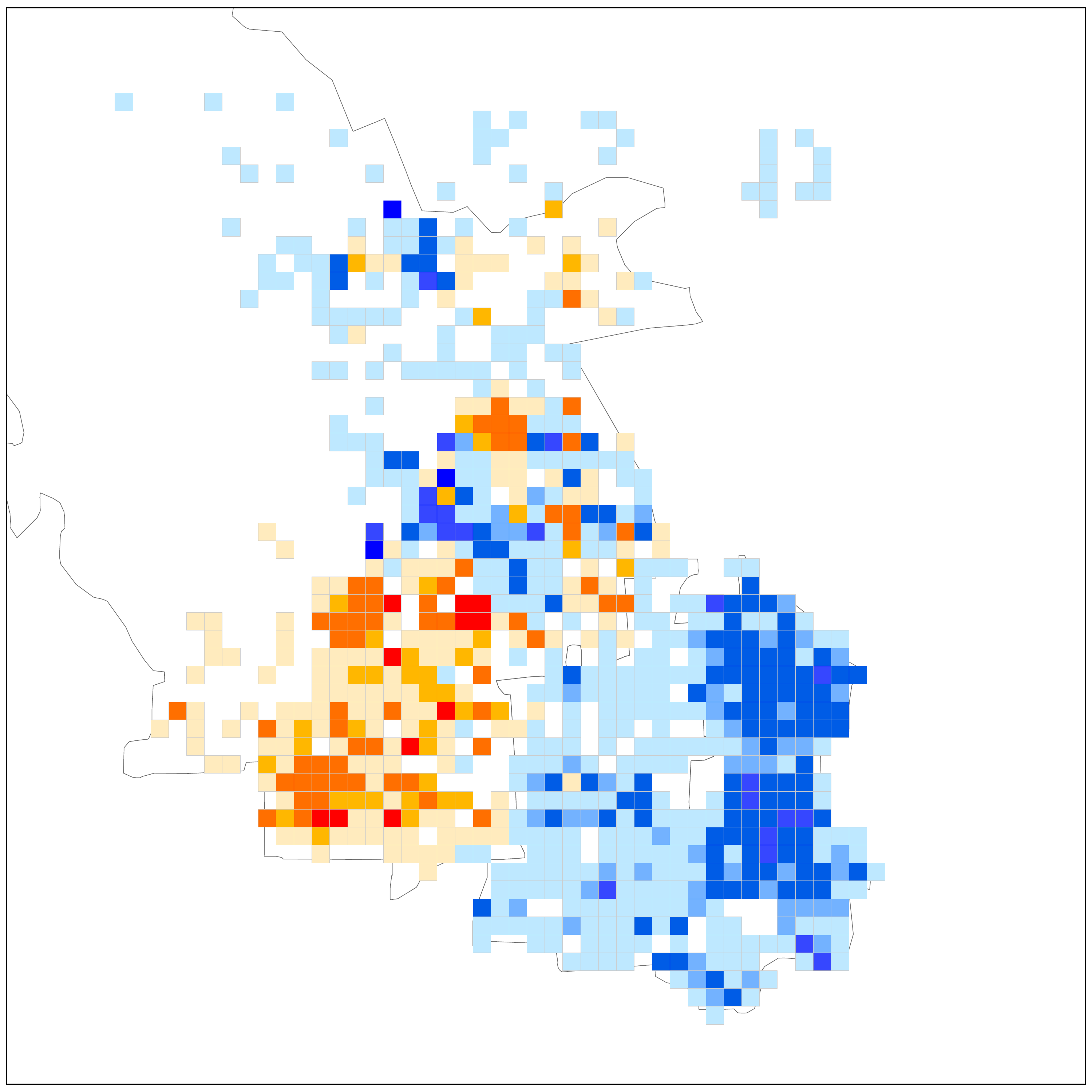}};
            \legendBar{Scalebar_basic1.pdf}
        \end{tikzpicture}
        \caption{Difference between \textbf{a} and \textbf{d}}
        \label{fig:p_a_dif}
    \end{subfigure}
        \begin{subfigure}{.28\textwidth}
        \begin{tikzpicture}[inner sep = 0pt]
            \mainBasic{Main_basic1.png}
            \node[opacity=0.8] at (a) {\includegraphics[width=\textwidth]{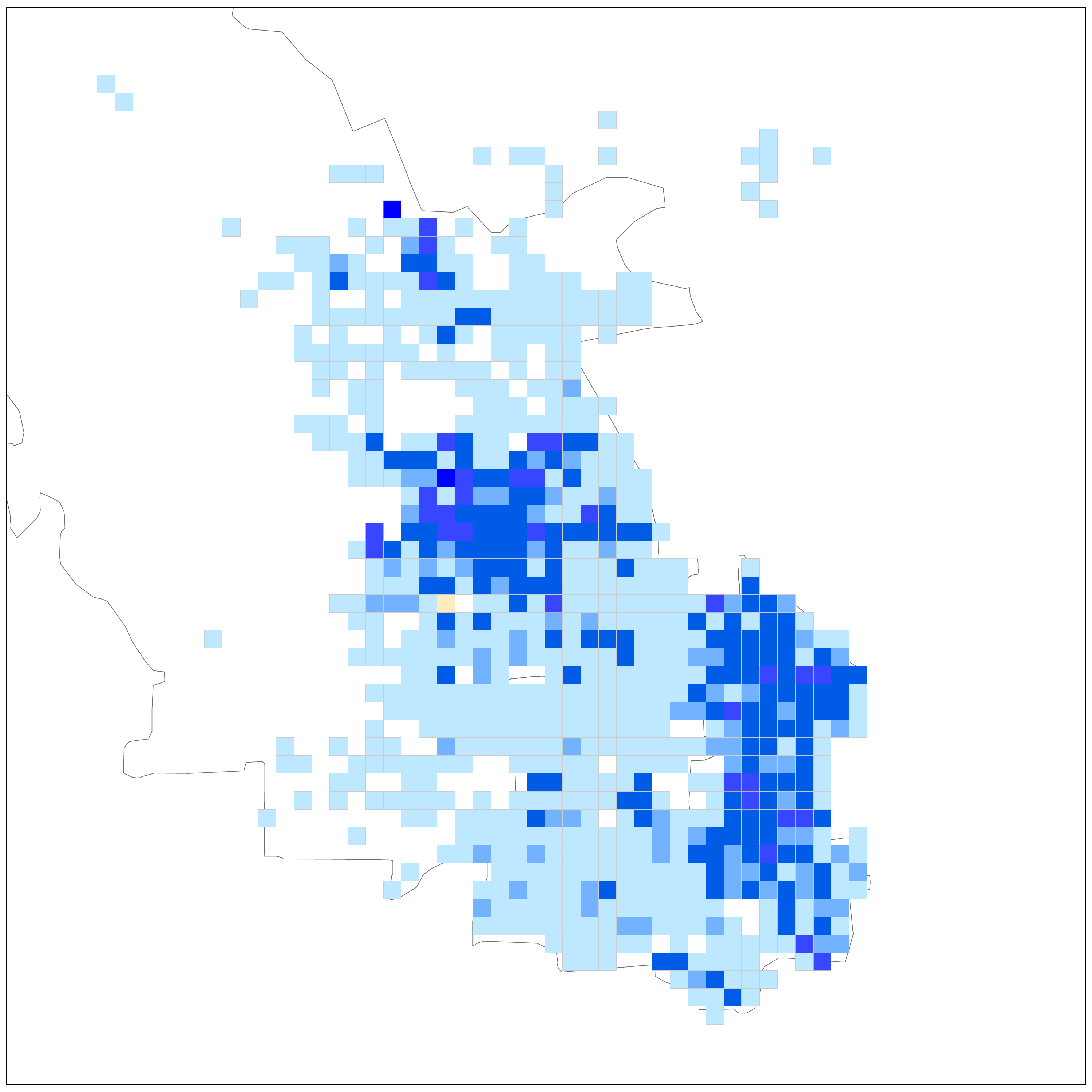}};
            \legendBar{Scalebar_basic1.pdf}
        \end{tikzpicture}
        \caption{Difference between \textbf{b} and \textbf{e}}
        \label{fig:p_b_dif}
    \end{subfigure}
        \begin{subfigure}{.28\textwidth}
        \begin{tikzpicture}[inner sep = 0pt]
            \mainBasic{Main_basic1.png}
            \node[opacity=0.8] at (a) {\includegraphics[width=\textwidth]{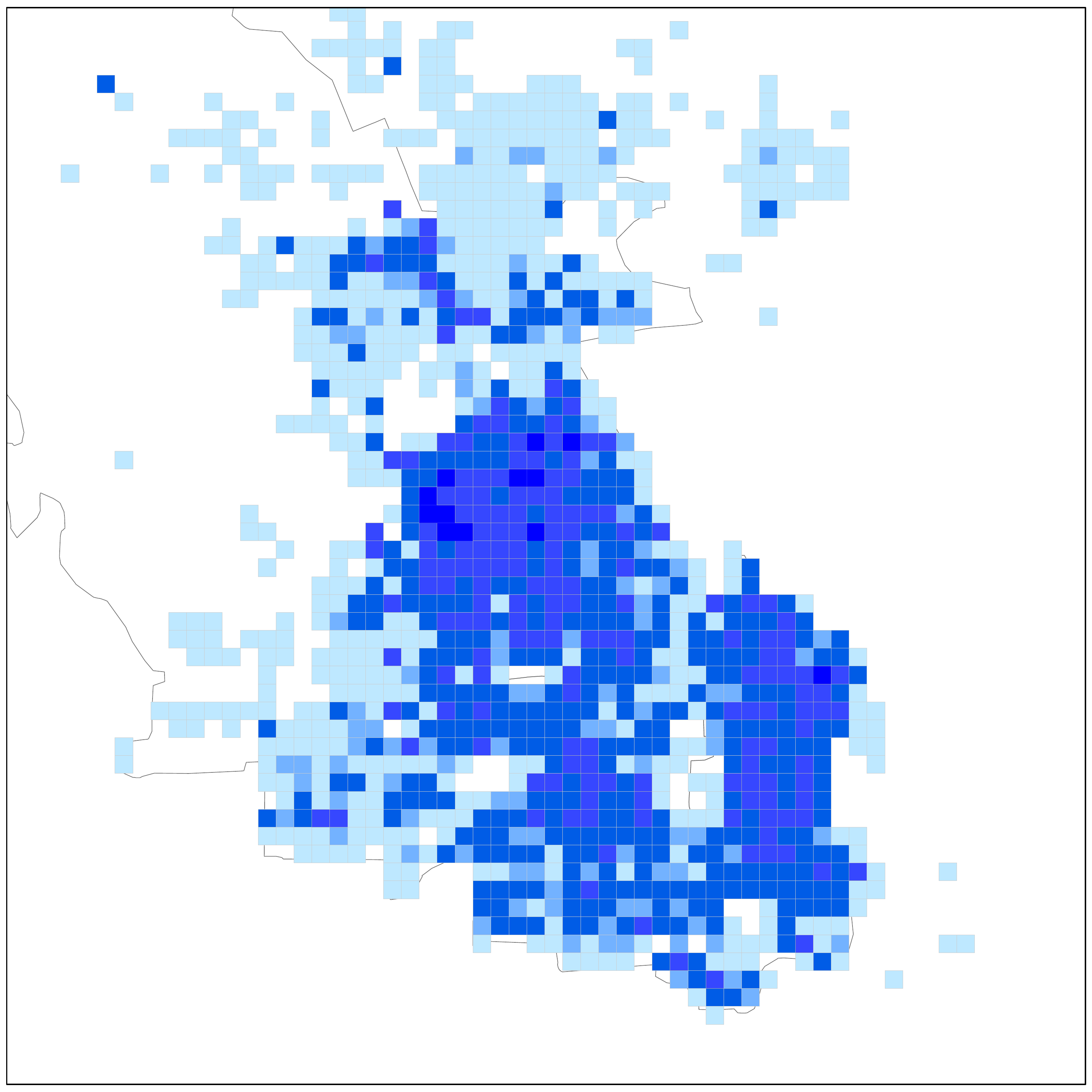}};
            \legendBar{Scalebar_basic1.pdf}
        \end{tikzpicture}
        \caption{Difference between \textbf{c} and \textbf{f}}
        \label{fig:p_c_dif}
    \end{subfigure}
    \begin{subfigure}{.14\textwidth}
        \includegraphics[width=\textwidth]{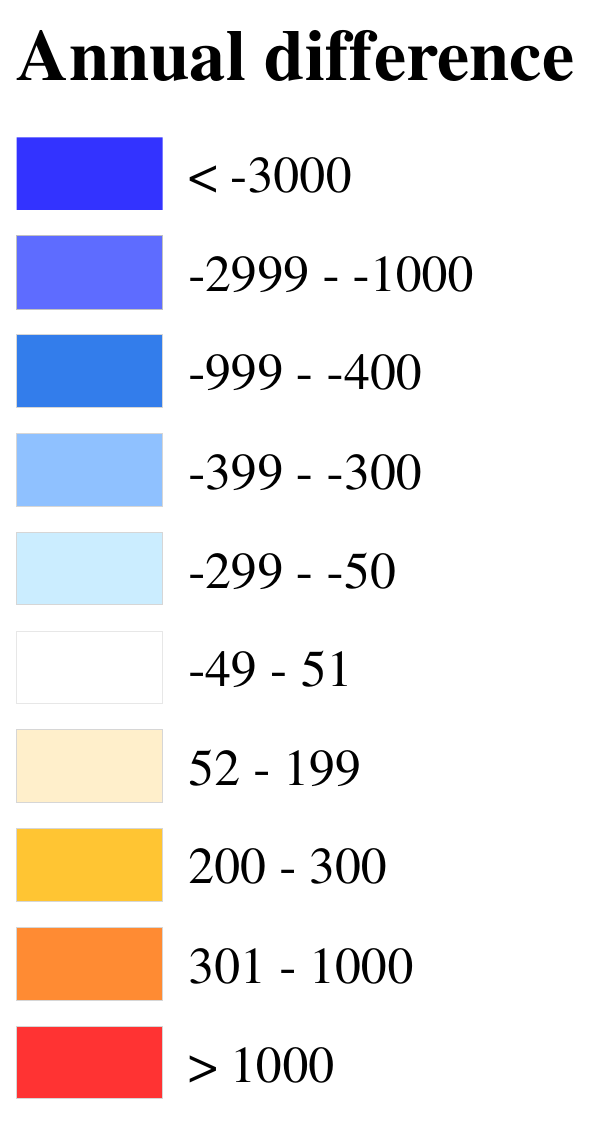}
    \end{subfigure}
    \caption[Difference of share bike usage of 2020 and 2019]{Comparison of share bikes usage between 2019 and 2020 in different phases.}
    \label{fig:compare_2019_and_2020}
\end{figure}

Figure~\ref{fig:compare_2019_and_2020} delineates the sum of aggregated share bike usage in \textbf{phase a, b, c} of 2020 (row 1), the share bike usage in the corresponding period of 2019 (row 2), and the difference of the former two results (row 3).
Figure~\ref{fig:p_a_2019} summarizes the average share bike usage intensity during the same time interval as \textbf{phase a} in 2019, showing similar spatial patterns as in 2020. 

The significant discrepancies in share bike usage between 2019 and 2020 during \textbf{phase b} and \textbf{phase c} are considered the consequence of the pandemic of COVID-19. 
According to Figure~\ref{fig:p_b_2019}, the human mobility decreased due to Chinese New Year in 2019 and high intensity remained in certain areas. 
However, the human mobility was in a state of complete suppression in 2020. 
The difference map in Figure~\ref{fig:p_b_dif} suggests that under the impact of COVID-19, trips were reduced for safety purposes.

Figure~\ref{fig:p_c_2019} indicates that the mobility would instantly return to pre-festival levels or more wide-spread as a result of the influx of migrant employees after the holiday in 2019.
According to Figure~\ref{fig:p_c_2020}, the rehabilitation was in progress, but much slower during the epidemic mitigated period.
The remarkable difference inferred from Figure~\ref{fig:p_c_dif} verifies this sustained impact.

\subsection{Estimation of Pandemic Impact \textit{via} DID Analysis}\label{sec:did}
We applied DID (Difference-In-Differences) technique to \textit{quantify} the impact of the pandemic by removing the influence of the Chinese New Year, weather factors, and the common trend of share bike use.
We combined \textbf{phase b} and \textbf{c} in our DID model because the pandemic lasted at least to the end of our study period.

Equation~\eqref{did_eq} in Section~\ref{sec:did_method} was applied to fit the share bike usage and weather data in 62 days from 01 Jan of 2019 and 2020, respectively. 
We use the year 2020 as the treatment group and 2019 as the control group (no pandemic).
$\mathit{Before_{2020,t}}=1$ if $t$ is between Jan 1 to Jan 10, 2020.
$\mathit{During_{2020,t}}=1$ if $t$ is in \textbf{phase b} or \textbf{c}.
The date range for $\mathit{Before_{2020,t}}=1$ is set to be unequal to \textbf{phase a} to avoid multi-collinearity problem.
Otherwise, as the two variables cover all the study period, $\mathit{Before_{2020,t}} = 1- \mathit{After_{2020,t}}$, they are linearly dependent, which cripples DID analysis.

\begin{table}[ht]\small
\resizebox{\textwidth}{15mm}{
    \begin{tabular}{|*{11}{l|}}
        \hline
        &Overall&RA&HC&OC&SS&SP&SM&TH&IRA&SRA\\
        \hline
        \multirow{2}{*}{$\beta_1$}  &0.033 & 0.011 & 0.051 & 0.023 & 0.027 & 0.005 & -0.01 & 0.017& -0.066 & -0.074\\
                                    &(0.066) & (0.069) & (0.114) & (0.1) & (0.084) & (0.166) & (0.083) & (0.067) & (0.076) & (0.067)\\
        \hline
        \multirow{2}{*}{$\beta_2$}  &-1.044 & -0.889 & -1.355 & -1.183 & -1.51 & -1.331 & -0.985 & -0.886 & -0.824 & -0.874\\
                                    &(0.125) & (0.136) & (0.214) & (0.189) & (0.156) & (0.249) & (0.143) & (0.165) & (0.136) & (0.137)\\
        \hline
        $R^2$&0.921 & 0.906 & 0.894 & 0.892 & 0.911 & 0.687 & 0.824 & 0.916 & 0.884 & 0.902\\
        \hline
        {\scriptsize $1\!-\!e^{\beta_2}$}&64.80\% & 58.89\% & 74.21\% & 69.36\% & 77.91\% & 73.58\% & 62.66\% & 58.77\% & 56.13\% & 58.27\%\\
        \hline
    \end{tabular}}
\caption[DID analysis]{The net effects of COVID-19 on share bike usage obtained \textit{via} DID, with all the value of $\mathit{During_{2020}}(\beta_2)$ at 5\% significance level. The result is in the form of $\bar{x}(\sigma)$}\label{tab:did}
\end{table}

Table~\ref{tab:did} shows the regression result of Equation~\eqref{did_eq}. 
We ignored the constant term $\alpha$ in the whole analysis because we are more interested in the \textit{change} of the share bike usage which reflects the human mobility of citizens. 
$|\beta_1|$ is small enough, suggesting that the effects of Chinese New Year and that of the share bike usage trend are well absorbed by $\theta$.
$\beta_2$ is negative and statistically significant, suggesting that the pandemic reduced the human mobility when compared to 2019.
We estimated the percentage of share bike usage decrease due to the net impact of the pandemic \textit{via} $1-\exp(\beta_2)$ (see Appendix~\ref{sec:pandemic_effect}).
According to Table~\ref{tab:did}, we found the coefficient $\beta_2$ for all the POI categories to be -1.044, which implied that the pandemic reduces the overall bike usage by 64.8\%. 
This percentage is slightly lower than 69.49\% estimated for Wuhan (the outbreak zone), as reported by Fang \textit{et al.}\ \cite{Fang2020human}. 
Mu \textit{et al.}\ \cite{mu2020interplay} estimated the intra-city mobility reduction in Beijing between 56.5\% and 65.2\%, which is consistent with our result (64.8\%).

The same analysis was conducted with the seven POI categories.
With all $\beta_2$ values at a 5\% significance level, we considered the COVID-19 has a negative impact on mobility close/belonging to these types of urban functional areas. 
The main reason could be attributed to the quarantine restrictions.
The estimated mobility reduction \textit{due to the pandemic} of SS (77.91\%), HC (74.21\%), and SP (73.58\%) are higher than that of other categories. 

\section{Co-location Analysis between Share Bike Usage and POIs}\label{sec:discussions}

To further explain the difference in the estimated mobility reduction of various urban functions above, we consequently studied the relationships between the locations of different types of POIs and adjacent share bike usage with co-location analysis.

\subsection{Mapping POIs with Share Bikes}

\begin{figure}[ht]
    \centering
    \begin{subfigure}{.3\textwidth}
        \begin{tikzpicture}[inner sep = 0pt]
            \mainBasic{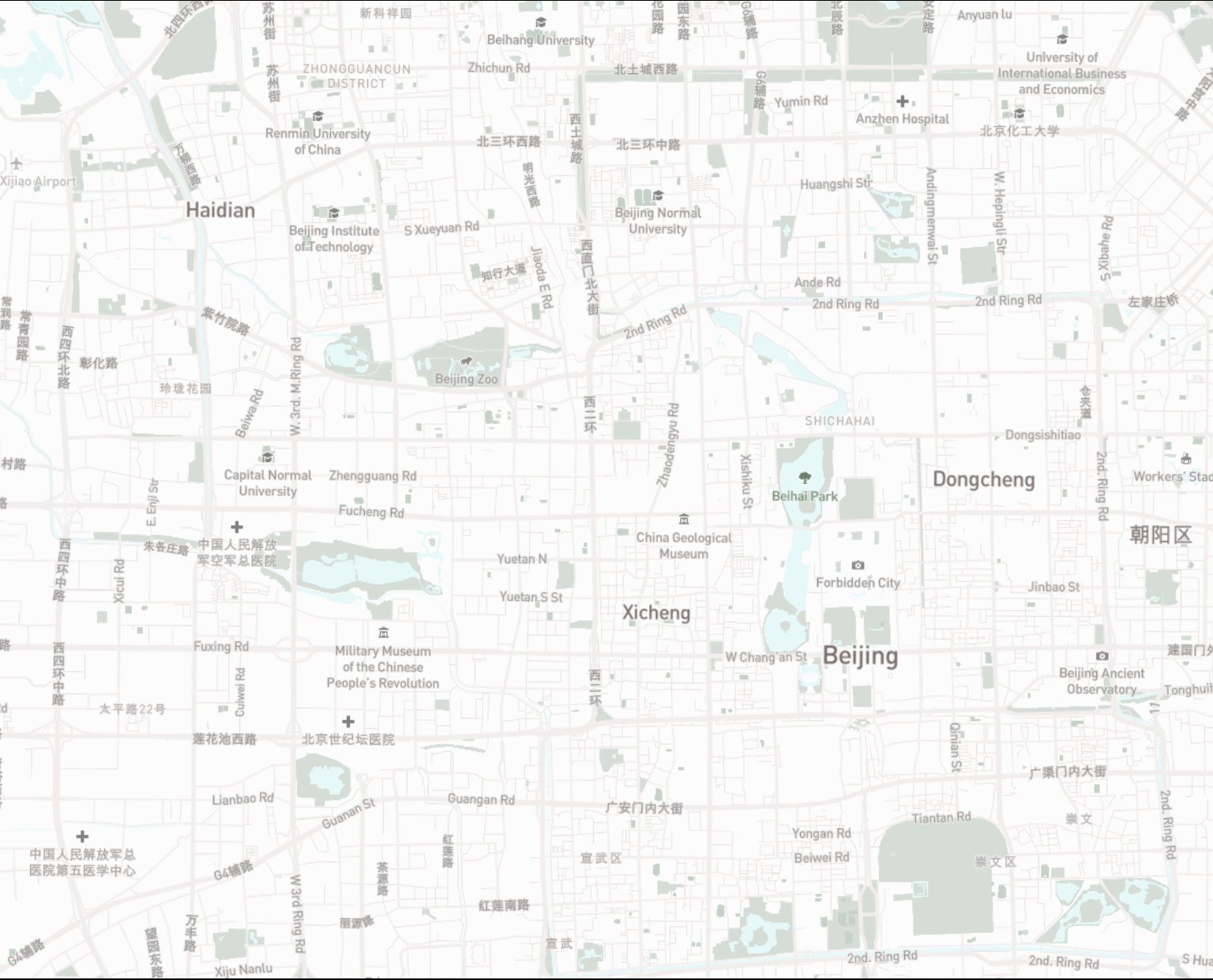}
            \node[opacity=0.5] at (a) {\includegraphics[width=\textwidth]{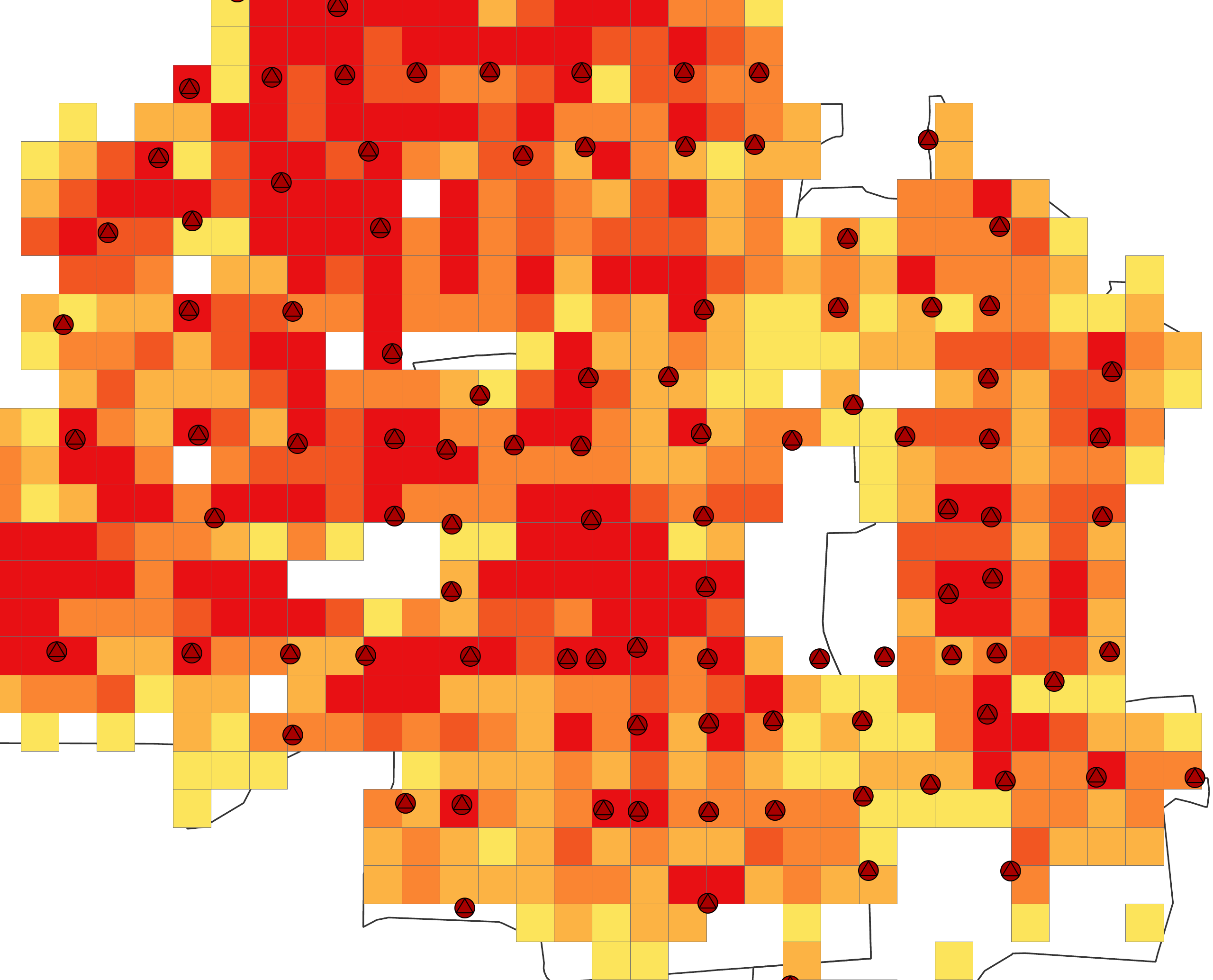}};
            \legendBar{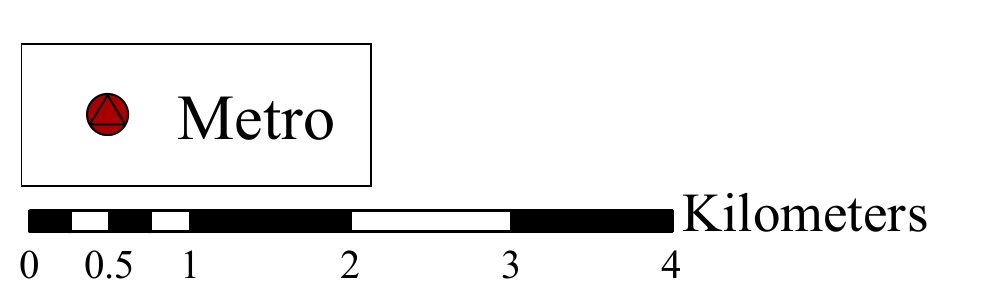}
        \end{tikzpicture}
        \caption{16 Jan}\label{fig:BSS_metro_c_16_Jan}
    \end{subfigure}
    \begin{subfigure}{.3\textwidth}
        \begin{tikzpicture}[inner sep = 0pt]
            \mainBasic{Appendix_basic.png}
            \node[opacity=0.5] at (a) {\includegraphics[width=\textwidth]{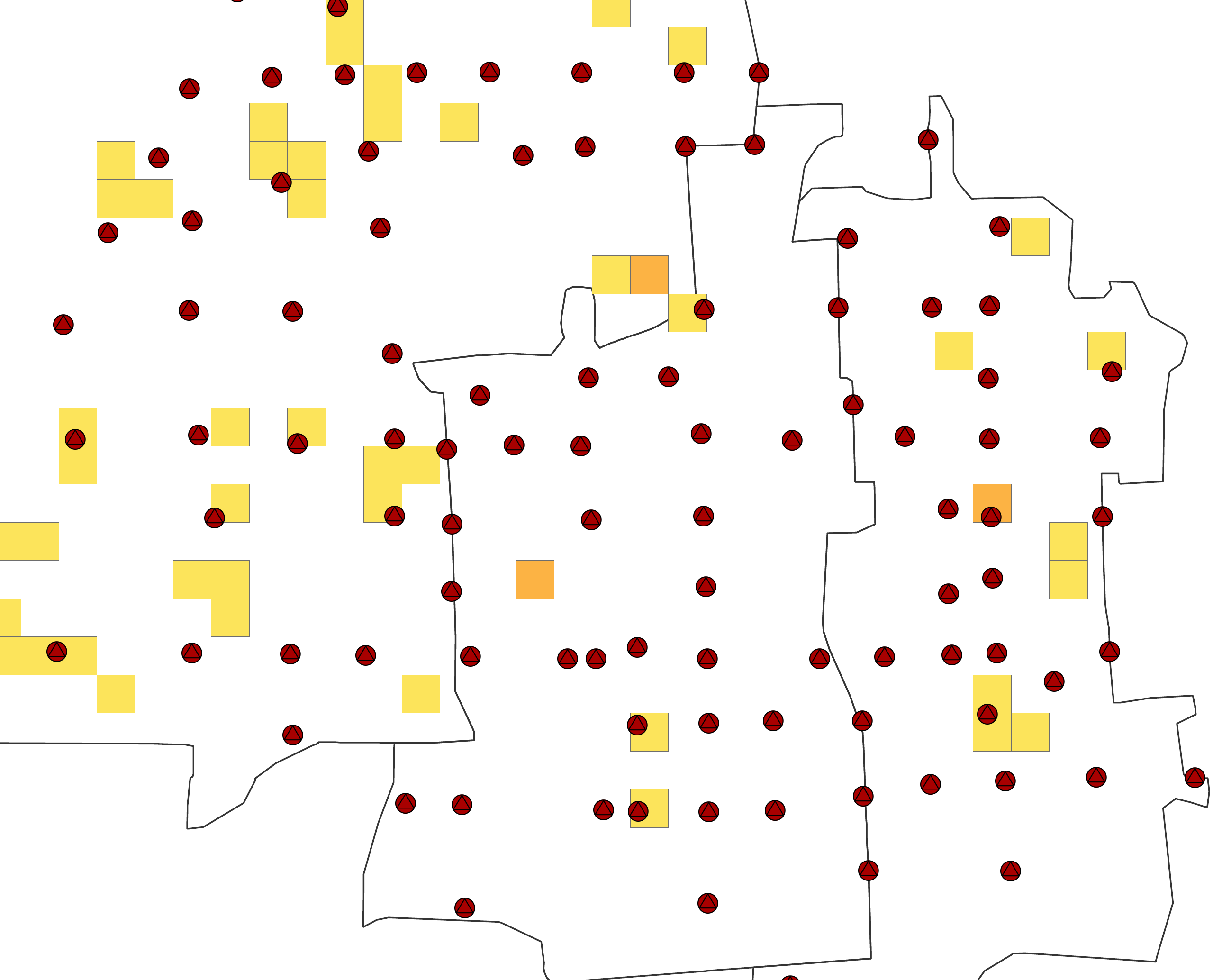}};
            \legendBar{Metro_legend.pdf}
        \end{tikzpicture}
        \caption{15 Feb}
    \end{subfigure}
    \begin{subfigure}{.3\textwidth}
        \begin{tikzpicture}[inner sep = 0pt]
            \mainBasic{Appendix_basic.png}
            \node[opacity=0.5] at (a) {\includegraphics[width=\textwidth]{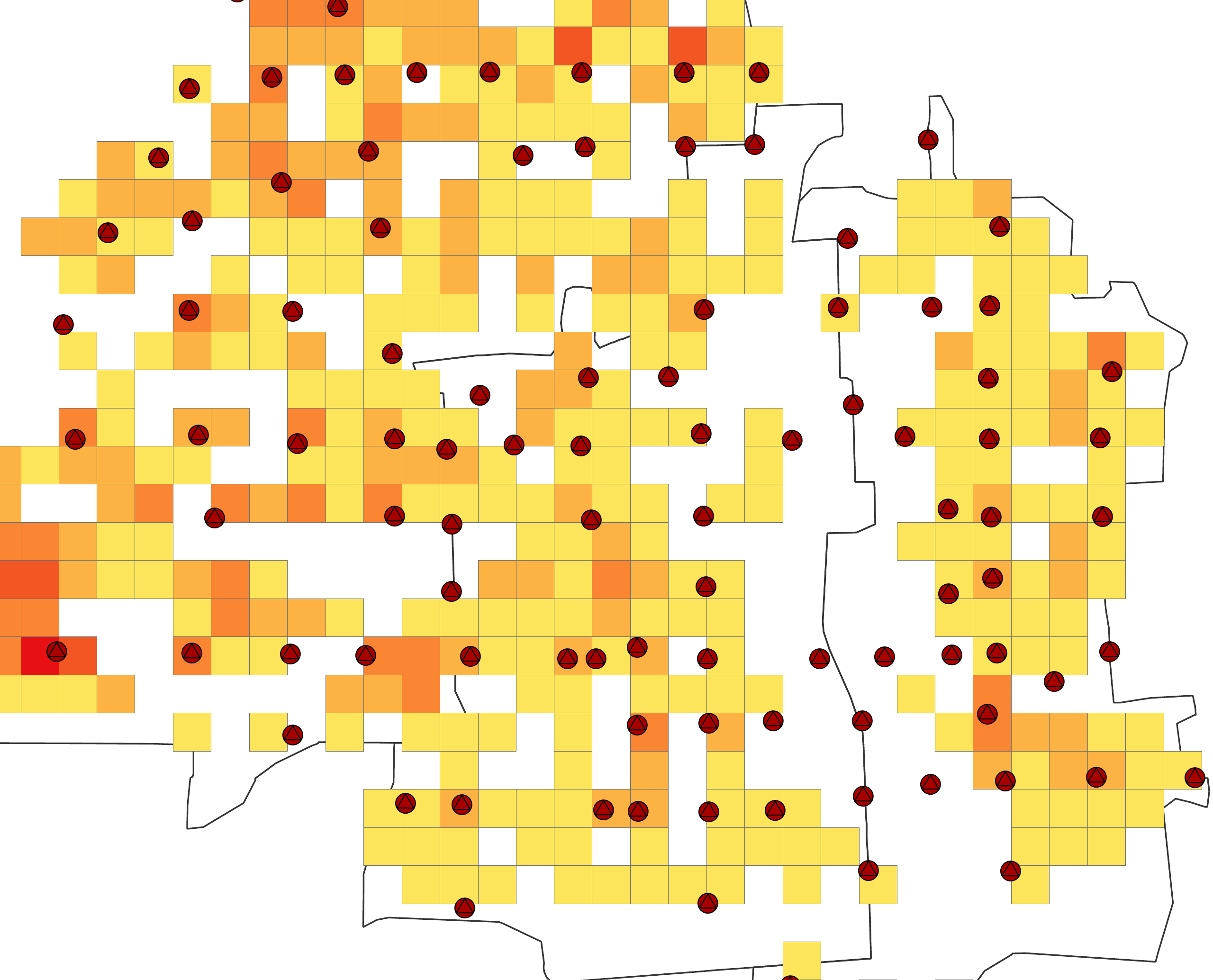}};
            \legendBar{Metro_legend.pdf}
        \end{tikzpicture}
        \caption{02 Mar}
        \label{fig:BSS_metro_c}
    \end{subfigure}

    \begin{subfigure}{.7\textwidth}
        \includegraphics[width=\textwidth]{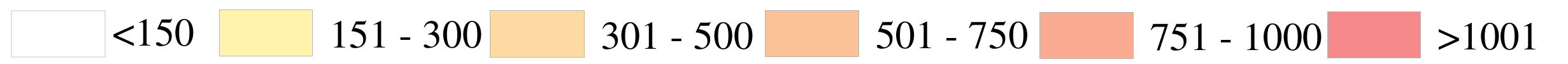}
    \end{subfigure}
    \caption{Share bikes and subway stations (SS).}
    \label{fig:BSS_metro}
\end{figure}

\begin{figure}[ht]
    \centering
    \begin{subfigure}{.3\textwidth}
        \begin{tikzpicture}[inner sep = 0pt]
            \mainBasic{Appendix_basic.png}
            \node[opacity=0.5] at (a) {\includegraphics[width=\textwidth]{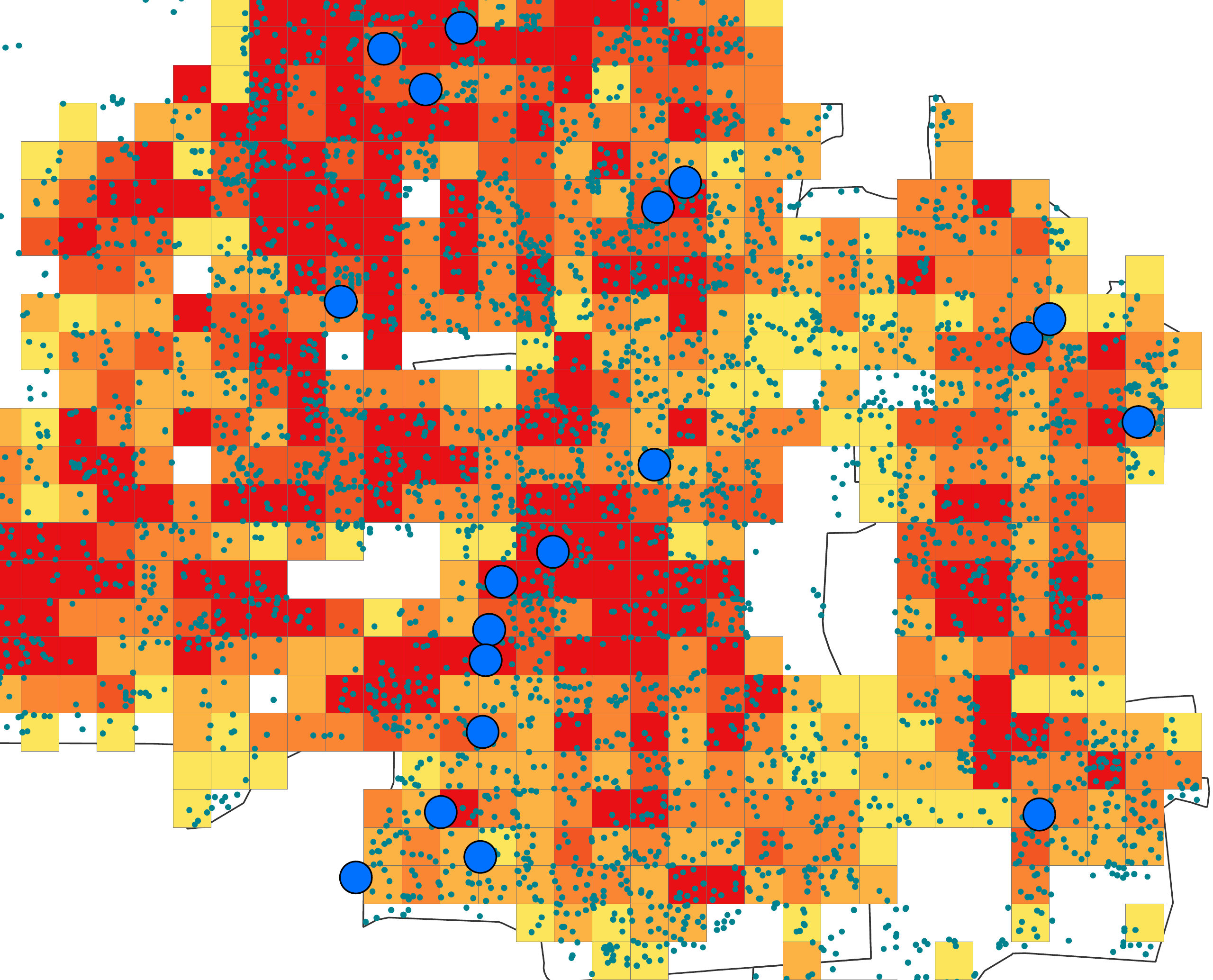}};
            \legendBar{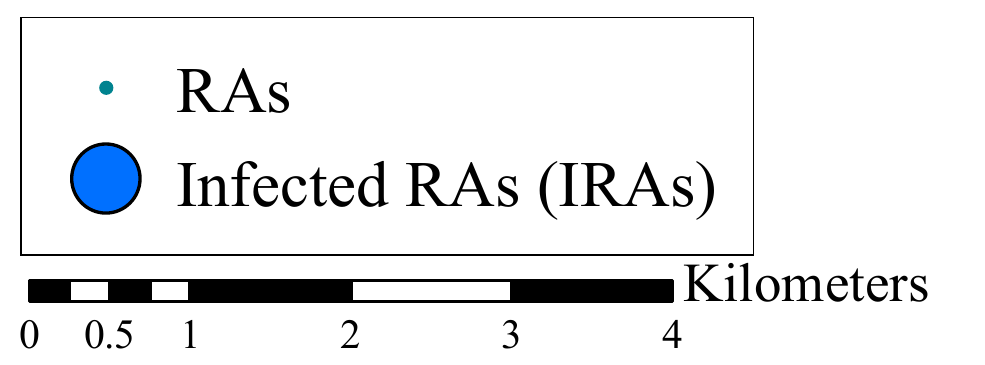}
        \end{tikzpicture}
        \caption{16 Jan}\label{fig:BSS_communities_16_Jan}
    \end{subfigure}
    \begin{subfigure}{.3\textwidth}
        \begin{tikzpicture}[inner sep = 0pt]
            \mainBasic{Appendix_basic.png}
            \node[opacity=0.5] at (a) {\includegraphics[width=\textwidth]{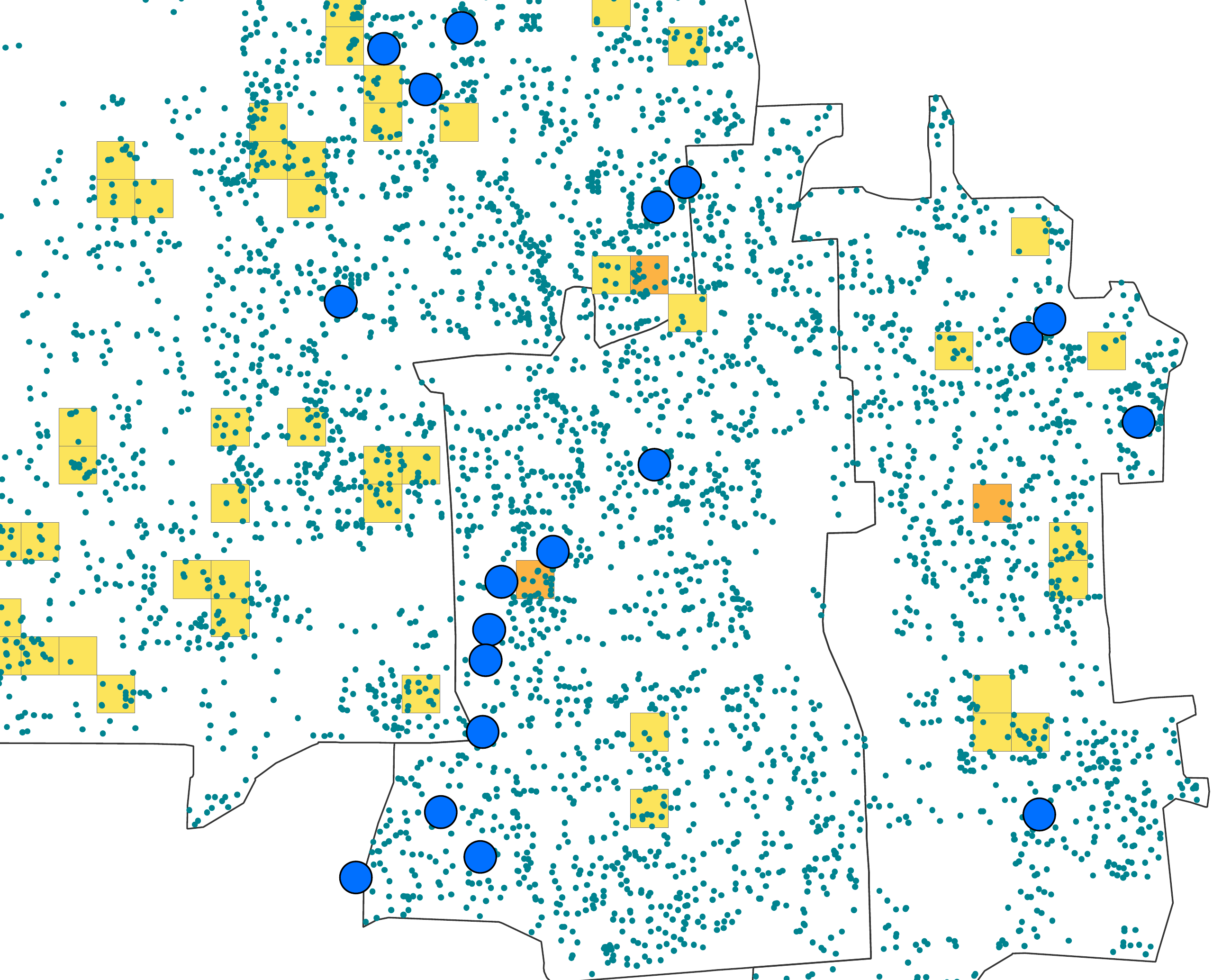}};
            \legendBar{RA_legend.pdf}
        \end{tikzpicture}
        \caption{15 Feb}
    \end{subfigure}
    \begin{subfigure}{.3\textwidth}
        \begin{tikzpicture}[inner sep = 0pt]
            \mainBasic{Appendix_basic.png}
            \node[opacity=0.5] at (a) {\includegraphics[width=\textwidth]{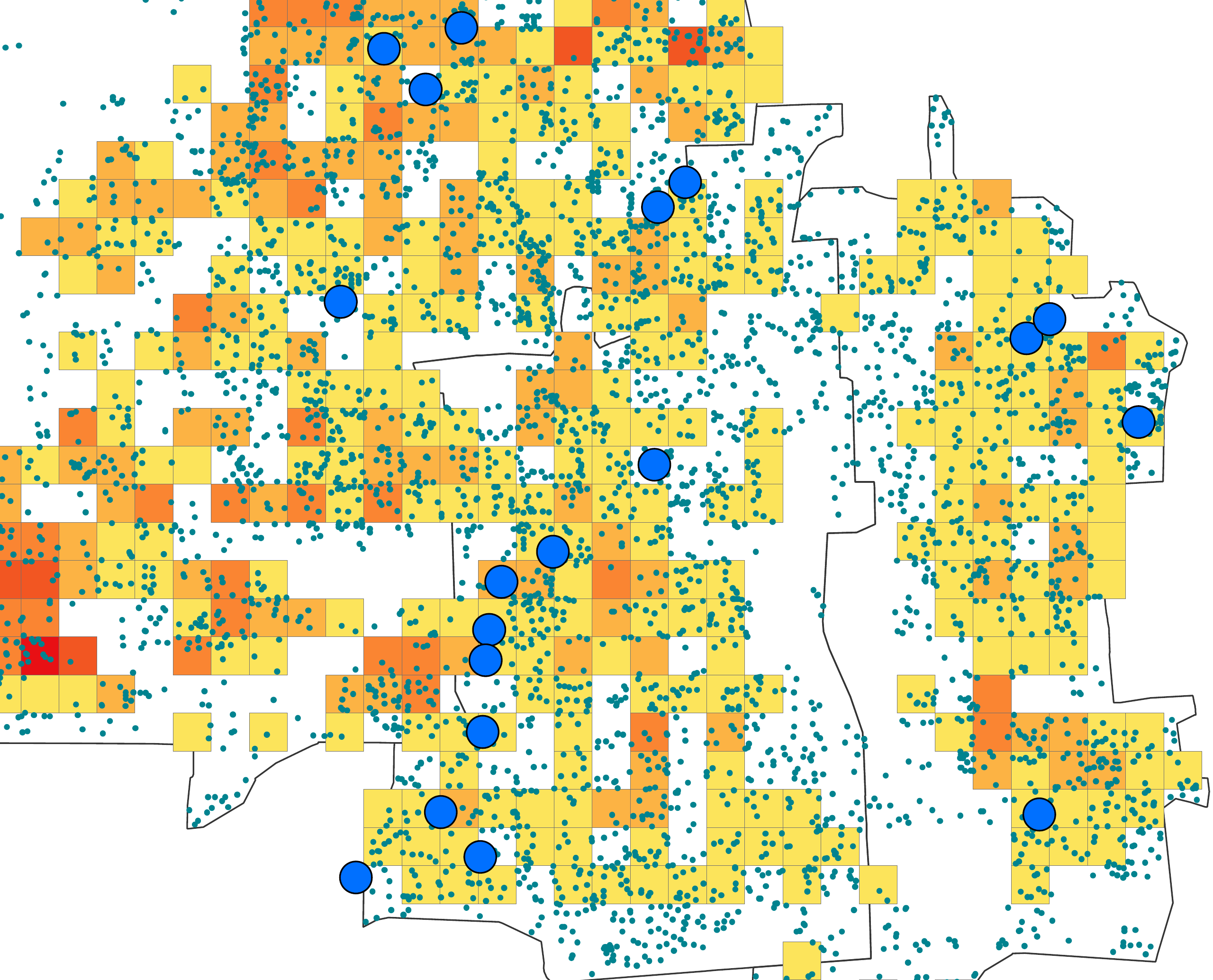}};
            \legendBar{RA_legend.pdf}
        \end{tikzpicture}
        \caption{02 Mar}
    \end{subfigure}

    \begin{subfigure}{.7\textwidth}
        \includegraphics[width=\textwidth]{Figures/AppendixLegend2.pdf}
    \end{subfigure}
    \caption{Share bikes and residential areas (RAs).}
    \label{fig:BSS_communities}
\end{figure}

Figure~\ref{fig:BSS_metro}-\ref{fig:BSS_communities} take 16 Jan, 15 Feb, and 02 Mar as the profiles of \textbf{phase a, b, c}, showing the position of our selected POI categories and the share bike usage in different pandemic phases.
We selected SS and RA as typical POI categories for the following reasons:
\begin{itemize}
    \item Share bikes fulfill the trip of ``first/last 1 km to/from subways'' in ordinary days. 
    \item However, during the pandemic, the flow of public transit was strictly limited and people were afraid of being infected, which may cause a drastical decrease in subway usage.
    \item Moreover, to suppress disease transmission, intervention strategies including work restriction, home quarantine, and facility closure were imposed and kept people staying at home, resulting in mobility gathering around residential areas.
\end{itemize}

As shown in Figure~\ref{fig:BSS_metro}, SS were always surrounded by massive share bikes.
Although a drastic decrease in share bike usage can be observed during the quarantine period, some share bike usage remained around subway stations, which is consistent with Table~\ref{tab:bike_usage_poi}.
According to figures corresponding to 02 Mar, the mobility started to recover after the pandemic was mitigated.
As can be seen in Figure~\ref{fig:BSS_communities}, the infected RAs (IRAs) were mostly situated in the city center on 16 Jan.
On 15 Feb, the share bike usage around these IRAs was limited.
On 02 Mar, share bike usage has recovered nearly uniformly in the whole city, including IRAs and their surrounding areas.

To confirm the above observation, we performed a categorized co-location analysis with POIs as follows.

\subsection{Estimating Pandemic Imapct and Rehabilitation Progress with Categorized Co-location Analysis}\label{sec:colo-poi}

As different \textit{POI categories} may co-locate with each other, \textit{e.g.} RA and SM.
To distinguish the type of share bike usage, we study the critical time slots, \textit{i.e.}, the main share bike use of each POI category.
For instance, we selected 8:00-10:00 as the critical time slot for HC and OC.

Table~\ref{tab:bike_usage_poi} shows the share bike usage per day of around each POI category with displacement of less than 100 m during its critical time slot, where the first three rows are cluster-based co-location, and the rest are point-based co-location.
$U_a$, $U_b$, $U_c$ stand for the share bike usage during \textbf{phase a, b, c}.
Two associated ratios measure the extent to which the COVID-19 epidemic has led: $U_b/U_a$ quantifies the decline in share bike usage during the Chinese New Year, which coincides with the quarantine period; $U_c/U_a$ reflects the recovery progress (\% of pre-pandemic).

\begin{table}[ht]\small
    \centering
    \begin{tabular}{|*{8}{l|}}
        \hline
        \multirow{2}{*}{\makecell[c]{POI\\category}} &\multirow{2}{*}{\makecell[c]{Time \\slot}}& \multirow{2}{*}{\makecell[c]{\#POI \\(\#cluster)}} & \multicolumn{3}{c|}{Bike usage per POI($\times 10^3$)}&\multicolumn{2}{c|}{Ratios}\\
        \cline{4-8}
        & & & $U_a$ & $U_b$ & $U_c$ & $U_b/U_a$ & $U_c/U_a$\\
        \hline
        RA & all-day&$5657(1204)$ & $254.39\pm133.76$ & $61.65\pm35.18$ & $99.30\pm5.59$ & $24.2\%$ & $39.0\%$\\
        \hline
        HC &8-10h & $3858(81)$ & $4.06\pm3.52$ & $0.42\pm0.56$ & $1.16\pm0.15$ & $10.4\%$ & $28.6\%$\\
        \hline
        OC &8-10h & $32301(577)$ & $32.69\pm26.68$ & $4.19\pm4.54$ & $9.95\pm1.11$ & $12.8\%$ & $30.4\%$\\
        \hline
        \hline
        SS &8-22h & $137$ & $34.07\pm21.57$ & $4.25\pm3.15$ & $8.33\pm0.61$ & $12.5\%$ & $24.5\%$\\
        \hline
        SP &18-20h & $217$ & $3.70\pm2.01$ & $0.79\pm0.62$ & $1.38\pm0.15$ & $21.3\%$ & $37.2\%$\\
        \hline
        SM &18-20h & $1076$ & $13.59\pm7.63$ & $3.43\pm2.44$ & $6.27\pm0.58$ & $25.2\%$ & $46.1\%$\\
        \hline
        TH &all-day & $86$ & $13.69\pm7.14$ & $3.38\pm1.79$ & $5.00\pm0.19$ & $24.7\%$ & $36.5\%$\\
        \hline
        Overall& all-day & \hrulefill & $547.32\pm301.74$ & $124.02\pm66.31$ & $195.75\pm10.43$ & $22.7\%$ & $35.8\%$\\
        \hline
    \end{tabular}
    \caption[Co-location analysis of different POI categories]{Bike usage in \textbf{phase a, b, c} around the chosen POIs, shown in the form of $\bar{x}\pm2\sigma$ (see abbreviations in Section \ref{subsec:dataset}).}
    \label{tab:bike_usage_poi}
\end{table}

$U_b/U_a$ shows that the bike usage of all categories decreased to one quarter of that before the pandemic, and HC dropped the most in both values (from 4.06 to 0.42) and proportion (10.4\%).
One possible explanation is that the staff in HC suit best ``work from home''.
During this period, the intensive intra-city public traffic (\textit{e.g.}, buses) was restricted to avoid contact infection, and share bikes is almost the only approach to fulfill short-term travel demand under this situation. 
$U_b/U_a$ of SM (25.2\%), TH (24.7\%) and RA (24.2\%) are higher than that in other categories. 
One possible explanation is: for residents, the need for essential goods (\textit{e.g.}, food, medicine) cannot be reduced, even though outdoor activities were voluntarily reduced during the quarantine.

$U_c/U_a$ shows that after the partial restart of productive and social activities, the bike usage of all categories has recovered to some extent but far from the situation of usual time.
The average increase of the seven chosen categories is 15.9\% compared to \textbf{phase b}.
Partial restart led to a slight increase in the daily average share bike usage near HC and OC. 
$U_c/U_a$ being 28.6\% for HC and 30.4\% for OC suggests that there is still a long way from full recovery.
During the mitigation period, SM has the highest recovery level, with the $U_c/U_a$ up to 46.1\%, because it served as the main subsistence suppliers.

In conclusion, HC and OC were greatly impacted because the beginning of the pandemic coincides with the Chinese New Year and people stopped working.
SS was influenced to the same degree due to travel limitations and the fear of COVID-19. 

Even if people had already stored sufficient supplies (usually the food storage for one week or longer) to prepare for the Chinese New Year, they were running out of storage and had to procure their necessities.
Thus, RA, SM were less impacted after one week of strict quarantine.
SP was more influenced than RA and SM because the goods provided by SP are mainly not necessities.
TH has the smallest $U_c/U_a - U_b/U_a$ value among the POIs, suggesting THs were providing continuous services to the public during the pandemic.

In Figure~\ref{fig:study_area}, we noticed that being the center of Beijing, Dongcheng and Xicheng district contained most of the IRAs, and the confirmed time lies between 05 and 12 Feb.
We carried out the same analysis on the sub-categories of RAs to assess the impact of confirmed cases, with results in Table~\ref{tab:bike_usage_ras}.
The IRAs and surrounding RAs (SRA, the RAs surrounding IRAs) are \textit{surprisingly} not the most affected category, implying that residents were indifferently in panic, and limited their travel purpose to their basic needs.
According to the ratio $U_c/U_a$, the share bike usage of IRAs (39.9\%) and their SRAs (39.0\%) have recovered roughly to the same level compared to other RAs (39.0\%) from the pandemic.

\begin{table}[ht]
    \centering
    \begin{tabular}{|*{8}{l|}}
        \hline
        \multirow{2}{*}{\makecell[c]{POI\\category}} &\multirow{2}{*}{\makecell[c]{Time \\slot}}& \multirow{2}{*}{\makecell[c]{\#POI}} & \multicolumn{3}{c|}{Bike usage per POI($\times 10^3$)}&\multicolumn{2}{c|}{Ratios}\\
        \cline{4-8}
        & & & $U_a$ & $U_b$ & $U_c$ & $U_b/U_a$ & $U_c/U_a$\\
        \hline
        IRA & all-day & $14$ & $5.86\pm2.95$ & $1.50\pm0.82$ & $2.34\pm0.30$ & $25.5\%$ & $39.9\%$\\
        \hline
        SRA & all-day & $169$ & $34.28\pm17.32$ & $8.30\pm4.77$ & $13.38\pm1.33$ & $24.2\%$ & $39.0\%$\\
        \hline
        RA & all-day & $5657$ & $254.39\pm133.76$ & $61.65\pm35.18$ & $99.30\pm5.59$ & $24.2\%$ & $39.0\%$\\
        \hline
    \end{tabular}
    \caption[Co-location analysis of residential areas]{Bike usage in \textbf{phase a, b, c} around different RAs, shown in the form of $\bar{x}\pm2\sigma$. (see abbreviations in Section \ref{subsec:dataset})}
    \label{tab:bike_usage_ras}
\end{table}

\section{Conclusion}\label{sec:conclusion}
The wide-spread BSS is an alternative data source for characterizing the spatial and temporal details of the human mobility.
This paper introduces a human mobility analyzing framework aiming at extracting and analyzing the impact of the pandemic on the human mobility and the rehabilitation process at city-scale.
The results reveal the period-wise spatiotemporal characteristics and co-location patterns of human mobility presented by share bike usage before \& during the pandemic in Beijing, China.
Based on our analysis, we make the following conclusions:

\begin{enumerate}
    \item Apart from the timeline of the outbreak, we identified two key time nodes: 23 Jan and 24 Feb, and segmented the study period (two months) into three pandemic phases: \textbf{phase a} before the pandemic, \textbf{phase b} during pandemic, \textbf{phase c} when pandemic mitigated.
    \item We used DID model to assess \textit{quantitatively} the net impact of the pandemic on various aspects of daily life.
    The simple but effective configuration (we considered only pandemic weather, and holidays) suggests the effect of the pandemic is important enough to overwhelm other candidate factors.
    After removing the factors of the Chinese New Year holiday and weather, our results confirmed that the activities of residents were hugely affected by the COVID-19, reflected in the drastic decrease of share bike usage by 64.8\%. 
    Reductions in mobility close to Subway Station (77.91\%), High-tech Company (74.21\%), and Shopping Plaza (73.58\%) are higher than other POI categories, which could be attributed to quarantine restrictions and ``work from home''. 
    Residential areas, supermarkets, tertiary hospitals were less impacted as people have basic food supplies and health care needs.
    \item As the pandemic mitigated, an increase in human mobility was observed since 24 Feb. 
    The rehabilitation progress was assessed by a category-wise co-location analysis. 
    The average increase of share bike usage among the seven chosen POIs categories is 15.9\%, suggesting that the mobility belonging to these types of functional areas has recovered to some extent but far from the situation of usual time after the partial restart. 
    The increase of 18.2\% for High-tech Company and 17.6\% for Ordinary Company are higher than other categories.
    \item In addition, our results imply that the infected Residential Areas and other Residential Areas are not the most affected POI categories, and these two categories have recovered roughly to the same level (about 39.0\%) from the pandemic.
\end{enumerate}

The present case study was implemented in Beijing, a typical metropolis affected by COVID-19 other than the epidemic center.
As the situation develops globally, our results could be a \textit{generalized} reference to epidemiological research and policymaking in the context of the current COVID-19 outbreak and be helpful for pandemic emergency response in the future.

\section{Future Work}\label{sec:future}

Since the municipal government implemented multiple interventions simultaneously or in a short timeframe to control the outbreak, the effect of individual strategies could not be evaluated.
The multi-dimensional principal component analysis might be helpful in extracting the modes contributed by each preventive measure.

Also, our current work has extracted the features of mobility from BSS and POI.
Another possible extension is to build a dynamic model based on the features predicting the share bike usage and potential social event/reaction of the public during the pandemic or emergencies.
\medskip

\textbf{Acknowledgments:} This study was supported by the National Key R\&D Program of China (2017YFB0503700 and 2018YFB2100701), the Research Program of Beijing Advanced Innovation Center for Future Urban Design (UDC2019031321), The Pyramid Talent Training Project of Beijing University of Civil Engineering and Architecture (JDYC20200322), and the National Natural Science Foundation of China (41601389).
\medskip

\noindent\textbf{Abbreviations:}

\vspace{5pt}

\noindent 
\begin{tabular}{@{}ll}
BSS & Bike Sharing System\\
GIS & Geographic Information System \\
POI & Point Of Interest \\
VGI & Volunteered Geographical Information\\
\hline
\hline
RA & Residential Area\\
HC & High-tech Company\\
OC & Ordinary Company\\
SS & Subway Station\\
SP & Shopping Plaza\\
SM & Supermarket\\
TH & Tertiary Hospital\\
IRA & Infected RA\\
SRA & Surrounding RA
\end{tabular}

\bibliographystyle{plain}
\bibliography{bib}

\appendix

\section{Calculation of Net Pandemic Effect}\label{sec:pandemic_effect}
From Equation~\ref{did_eq}, in the pandemic phase, $\mathit{Before}_{2020}=0$, $\mathit{During}_{2020}=1$, we have 

$$U=\exp(\alpha+\theta_t+\epsilon_t)\cdot\exp(\beta_2)$$

Let $\exp(\alpha+\theta_t+\epsilon_t)=C$, we have $U=C\cdot\exp(\beta_2)$. 
The proportion of share bike usage reduction done by pandemic effect is 

$$\frac{C-U}{C}=1-\exp(\beta_2)$$
\section{Figures}\label{sec:figures}
\centering
\includegraphics[width=0.8\textwidth]{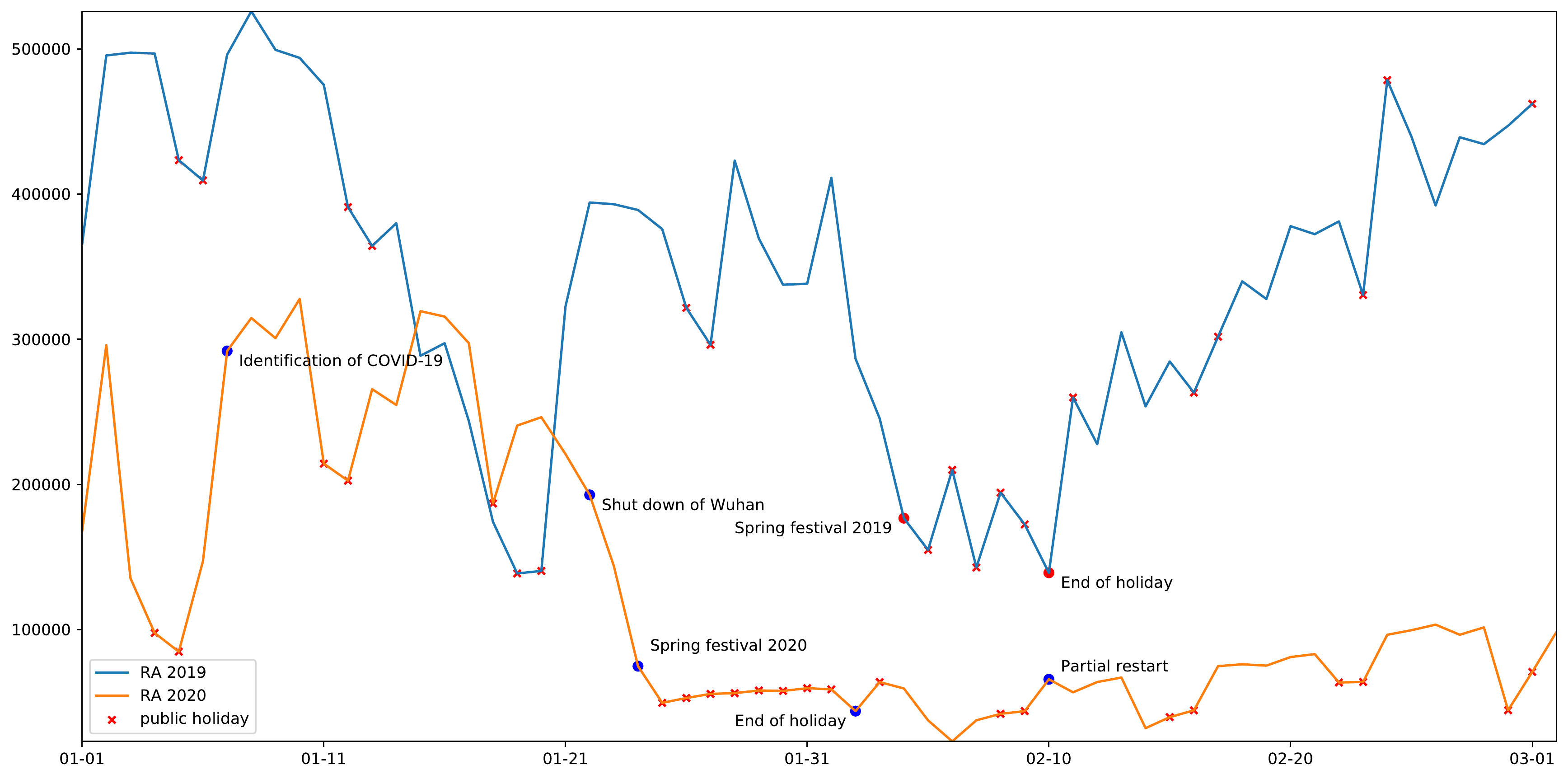}
\newgeometry{left=1cm, bottom=1cm, right=1cm, top=1cm, includehead, includefoot}
\begin{sidewaysfigure}[ht]
	\includegraphics[width=0.48\textwidth]{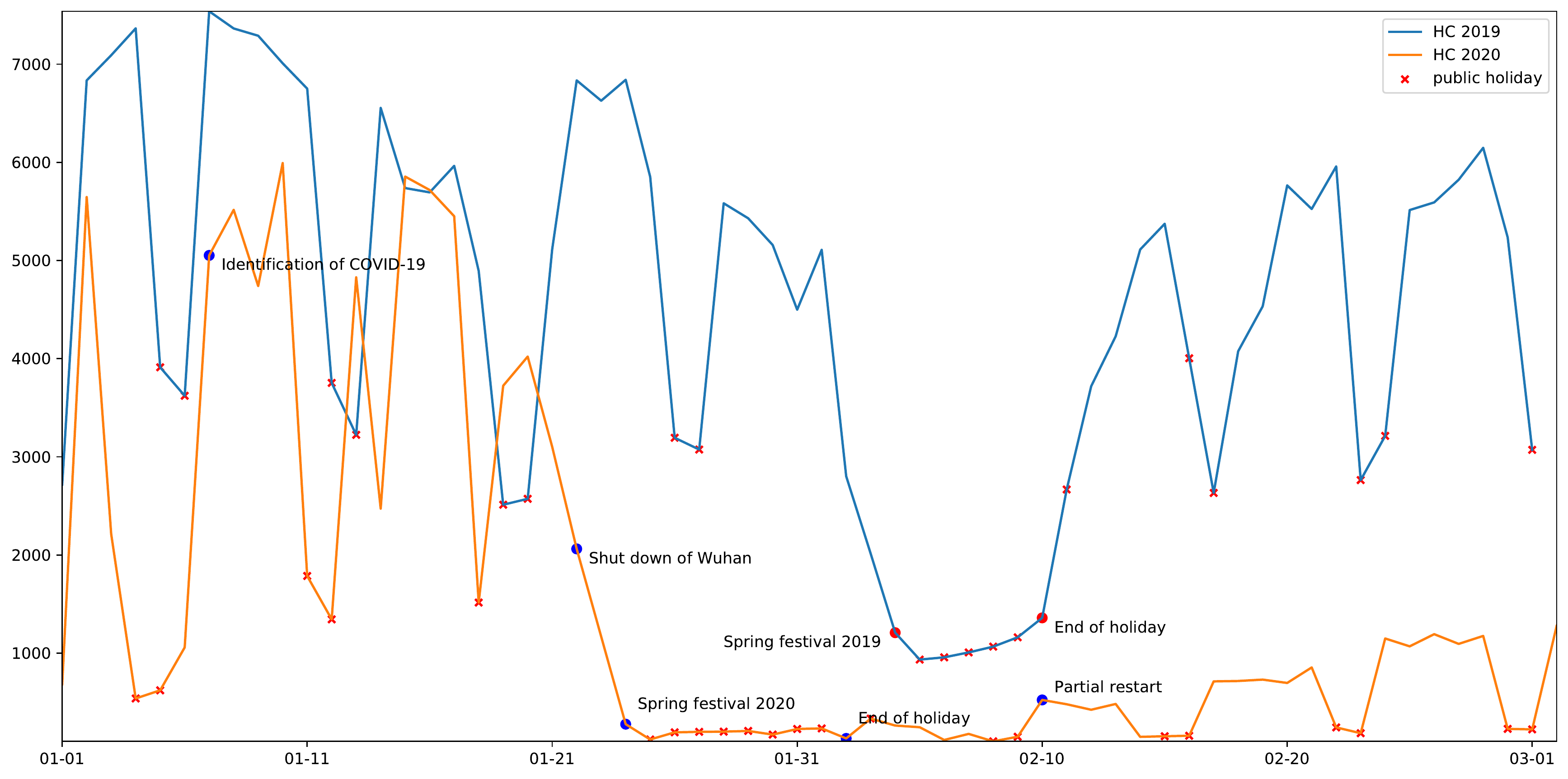}
	\includegraphics[width=0.48\textwidth]{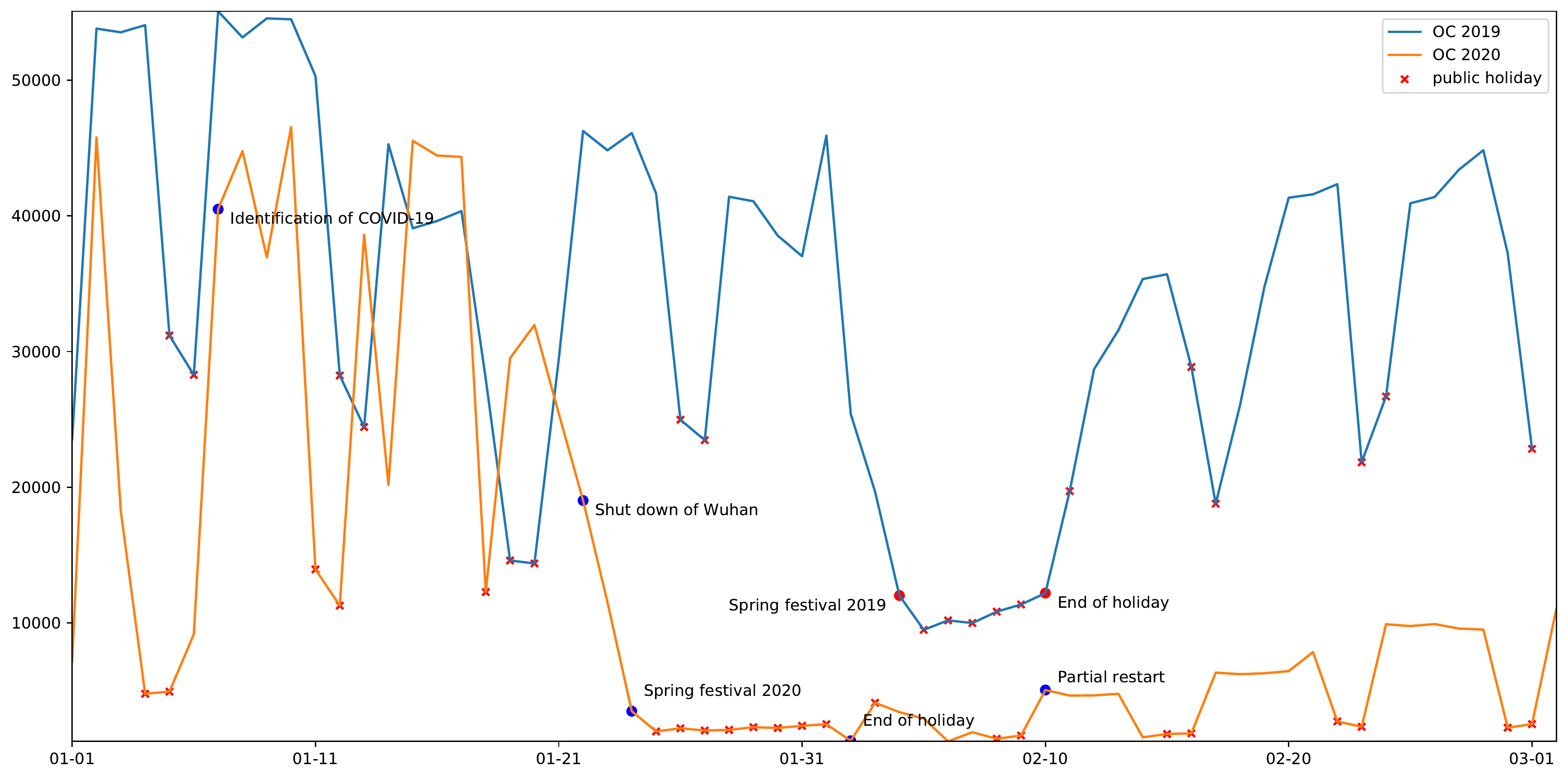}\\
	\includegraphics[width=0.48\textwidth]{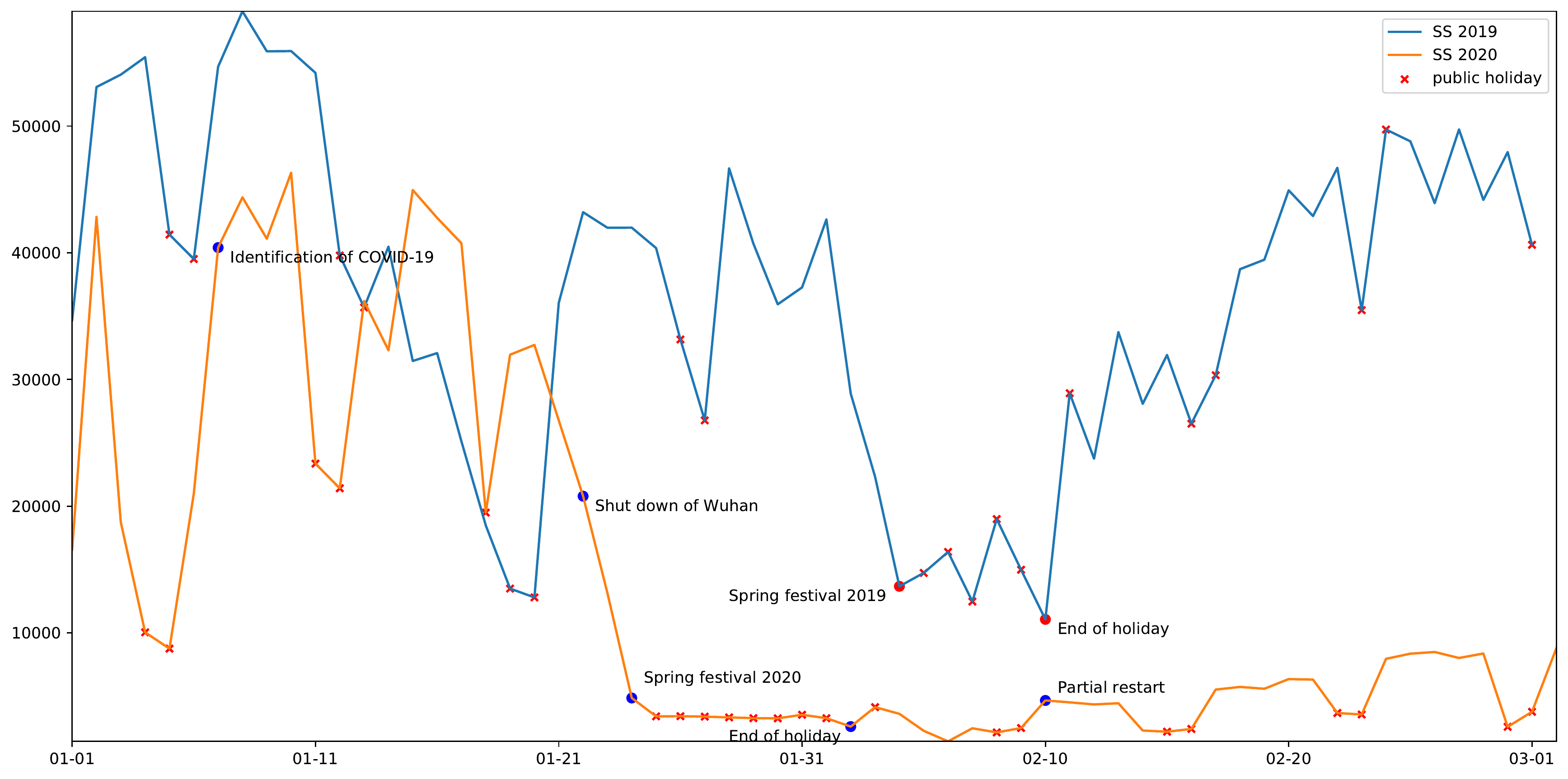}
	\includegraphics[width=0.48\textwidth]{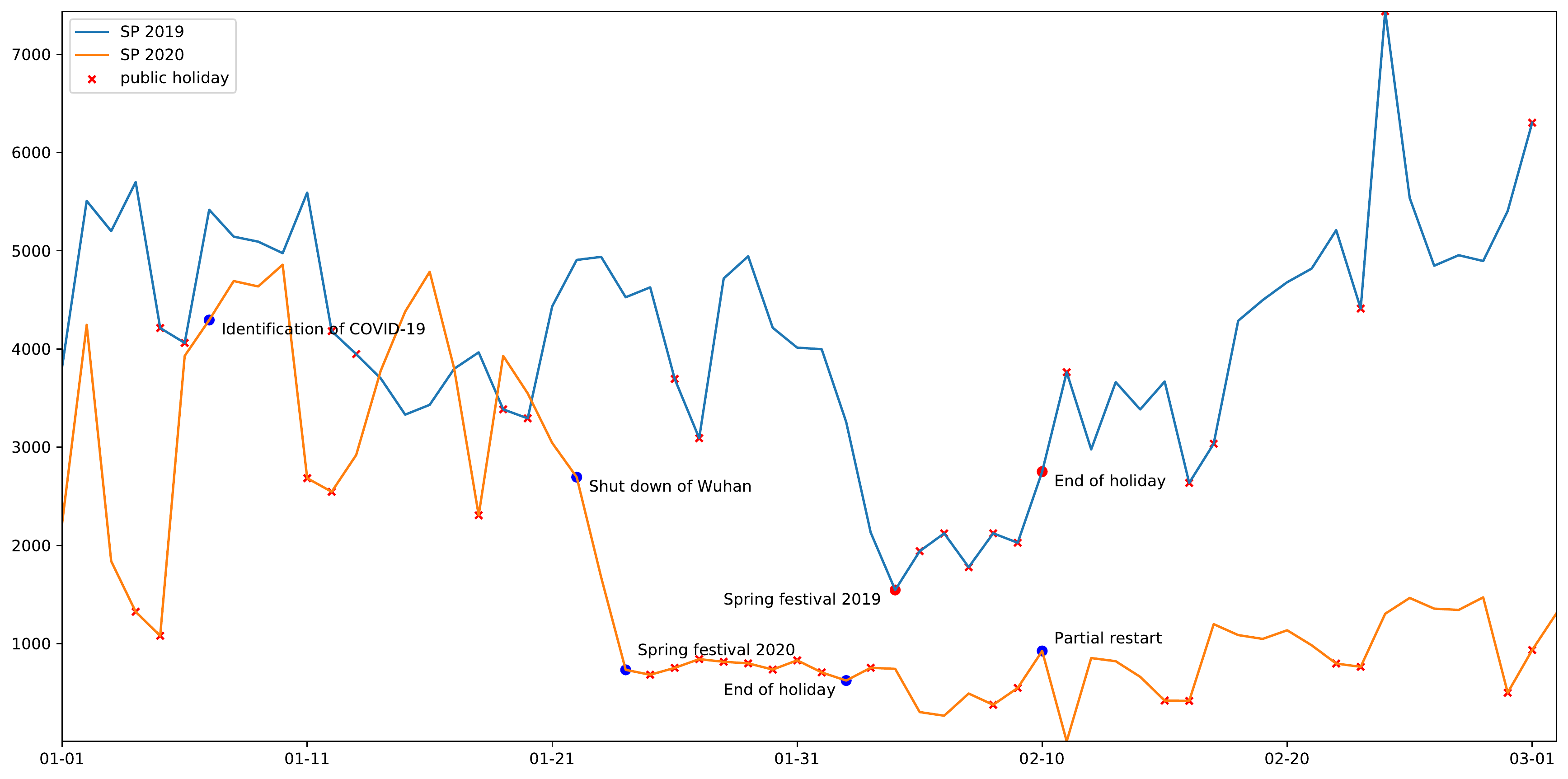}\\
	\includegraphics[width=0.48\textwidth]{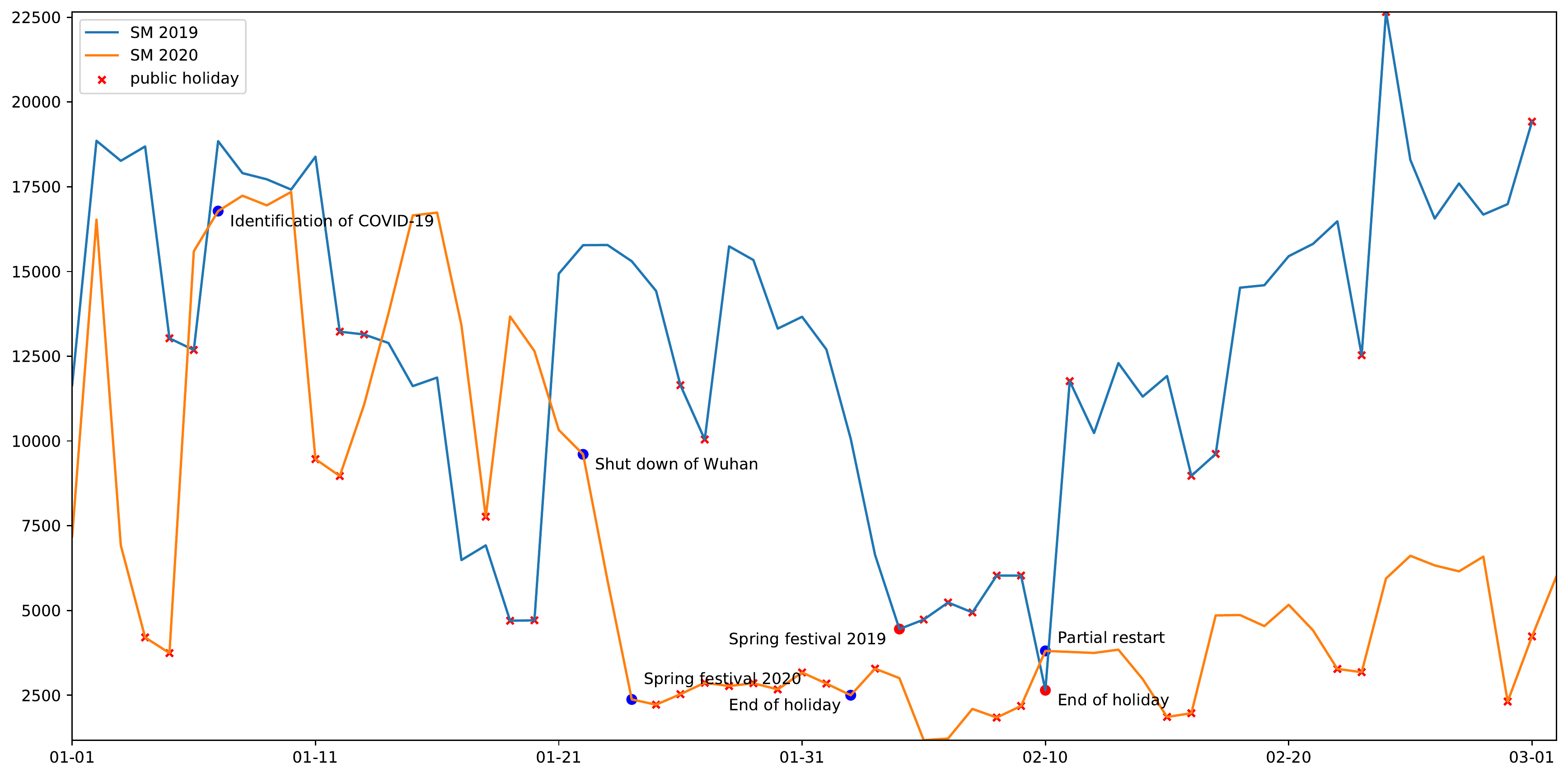}
	\includegraphics[width=0.48\textwidth]{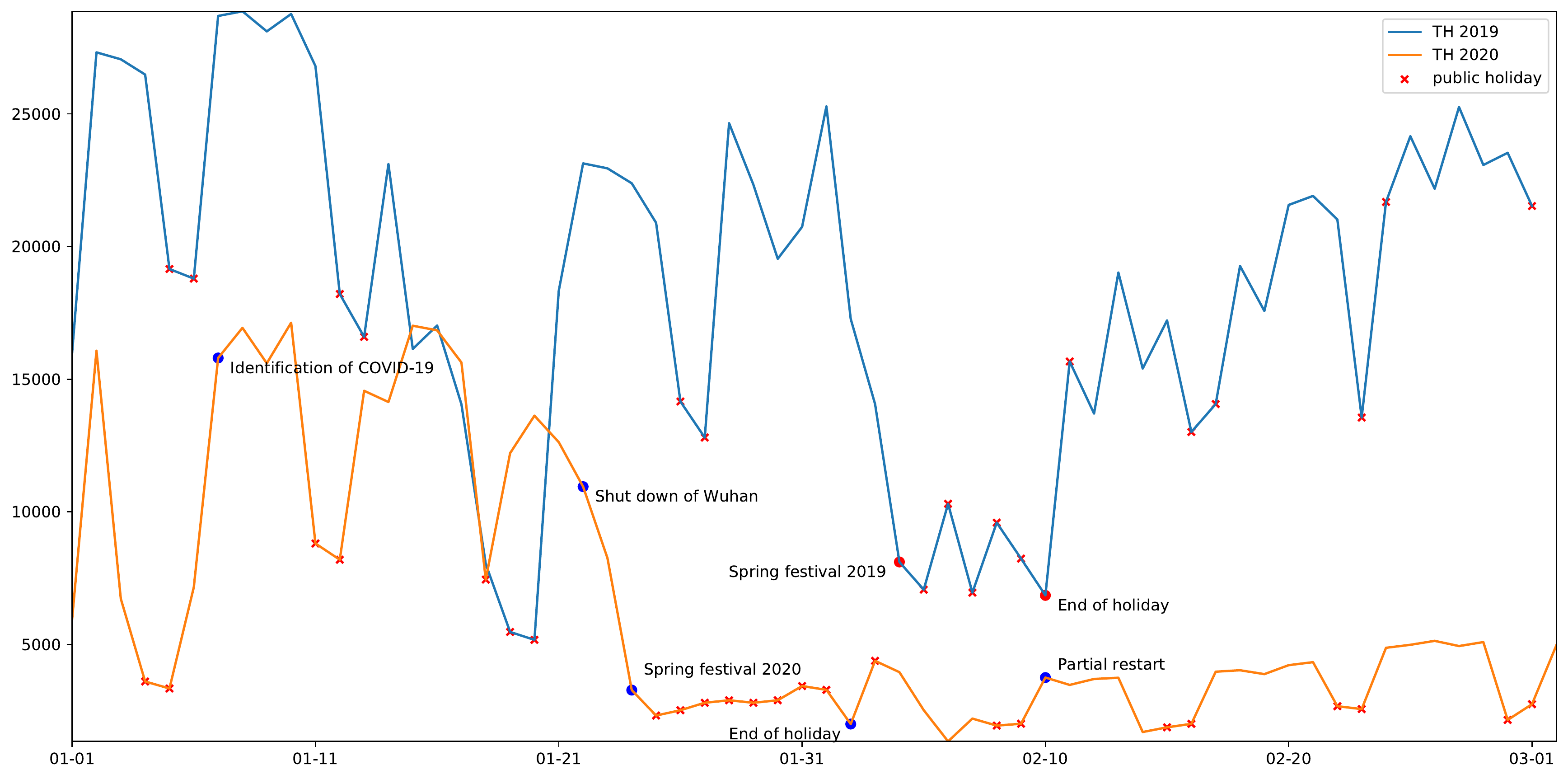}
	\caption{POI-wise evolution of share bike usage}\label{fig:poi_wise}
\end{sidewaysfigure}
\end{document}